\documentclass[english,12pt]{article}
\usepackage[T1]{fontenc}
\usepackage[utf8]{inputenc}

\usepackage{geometry}
\usepackage{epsfig}
\usepackage{cite}
\usepackage{color}
\usepackage{amsmath}
\usepackage{amssymb}
\usepackage{multirow}
\usepackage{bigdelim}
\usepackage{microtype}
\usepackage[linktocpage]{hyperref}
\usepackage{booktabs}
\usepackage{longtable}
\usepackage{soul}

\newcommand\sss{\mathchoice%
{\displaystyle}%
{\scriptstyle}%
{\scriptscriptstyle}%
{\scriptscriptstyle}%
}

\def\({\left(} 
\def\){\right)} 
\newcommand\pt{p_{\sss\rm T}}
\newcommand{\ii}{\mathord{\mathrm{i}}}
\newcommand{\VBFNLO}{\textsc{Vbfnlo}}
\newcommand{\msbar}{$\overline{\text{MS}}$}
\newcommand{\sect}[1]{Section~\ref{#1}}
\newcommand{\Openmpi}{\mbox{\textsc{Openmpi}}}

\newcommand{\rot}[1]{\rlap{\rotatebox{45}{#1}}\hspace*{1em}}

\makeatletter
\def\instring#1#2{TT\fi\begingroup
  \edef\x{\endgroup\noexpand\in@{#1}{#2}}\x\ifin@}
\makeatother
\newcommand{\makeflag}[3]{%
\if\instring{#1}{#3}{$\checkmark$}\else\if\instring{#2}{#3}{$\bigcirc$}\else{$-$}\fi\fi%
}
\newcommand{\bsmoptionsBLHA}[1]{%
\makeflag{B}{b}{#1} 
& 
\makeflag{L}{l}{#1} 
&
\makeflag{F}{f}{#1} 
&
\makeflag{V}{v}{#1} 
&
\makeflag{H}{h}{#1} 
&
\makeflag{T}{t}{#1} 
&
\makeflag{K}{k}{#1} 
&
\makeflag{S}{s}{#1} 
&
\makeflag{M}{m}{#1} 
}

\newcommand{\bsmgfoptionsBLHA}[1]{%
\makeflag{B}{b}{#1} 
&
\makeflag{G}{g}{#1} 
&
\makeflag{L}{l}{#1} 
&
\makeflag{H}{h}{#1} 
&
\makeflag{T}{t}{#1} 
&
\makeflag{M}{m}{#1} 
}

\begin{document}
\begin{titlepage}
\vspace{1cm}
\begin{center} {\Large \bf \textsc{Vbfnlo}: A parton level Monte Carlo
    } 
\end{center}
\begin{center}
 {\Large \bf  for processes with electroweak bosons --}
\end{center}
\begin{center}
 {\Large \bf  Manual for Version 3.0}
\end{center}
\vspace{0.6cm}
\begin{center}

{\renewcommand{\baselinestretch}{4}
J.~Baglio$^{1}$, 
F.~Campanario$^{2}$,
T.~Chen$^{3}$,
H.~Dietrich-Siebert$^{4}$,
T.~Figy$^{5}$,
M.~Kerner$^{4}$,
M.~Kubocz$^{6}$,
Duc Ninh Le$^{7}$,
M.~L\"oschner$^{8}$,
S.~Pl\"atzer$^{9,}$$^{10}$,
M.~Rauch$^{4}$,
I.~Rosario$^{2}$,
R.~Roth$^{4}$,
D.~Zeppenfeld$^{4}$
}
\end{center}
\vspace{0.4cm}
\begin{center}
  $^{1}$ QuantumBasel Schorenweg 44B, CH-4144 Arlesheim, Switzerland\\ \noindent
  $^{2}$ Theory Division, IFIC, University of Valencia-CSIC, Parque Científico, C/Catedrático José Beltrán, 2, E-46980 Paterna, Spain \\ \noindent
  $^{3}$ IT Research Cyberinfrastructure, University of Delaware Newark, DE 19716, USA\\ \noindent
  $^{4}$ Institute for Theoretical Physics, Karlsruhe Institute of Technology (KIT), 76128 Karlsruhe, Germany \\ \noindent  
  $^{5}$ Department of Mathematics, Statistics and Physics, Wichita State University, 1845 Fairmount Street, Wichita, KS\ 67002, USA\\ \noindent
  $^{6}$ Institut f\"{u}r Theoretische Teilchenphysik und Kosmologie, RWTH Aachen University, 52056 Aachen, Germany  \\ \noindent
  $^{7}$ Phenikaa Institute for Advanced Study, Phenikaa University, Hanoi 12116, Vietnam \\ \noindent
  $^{8}$ Deutsches Elektronen-Synchrotron DESY, Notkestr.\ 85, 22607 Hamburg, Germany \\ \noindent
  $^{9}$ Institute of Physics, NAWI Graz, University of Graz, Universit\"atsplatz 5, A-8010 Graz, Austria \\ \noindent
  $^{10}$ Particle  Physics,  Faculty  of  Physics,  University  of  Vienna,  Austria\\ \noindent  
\noindent
\end{center}
\vspace{0.6cm}

\newpage

\begin{abstract}

\noindent
  \textsc{Vbfnlo} is a flexible parton level Monte Carlo program for the
  simulation of vector boson fusion (VBF), QCD-induced single and double
  vector boson production plus two jets, and double and triple vector
  boson production (plus jet) in hadronic collisions at next-to-leading
  order~(NLO) in the strong coupling constant, as well as Higgs boson plus
  two jet production via gluon fusion at the one-loop level.  For the new
  version -- \textsc{Version 3.0} -- several major enhancements have
  been included. An interface according to the Binoth Les Houches Accord (BLHA) has been
  added for all VBF and di/tri-boson processes including fully
  leptonic decays.
  For all dimension-8 operators affecting vector boson scattering
  (VBS) processes, a modified T-matrix unitarization procedure has been
  implemented. Several new production processes have been added, namely
  the VBS $Z\gamma jj$ and $\gamma \gamma jj$ processes at NLO, $\gamma \gamma jj $, $WWj$ and
  $ZZj$ production at NLO including the loop-induced gluon-fusion
  contributions and the gluon-fusion one-loop induced $\Phi jjj$ ($\Phi$ is a CP-even or CP-odd scalar boson) process at LO,
  retaining the full top-mass dependence. Finally, the code has been parallelized using \Openmpi.
\end{abstract}
\vspace{2 cm}
\today
\end{titlepage}

\newpage

\tableofcontents

\newpage

\section{\textsc{Introduction}}

The physics potential of the \textsc{Lhc} depends heavily on our ability to
provide accurate cross section predictions for both signal and background
processes. The latter are often generated by parton radiation from processes
with weak bosons in the final state. A precise description of such hard QCD
production processes is needed, as well as a method for simulating the
measurable hadronic final states. Reaching these goals requires next-to-leading
order~(NLO) QCD calculations presented in the form of parton level Monte
Carlo~(MC) generators, which are an efficient solution when it comes to final
states characterized by a high number of jets and/or identified particles. When
kinematic cuts are imposed, as is mandatory for processes involving QCD
radiation, analytic phase-space integration becomes impractical and
implementation of results in the form of Monte Carlo programs becomes the method
of choice. 

\textsc{Vbfnlo} is a flexible MC program for vector boson fusion (VBF) \footnote{The terminology VBF includes also 
the class of vector boson scattering processes.},
QCD-induced single and double vector boson production plus two jets, and double
and triple vector boson (plus jet) production processes at NLO QCD accuracy. Furthermore,
the electroweak corrections to Higgs boson production via VBF (which are of the
same order of magnitude as the QCD corrections in the experimentally accessible
regions of phase-space) have been included.  Since real emission processes are part
of the NLO cross sections, \textsc{Vbfnlo} provides the means to calculate
cross sections for the corresponding process with  one additional jet at leading
order (LO) in the strong coupling. In addition, the simulation of
$\cal{CP}$-even and $\cal{CP}$-odd Higgs boson production in gluon fusion,
associated with two additional jets, is implemented at LO QCD.  The full top-
and bottom-quark mass dependence of the one-loop contributions in the Standard
Model (SM), in the Minimal Supersymmetric Standard Model (MSSM) and in a
generic two-Higgs-doublet model is included.
\textsc{Vbfnlo} can be run in the MSSM (with real or complex parameters), and
anomalous couplings of the Higgs boson and gauge bosons have been implemented
for a multitude of processes.
Additionally, some extra dimension models are
included  to simulate the production of
technicolor-type vector resonances in VBF and triple vector boson production.
Diboson plus two jets production via VBF can also be run in a spin-2
model and in a model with two Higgs resonances.

Arbitrary cuts can be specified as well as various scale choices. Any currently
available parton distribution function~(PDF) set can be used through the
\textsc{Lhapdf} library. In addition, some PDFs are hard-wired into the code for checking with 
the results in the regress directory, which are shipped with the program for testing purpose.
For most processes implemented at leading order the
program is capable of generating event files in the Les Houches Accord~(LHA) and the 
HepMC format.  When working in the MSSM, the SUSY parameters can be input via a
standard SLHA file.

This manual supersedes the previous versions~\cite{Arnold:2008rz,Arnold:2011wjv1,Arnold:2011wjv2,Arnold:2011wj} released with
\textsc{Vbfnlo Version~2.0}, \textsc{Version~2.5.0}, \textsc{Version~2.6.0} and \textsc{Version~2.7.0}.

\noindent The version of \textsc{Vbfnlo~3.0} extends the following parts of \textsc{Version~2.7}:
\begin{itemize}
 \item The following new processes have been added: \\
       $WWj$ and $ZZj$ production including the loop-induced gluon-fusion
       contributions of $\mathcal{O}(\alpha_s^3)$. VBF $\gamma\gamma jj$, QCD-induced $\gamma jj$, and QCD-induced $\gamma\gamma jj$ processes. 
 \item An interface according to the Binoth Les Houches Accord
       (BLHA) has been added for all VBF processes
       including fully leptonic decays.
 \item The dimension~8 operator $\mathcal{O}_{S,2}$ as well as
       several transverse dimension-8 operators have been added for
       VBS processes.
       
 \item Several unitarization procedures have been implemented for VBS
       processes.
       
 \item The $T_u$-model unitatisation procedures has been implemented for all VBF dimension-8 operators listed in sec. \ref{listofoperators}.
     
 \item Several additional types of form factors, namely complex dipole, step and kink, have been added.
\end{itemize}
Additionally it contains the following improvements also released as \textsc{Vbfnlo Version~2.7.1}:
\begin{itemize}
 \item The following new processes have been added: \\
       QCD-induced $ZZjj$ and $Z\gamma jj$, $Z\gamma jj$ in VBF, $ZZ\gamma$ production.
 \item Higgs decays have been implemented for the VBF-HHjj process ($b\bar b \tau^+\tau^-$ and $b\bar b \gamma\gamma$ final states).
 \item Linking with LHAPDF v6 has been enabled.
\end{itemize}
Several bugfixes and smaller improvements have also been included. The
complete list can be found in the CHANGELOG.md file included in the tarball.

The \textsc{Vbfnlo} webpage -- 
\url{https://ific.uv.es/vbfnlo/} -- contains, in addition to
the latest version of the code, extra information such as the explicit
implementation of the electroweak parameters and couplings. To enable a simple installation test
\textsc{Vbfnlo} is shipped with a complete set of example results, together
with input files, in the {\tt regress} directory.  On the webpage, users can subscribe to
a low-traffic mailing list, where new versions of \textsc{Vbfnlo} are announced.


\newpage

\section{\textsc{Installing VBFNLO}}

The source code of the current version of \textsc{Vbfnlo} can be
downloaded from the \textsc{Vbfnlo} webpage
\begin{center}
\url{https://ific.uv.es/vbfnlo/}
\end{center}
and includes a GNU conforming build system for portability and an
easy build and installation procedure.

\subsection{Prerequisites}

The basic installation requires GNU {\tt make}, a {\sc Fortran} 95 \footnote{{\tt gfortran}
and {\tt ifort} have been tested. Pure FORTRAN77 compilers like {\tt g77} are no longer supported.} and a C++ compiler.
\textsc{Vbfnlo} offers the possibility of using the \textsc{Lhapdf}\footnote{\tt
\url{https://lhapdf.hepforge.org/}}~\cite{Whalley:2005nh} library (versions 5 and 6) for parton
distribution functions. In order to include the electroweak corrections to VBF
Higgs production, the program \textsc{LoopTools}\footnote{\tt
\url{http://www.feynarts.de/looptools/}}~\cite{Hahn:1998yk,Hahn:2006qw} is required.
Additionally, \textsc{FeynHiggs}\footnote{\tt \url{http://www.feynhiggs.de/}}
~\cite{Frank:2006yh,Degrassi:2002fi,Heinemeyer:1998np,Heinemeyer:1998yj} can be
linked to the code in order to calculate the Higgs boson sector of the MSSM,
although a SLHA file can be used as an alternative.  If the simulation of
Kaluza-Klein resonances is enabled, an installation of the GNU Scientific
Library~(GSL)\footnote{\tt \url{https://www.gnu.org/software/gsl/}} is required.
\textsc{Vbfnlo} can also be linked to \textsc{Root}\footnote{\tt
\url{https://root.cern.ch/}} and \textsc{HepMC}\footnote{\tt
\url{http://lcgapp.cern.ch/project/simu/HepMC/}} to produce histograms and event files
in those formats. 


\subsection{Compilation and installation}
\label{compilation}

After unpacking the source archive and entering the source directory, the {\tt
configure} script \footnote{If the {\tt configure} file is absent, it can be 
generated using the {\tt autoreconf -i} command.} can be invoked with several options, a complete list of which are
available via {\tt ./configure~-{}-help}.  Among these, the most important ones
are:

\begin{itemize}
\item {\tt -{}-prefix=[path]} \\ Install \textsc{Vbfnlo} in the location given by {\tt
[path]}.  If not specified, \textsc{Vbfnlo} is installed in the {\tt root} directory.
\item {\tt -{}-enable-processes=[list]} \\ By default, the code for all
available processes except those involving hexagon contributions (i.e.\ triboson
plus jet and QCD-induced $Vjj$/$VVjj$ processes) is compiled\footnote{This is simply due to the relatively
long time required to compile the hexagon routines.}. Optionally, {\tt [list]}
gives a comma-separated list of selected process classes to be compiled. Possible
choices are:

\begin{tabular}{lp{0.65\textwidth}}
{\tt vbf} & Vector boson fusion processes \\ 
{\tt qcdvjj} & QCD-induced vector boson plus two jet production \\
{\tt qcdvvjj} & QCD-induced vector boson pair plus two jet production \\
{\tt diboson} & Double gauge boson production, including $W$ and $WH$ production \\ 
{\tt triboson} & Triple gauge boson production \\ 
{\tt dibosonjet} & Double gauge boson production with a hadronic jet, including $Wj$ and $WHj$ production \\ 
{\tt tribosonjet} & Triboson production in association with a hadronic jet \\ 
{\tt hjjj} & Higgs boson plus three jet production via vector boson fusion \\ 
{\tt ggf} & Higgs boson plus two and three jet production via gluon fusion \\
{\tt all\_except\_hexagons} & All the above processes except those with
                              hexagon contributions (default) \\
{\tt all} & All the above processes
\end{tabular}

\item {\tt FC=/path/to/fortran/compiler} \\ With the {\tt FC} option a specific {\sc Fortran}~95 compiler
      can be requested. Otherwise a default compiler will be used, which will be most probably {\tt gfortran}.
\item {\tt -{}-disable-NLO} \\ Disable the next-to-leading order QCD
corrections.  
\item {\tt -{}-enable-kk} \\ Enable simulation of Kaluza-Klein resonances.
Disabled by default, the Kaluza-Klein option requires the installation of the
GNU Scientific Library, which can be specified via {\tt -{}-with-gsl}. 
\item {\tt -{}-enable-quad} \\ Enable quadruple precision for difficult phase space points.
Enabled by default, if QCD-induced vector boson (pair) plus two jet 
and Higgs plus three jet gluon-fusion-induced processes are requested, otherwise disabled by default.
\item {\tt -{}-enable-MPI} \\ Enables support for parallel runs using MPI. See
  Sec.~\ref{sec:MPI} for additional requirements for using MPI.
Disabled by default.
\item {\tt -{}-with-gsl=[path]} \\ Enable the use of the GNU
Scientific Library. {\tt [path]} specifies the location of the GSL installation.
If the GSL is available directly from the system libraries this flag can be omitted.
\item {\tt -{}-with-LHAPDF=[path]} \\ Enable the use of \textsc{Lhapdf} in addition to the
built-in PDF sets.  Disabled by default. {\tt [path]} specifies the location of
the \textsc{Lhapdf} installation. 
\item {\tt -{}-with-LOOPTOOLS=[path]} \\ Enable the use of
\textsc{LoopTools} in order to calculate the electroweak corrections. 
If this option is not specified, the electroweak corrections cannot be included.
Disabled by default. {\tt [path]} specifies the location of the \textsc{LoopTools}
installation. 
\item {\tt -{}-with-FEYNHIGGS=[path]} \\ Enable the use of
\textsc{FeynHiggs}\footnote{Note that different versions of \textsc{FeynHiggs}
give slightly different results as more corrections are added to the
calculations.  \textsc{Vbfnlo} has been tested with \textsc{FeynHiggs} versions
2.6.x, 2.7.x, 2.8.0, 2.8.3 and 2.12.0} to calculate the MSSM Higgs sector parameters.
Disabled by default. {\tt [path]} specifies the location of the
\textsc{FeynHiggs} installation. 
\item  {\tt -{}-with-root=[path]} \\ Enable the use of \textsc{Root} for histograms.  {\tt
[path]} specifies the location of the \textsc{Root} installation.
\item {\tt -{}-with-hepmc=[path]} \\ Enable the production of \textsc{HepMC}
format event files.  {\tt [path]} specifies the location of the \textsc{HepMC}
installation.

\end{itemize}

Note that, by default, both \textsc{LoopTools} and \textsc{FeynHiggs} are
installed as static libraries.  If this is the case, {\tt configure} must be run
with the option {\tt -{}-enable-shared=no}.  Also note that, in order to link to
an external program such as \textsc{Lhapdf}, the external program needs to have
been compiled using the same compiler (e.g.\ {\tt gfortran}) as \textsc{Vbfnlo}.
If no {\tt path} is specified, \textsc{Vbfnlo} will attempt to find the desired
program in the system directories.  Once {\tt configure} has finished
successfully, the {\tt make} and {\tt make install} commands will compile and
install \textsc{Vbfnlo}, respectively.

To test the installation against a set of typical run files, {\tt make check} can be used.
It requires {\tt py.test} to be installed and takes a few minutes to run.
Due to numerical checks it will fail if any compiler optimization beyond the default
settings was used.

If at any point include files, with a file name suffix {\tt .inc}, are
changed, the command {\tt make clean} must be run so that the changes
are picked up correctly. For normal program usage, this will not be
necessary, but cases where options can be altered in the file {\tt
global.inc} will be mentioned later.
When changing only Fortran source code files, running {\tt make clean}
first is not mandatory. In both cases, {\tt make all install} must be
run afterwards in the main directory to recompile and reinstall the
altered code.


\subsection{Source and installation directory layout}

The \textsc{Vbfnlo} source tree contains the following subdirectories:
\begin{itemize}
\item {\tt amplitudes/:} Routines to calculate matrix elements for the processes
provided.
\item {\tt doc/:} The source of this manual.
\item {\tt helas/:} \textsc{Helas}~\cite{Murayama:1992gi} subroutines used
to calculate helicity amplitudes.
\item {\tt loops/:} One-loop tensor integrals up to six-point functions~\cite{Campanario:2011cs}.
\item {\tt PDFsets/:} Built-in parton distributions
(CT18~\cite{Hou:2019efy}, as well as MRST2004qed~\cite{Martin:2004dh} and MSTW2008~\cite{Martin:2009iq}). 
\item {\tt phasespace/:} Specialized phase-space generators for the processes
provided.
\item {\tt regress/:} Folder containing example results, together with input
files, for all processes.
\item {\tt src/} and {\tt lib/:} Source code of the main programs and input files. 
\item {\tt utilities/:} Routines for administrative tasks, cuts, scale choices
and interfaces.
\end{itemize} 

The source does not need to be modified to change the simulation parameters.
\textsc{Vbfnlo} offers several kinematic cuts and scale choices. This is
illustrated in Sec.~\ref{sec:input}. In addition, it provides a few basic
histograms. Cuts, histograms and scale choices not already provided may be
added in the {\tt utilities/cuts.F}, {\tt utilities/histograms.F} and {\tt
utilities/scales.F90} files.

The installation is performed in a standard \textsc{Unix}-layout, i.e.\ the directory
specified with the {\tt -{}-prefix} option of the {\tt configure} script contains the
following directories:
\begin{itemize}
\item {\tt bin/:} {\tt vbfnlo} executable.
\item {\tt include/VBFNLO/:} \textsc{Vbfnlo} header files. 
\item {\tt lib/VBFNLO/:} \textsc{Vbfnlo} modules as dynamically loadable
  libraries. These can also
be used independently from one of the main programs.
\item {\tt share/VBFNLO/:} Input files and internal PDF tables.
\end{itemize}


\subsection{Running the program}

The {\tt vbfnlo} executables contained in the {\tt bin}
directory of the installation path looks for input files in the current working
directory. An alternative path to input files may be specified explicitly by
passing the {\tt -{}-input=[path]} argument to the programs, with {\tt path}
denoting the full path where input files are located.  I.e.\ in order to run
\textsc{Vbfnlo} from the installation ({\tt prefix}) directory, the command is
\begin{center}
{\tt ./bin/vbfnlo -{}-input=[path]}
\end{center}
The input files contained in the {\tt share/VBFNLO} directory are meant to
represent default settings and should not be changed. We therefore recommend
that the user copies the input files to a separate directory. Here, special
settings may be chosen in the input files and the program can be run from that
directory without specifying further options.  

\textsc{Vbfnlo} outputs a running `log' to the terminal, containing information
about the settings used.  In addition, a file (named {\tt xsection.out}) is produced,
which contains only the LO and NLO cross sections, with the associated statistical errors. 
Histograms and event files, in various forms, can be output as described later.

\noindent The version number of \textsc{Vbfnlo} can be obtained by appending the
argument {\tt -{}-version}. This will print out the string {\tt VBFNLO}
followed by the release number and then exit the program.


\subsubsection{Parallel jobs and optimised grids}
Owing to the complexity of the calculations involved, some of the processes
implemented in \textsc{Vbfnlo} (in particular the spin-2, triboson
plus jet and QCD-induced diboson plus two jets processes) 
require a significant amount of time in order to obtain
reasonable results.  There are, however, methods which can be used in order to
reduce the necessary run time.

By using an optimised grid, the number of iterations needed in order to improve
the efficiency of the MC integration can be reduced.  Optimised grids can be easily 
generated by the user (see Sec.~\ref{sec:general}). The optimised grids previously
provided on the \textsc{Vbfnlo} webpage are now no longer sustained, 
as they were tailored for a specific set of
kinematic cuts and input parameters. Some old grids are
still available on the \textsc{Vbfnlo} webpage for all processes using standard cuts
and parameter settings. The variables used to set the input grid files and
number of iterations are described in Sec.~\ref{sec:general}. 

Another method of improving the run time is to run several jobs in parallel and
then combine the results.  In order to do this, several input directories need to
be set up containing all the necessary {\tt .dat} input files for the process. 
The variable {\tt SEED} in {\tt random.dat} (Sec.~\ref{sec:random}) needs to be
set to a different integer value in each directory.  A short example of the
results of a parallel run, together with their combination, is provided in the
regress directory {\tt regress/100\_Hjj\_parallel}.  On the \textsc{Vbfnlo}
website there is a shell script which can be used to combine the cross sections
and histograms from parallel runs.


\subsection{MPI}
\label{sec:MPI}

MPI allows to run code in parallel. VBFNLO provides experimental support for usage with MPI.
Integration as well as grid and histogram generation are available with MPI parallelization, while for event
outputs only single-core runs are supported.

Compilation und running with MPI is so far only tested with \textsc{GNU} OpenMPI and
GFortran. 
To use this combination load openmpi in your environment (e.g. {\tt module load openmpi}),
then configure with {\tt -{}-enable-MPI} and the OpenMPI wrapper {\tt FC=mpifort}.

To run the resulting binary start {\tt mpirun -np 8 /prefix/bin/vbfnlo}, replacing 8 with
the number of requested parallel runs.

Due to the experimental nature, please test that for your configuration and choice of
process the compilation with and without MPI leads to the same numerical output. For
single-core usage MPI and non-MPI runs should agree up to machine precision, while for
multi-core runs, numerical differences up to the stated statistical uncertainty are expected.

Use the xoroshiro random number generator ({\tt RTYPE=3} in {\tt random.dat}) for best
performance with MPI. For details see Section~\ref{sec:random}.

\subsection{MacOSX}

Compilation of \textsc{Vbfnlo} on MacOSX only works with static libraries.
This can be requested with the configure-option {\tt -{}-enable-shared=no}.
With this option the current version of \textsc{Vbfnlo} has been compiled successfully 
on different versions of MacOSX  between 10.6.8 and 10.12.3
using GFortan 5 or 6 from Homebrew.
However, linking with the library {\sc LoopTools}, which is only
needed for the calculation of the electroweak corrections in the {\tt Hjj}
process, is known to fail.


\subsection{Bug reports}

Please report any problems to
\begin{verbatim}
  vbfnlo@ific.uv.es
\end{verbatim}
with a short report including the {\tt configure} options used to build
\textsc{Vbfnlo}, as well as the versions of compilers and external libraries
used.

You can also open issues in github: \url{https://github.com/vbfnlo/vbfnlo/issues}

\subsection{License}

\textsc{Vbfnlo} is distributed under the GNU General Public License~(GPL)
version~2. This ensures that the source code will be available to users,
grants them the freedom to use and modify the program and sets
out the conditions under which it can be redistributed. However, it was
developed as part of an academic research project and is the result of
many years of work by the authors, which raises various issues that are
not covered by the legal framework of the GPL. It is therefore
distributed together with a set of guidelines\footnote{These guidelines
are contained in the \texttt{GUIDELINES} file distributed with the
release.}, which were originally formulated and agreed on by the
MCnet collaboration for event generator software.
In particular, the original literature on which the program is based
should be cited along with the reference to \textsc{Vbfnlo}.

Some parts of the \textsc{Vbfnlo} matrix elements are based on automatically
generated code from {\sc MadGraph}~\cite{Stelzer:1994ta, Alwall:2007st}
and {\sc FeynArts/FormCalc}~\cite{Kublbeck:1990xc, Denner:1992vza, 
Kublbeck:1992mt, Hahn:2000kx, Hahn:2001rv, Hahn:1998yk, Hahn:2006zy, Hahn:2006qw}.
The routine which reads in SLHA~\cite{Allanach:2008qq} files is largely based 
on {\sc SLHALib-2.2}~\cite{Hahn:2006nq}.
\textsc{Vbfnlo} ships with a copy of {\sc FF~2.0}~\cite{vanOldenborgh:1989wn}
for the evaluation of one-loop integrals, and includes the file {\sc
CPS\_HIGGSTO.F}~\cite{Passarino:2010qk,Goria:2011wa} for calculating the
Higgs width in the CPS scheme.

\newpage

\section{\textsc{Processes}}
\label{sec:proc}
In the following sections, we describe all production processes and decay
modes implemented in \textsc{Vbfnlo}, together with references to more
detailed discussions of the underlying calculations.

In the phase-space regions that are accessible at hadron colliders, VBF
reactions are dominated by $t$-channel electroweak gauge boson exchange. In this
class of processes in \textsc{Vbfnlo} $s$-channel exchange contributions and
kinematically suppressed interference contributions for identical
quarks~\cite{CO,Andersen:2007mp,Bredenstein:2008tm} are therefore
disregarded. While numerically small contributions from Pauli-interference
effects for identical fermions from weak boson decays are neglected,
off-shell contributions to leptonic final states are always included.
For final-state identified photons we employ
Frixione's smooth-cone isolation criterion \cite{Frixione:1998jh}. This ensures that
divergences from the collinear emission of a photon from a final-state massless
quark or anti-quark are avoided, while the cancellation of the infrared
divergences between the virtual and the real part is kept intact.
Due to the inclusion of semileptonic diboson and triboson production processes,
the missing $s$-channel gauge boson exchanges to several VBF processes
can be calculated within \textsc{Vbfnlo} as well. This topic is described
briefly in Section~\ref{sec:schan}.

Higgs contributions (and interference effects) are included in all appropriate
processes, using a fixed Higgs width.  The Higgs propagator is given by
\begin{equation}
 \frac{1}{s - M_{H}^{2} + i M_{H} \Gamma_{H}} .
\end{equation}

In \textsc{Vbfnlo} external quarks are treated as massless.  By default,
external bottom quarks are excluded in those processes where their inclusion
could lead to the presence of a top quark.  For example, in $WW$
diboson production, external bottom quarks are not allowed, as this would lead
to a (massless) $t$-channel top quark, but in $ZZ$ diboson production external
bottom quarks are included.  In the VBF processes such as $pp \rightarrow Hjj$
there are both neutral and charged current components (when the external quark
lines emit $Z$ and $W$ bosons respectively) -- by default, external bottom
quarks are excluded from this class of processes.  A message is printed in the
log file stating whether external bottom quarks are included or excluded.  In
neutral current processes, bottom quarks in the final state are at present
treated no differently from other final state quarks.  The number of external
quark flavours considered is set in the file {\tt utilities/global.inc}.  By
changing the parameter {\tt nfl} from {\tt 5} to {\tt 4}, bottom quarks will
be excluded in any process.  By changing the parameter {\tt vbfNFLb} from {\tt
false} to {\tt true}, external bottom quarks will be included in the neutral
current diagrams in VBF processes if {\tt nfl} is set to {\tt 5}\footnote{Note
that, if these parameters are changed, {\tt make clean} must be run in
the folders {\tt utilities} and {\tt amplitudes}, and then {\tt make all
install} must be run for the \textsc{Vbfnlo} code.}. 

VBF processes with a $t$-channel virtual photon produce singularities at low
photon virtuality, $q^2$, at NLO, when the out-going quark is not tagged.
To remove this photon singularity at $q^2 = 0$, \textsc{Vbfnlo} imposes a default
technical cut of $|q^2| > 4\, \text{GeV}^2$ for the EW processes 
where the singularity occurs. This is expected to be a very good
approximation for measurements at high energies.    

In addition, the program can also calculate some semi-leptonic final-state
processes, based on the following approximation. Starting from the above 
leptonic final-state processes, massless leptons are replaced by massless
quarks, properly modifying color factors, couplings to $Z$ bosons and
virtual photons and including photon exchange, whenever a neutrino is replaced
by an up-quark. To distinguish them from the other quarks (e.g. tagging quarks)
in the final state, we call the replacement quarks decay quarks. 
In this way, NLO QCD corrections to the production part can be taken over
from the corresponding leptonic final state calculation. 
QCD corrections from gluon exchange between the other partons and decay
quarks are neglected because these 
effects are very small, even vanishing in many cases due to color factors. 
Corrections from final state radiation off decay quarks (including virtual
corrections) can be approximated by adding a correction factor 
$(1+\alpha_s(Q^2)/\pi)$ for the decay $V\to \bar{q} q^\prime$ with
$V=W^\pm, Z, \gamma$, $Q = p_V$ being the four-momentum 
of the di-quark system. This correction factor, being about $1.04$ at
$Q=M_Z$, is off by default but can be 
turned on by using the switch {\tt NLO\_SEMILEP\_DECAY} in the
{\tt vbfnlo.dat} input file (see Section~\ref{sec:vbfnlodat}). 
The key features of the semileptonic processes are:
\begin{itemize}
\item NLO QCD corrections to the production part are fully taken into account. 
\item Interference effects and gluon exchange between the decay quarks and the
  other quarks in the process are ignored. 
\item NLO QCD corrections involving decay quarks are not fully included. The
  real gluon-emission corrections cannot be resolved. 
\end{itemize}
Further details can be found in \cite{semilep}. Since the semileptonic processes 
are not fully calculated at NLO QCD as for the leptonic case, they will be marked 
with a star in the following process tables as a warning to the users about the limitations. 

Finally, we note that realistic simulations for experimental measurements
of semileptonic processes often require a sum over several VBFNLO processes.
For example, the final states of semileptonic $WZjj$ and $ZZjj$ production,
with a $Z$ boson decaying leptonically, are indistinguishable. A summation
of these two processes is therefore needed.  


\subsection{VBF Higgs boson production in association with two jets}
\label{sec:vbf-hjj}

$Hjj$ production via VBF mainly proceeds via electroweak quark-quark scattering
processes such as $qq'\to qq'H$ and crossing-related reactions. In
\textsc{Vbfnlo}, tree level cross sections, NLO QCD corrections and one-loop
electroweak corrections (in the Standard Model and the MSSM) to the $t$-channel
production process are provided. The subsequent decay of the Higgs boson is
simulated in the narrow width approximation~(NWA). For the $H\to W^+W^- $ and
the $H\to ZZ$ modes, full off-shell effects and spin correlations of the decay
leptons are included.   The available production process and decay modes are
listed with the corresponding process IDs in Table~\ref{tab:prc1}.  Anomalous
couplings between a Higgs boson and a pair of vector bosons are implemented in the
code and can be input via the file {\tt anom\_HVV.dat}.  Details of the 
calculations can be found in Refs.~\cite{Figy:2003nv,Hollik:2008xn,Figy:2010ct}. 

Furthermore, semileptonic decay modes of the vector bosons are implemented
for the WW and ZZ decay modes~\cite{semilep}. 
\begin{table}[h!]
\newcommand{\lstrut}{{$\strut\atop\strut$}}
\begin{center}
\small
\begin{tabular}{c|l|l}
\hline
&\\
\textsc{ProcId} & \textsc{Process} & \textsc{Bsm} \\
&\\
\hline
&\\
\bf 100 & $p \overset{\mbox{\tiny{(--)}}}{p} \to H \, jj$ & \ldelim \} {10}{0.8cm} \multirow{10}{*}{anomalous HVV couplings, MSSM} \\
\bf 101 & $p \overset{\mbox{\tiny{(--)}}}{p} \to H \, jj\to \gamma\gamma \, jj$ & \\
\bf 102 & $p \overset{\mbox{\tiny{(--)}}}{p} \to H \, jj\to \mu^+\mu^- \, jj$ & \\
\bf 103 & $p \overset{\mbox{\tiny{(--)}}}{p} \to H \, jj\to \tau^+\tau^- \, jj$ & \\
\bf 104 & $p \overset{\mbox{\tiny{(--)}}}{p} \to H \, jj\to b\bar{b} \, jj$ & \\
\bf 105 & $p \overset{\mbox{\tiny{(--)}}}{p} \to H \, jj\to W^{+}W^{-} \, jj\to \ell_{1}^+\nu_{\ell_{1}} \ell_{2}^- 
\bar{\nu}_{\ell_{2}} \,jj$ &  \\
\bf 106 & $p \overset{\mbox{\tiny{(--)}}}{p} \to H \, jj\to ZZ \, jj\to \ell_{1}^+ \ell_{1}^- \ell_{2}^+ 
\ell_{2}^- \,jj$ & \\
\bf 107 & $p \overset{\mbox{\tiny{(--)}}}{p} \to H \, jj\to ZZ \, jj\to \ell_{1}^+ \ell_{1}^- \nu_{\ell_{2}}  
\bar{\nu}_{\ell_{2}} \,jj$ & \\
&\\
\hline
&\\
\bf 108$^*$ & $p \overset{\mbox{\tiny{(--)}}}{p} \to H \, jj\to W^{+}W^{-} \, jj\to q\bar{q} \, \ell^- 
\bar{\nu}_{\ell} \,jj$ & \ldelim \} {4}{0.8cm} \multirow{4}{*}{anomalous HVV couplings, MSSM} \\
\bf 109$^*$ & $p \overset{\mbox{\tiny{(--)}}}{p} \to H \, jj\to W^{+}W^{-} \, jj\to \ell^+\nu_{\ell} \,
q\bar{q}  \,jj$ &  \\
\bf 1010$^*$ & $p \overset{\mbox{\tiny{(--)}}}{p} \to H \, jj\to ZZ \, jj\to q\bar{q} \, \ell^+ \ell^- \,jj$ & \\
&\\
\hline
\end{tabular}
\caption {\em  Process IDs for $p \overset{\mbox{\tiny{(--)}}}{p} \to Hjj$
production via vector boson fusion at NLO (QCD and electroweak) accuracy in the SM
and MSSM.  Anomalous couplings between the Higgs boson and vector bosons are implemented for all decay modes. 
Semileptonic processes are marked with a star to indicate that NLO QCD corrections are approximated as explained in the text.}
\vspace{0.2cm}
\label{tab:prc1}
\end{center}
\end{table}

\subsection{VBF Higgs boson production in association with three jets}

Adding an extra parton to the Higgs production processes of
Sec.~\ref{sec:vbf-hjj} gives rise to $Hjjj$ final states. The corresponding
cross sections are implemented at NLO QCD accuracy (with certain approximations)
in \textsc{Vbfnlo}.  A list of all available modes and corresponding process IDs
is given in Table~\ref{tab:prc2}.  Details of the calculation can be found in Ref.~\cite{Figy:2007kv}.
\begin{table}[t!]
\newcommand{\lstrut}{{$\strut\atop\strut$}}
\begin{center}
\small
\begin{tabular}{c|l}
\hline
&\\
\textsc{ProcId} & \textsc{Process}  \\
&\\
\hline
&\\
\bf 110 & $p \overset{\mbox{\tiny{(--)}}}{p} \to H \, jjj$ \\
\bf 111 & $p \overset{\mbox{\tiny{(--)}}}{p} \to H \, jjj\to \gamma\gamma \, jjj$ \\
\bf 112 & $p \overset{\mbox{\tiny{(--)}}}{p} \to H \, jjj\to \mu^+\mu^- \, jjj$ \\
\bf 113 & $p \overset{\mbox{\tiny{(--)}}}{p} \to H \, jjj\to \tau^+\tau^- \, jjj$ \\
\bf 114 & $p \overset{\mbox{\tiny{(--)}}}{p} \to H \, jjj\to b\bar{b} \, jjj$ \\
\bf 115 & $p \overset{\mbox{\tiny{(--)}}}{p} \to H \, jjj\to W^+W^- \, jjj\to \ell_{1}^{+}\nu_{\ell_{1}} \ell_{2}^- 
\bar{\nu}_{\ell_{2}} \,jjj$ \\
\bf 116 & $p \overset{\mbox{\tiny{(--)}}}{p} \to H \, jjj\to ZZ \, jjj\to \ell_{1}^+ \ell_{1}^- \ell_{2}^+ \ell_{2}^- \,jjj$ \\
\bf 117 & $p \overset{\mbox{\tiny{(--)}}}{p} \to H \, jjj\to ZZ \, jjj\to \ell_{1}^+ \ell_{1}^- \nu_{\ell_{2}}  
\bar{\nu}_{\ell_{2}} \,jjj$ \\
&\\
\hline
\end{tabular}
\caption {\em  Process IDs for $p \overset{\mbox{\tiny{(--)}}}{p} \to Hjjj$ production via VBF at NLO QCD accuracy.}
\vspace{0.2cm}
\label{tab:prc2}
\end{center}
\end{table}

\subsection{VBF Higgs boson production with a photon and two jets}

The emission of an additional photon in VBF Higgs boson production 
(Sec.~\ref{sec:vbf-hjj}) results in $H \gamma jj$ final states. These
are implemented at NLO QCD accuracy in \textsc{Vbfnlo}, with process IDs as given in 
Table~\ref{tab:prcHAjj}. Details of the calculation can be found in
Ref.~\cite{Arnold:2010dx}.
\begin{table}[t!]
\newcommand{\lstrut}{{$\strut\atop\strut$}}
\begin{center}
\small
\begin{tabular}{c|l}
\hline
&\\
\textsc{ProcId} & \textsc{Process}  \\
&\\
\hline
&\\
\bf 2100 & $p \overset{\mbox{\tiny{(--)}}}{p} \to H \gamma \, jj$  \\
\bf 2101 & $p \overset{\mbox{\tiny{(--)}}}{p} \to H \gamma \, jj\to \gamma\gamma \gamma \, jj$ \\
\bf 2102 & $p \overset{\mbox{\tiny{(--)}}}{p} \to H \gamma \, jj\to \mu^+\mu^- \gamma \, jj$ \\
\bf 2103 & $p \overset{\mbox{\tiny{(--)}}}{p} \to H \gamma \, jj\to \tau^+\tau^- \gamma \, jj$ \\
\bf 2104 & $p \overset{\mbox{\tiny{(--)}}}{p} \to H \gamma \, jj\to b\bar{b} \gamma \, jj$ \\
\bf 2105 & $p \overset{\mbox{\tiny{(--)}}}{p} \to H \gamma \, jj\to W^+W^- \gamma \, jj\to \ell_{1}^+\nu_{\ell_{1}} \ell_{2}^- 
\bar{\nu}_{\ell_{2}} \gamma \,jj$ \\
\bf 2106 & $p \overset{\mbox{\tiny{(--)}}}{p} \to H \gamma \, jj\to ZZ \gamma \, jj\to \ell_{1}^+ \ell_{1}^- \ell_{2}^+ 
\ell_{2}^- \gamma \,jj$ \\
\bf 2107 & $p \overset{\mbox{\tiny{(--)}}}{p} \to H \gamma \, jj\to ZZ \gamma \, jj\to \ell_{1}^+ \ell_{1}^- \nu_{\ell_{2}}  
\bar{\nu}_{\ell_{2}} \gamma \,jj$ \\
&\\
\hline
\end{tabular}
\caption {\em  Process IDs for $p \overset{\mbox{\tiny{(--)}}}{p} \to H \gamma jj$ production via
  VBF at NLO QCD accuracy. }
\vspace{0.2cm}
\label{tab:prcHAjj}
\end{center}
\end{table}

\subsection{VBF production of a single vector boson and two jets}

Vector boson fusion processes can also produce final states with two leptons
plus two jets, which are generically referred to as ``VBF $Zjj$ and $W^\pm jj$
production''. These reactions and the one with a photon plus two jets in the
final state are implemented to NLO QCD accuracy in \textsc{Vbfnlo}, with the
process IDs given in Table~\ref{tab:prc3}.  Anomalous triboson couplings are input via {\tt anomV.dat} and can be
included in all processes of this class.  Details of the calculations can be found in
Refs.~\cite{Oleari:2003tc,Jager:2010aj}.
\begin{table}[t!]
\newcommand{\lstrut}{{$\strut\atop\strut$}}
\begin{center}
\small
\begin{tabular}{c|l|l}
\hline
&\\
\textsc{ProcId} & \textsc{Process} & \textsc{Bsm}\\
&\\
\hline
&\\
\bf 120 & $p \overset{\mbox{\tiny{(--)}}}{p} \to Z \, jj \to \ell^{+} \ell^{-} \, jj$ & \ldelim \} {6}{0.8cm} \multirow{5}{*}{anomalous  couplings} \\
\bf 121 & $p \overset{\mbox{\tiny{(--)}}}{p} \to Z  \, jj\to \nu_\ell \bar{\nu}_\ell \, jj$ & \\
\bf 130 & $p \overset{\mbox{\tiny{(--)}}}{p} \to W^{+} \,  jj\to \ell^{+} \nu_\ell \, jj$ & \\
\bf 140 & $p \overset{\mbox{\tiny{(--)}}}{p} \to W^{-} \, jj\to \ell^{-} \bar{\nu}_\ell  \, jj$ &\\
\bf 150 & $p \overset{\mbox{\tiny{(--)}}}{p} \to \gamma \, jj$ &\\
&\\
\hline
\end{tabular}
\caption {\em  Process IDs for vector boson + 2 jet production 
via vector boson fusion at NLO QCD accuracy.}
\vspace{0.2cm}
\label{tab:prc3}
\end{center}
\end{table}

\subsection{VBF production of a spin-2 particle}


\textsc{Vbfnlo} can simulate the production of a spin-2 particle via VBF, which then decays into two photons, 
with the process ID 191 (see Table~\ref{tab:spin2}). This process 
will only run if the switch {\tt SPIN2} 
in {\tt vbfnlo.dat} is set to {\tt true}. Furthermore, spin-2 production
is also available in the $WW$ and $ZZ$ modes 
given in Table~\ref{tab:spin2} if the switch {\tt SPIN2} 
in {\tt vbfnlo.dat} is set to {\tt true}. If {\tt SPIN2} is set to {\tt false}, VBF Higgs production is 
simulated instead, similarly to the processes 105, 106 or 107 of Table~\ref{tab:prc1} for a SM Higgs. 
The parameters of the spin-2 model are input via {\tt spin2coupl.dat}. Details of the calculations can be 
found in Refs.~\cite{frank,Frank:2012wh,Frank:2013gca}.

\begin{table}[t!]
\newcommand{\lstrut}{{$\strut\atop\strut$}}
\begin{center}
\small
\begin{tabular}{c|l|l}
\hline
&\\
\textsc{ProcId} & \textsc{Process} & \textsc{Bsm}\\
&\\
\hline
&\\
\bf 191 & $p \overset{\mbox{\tiny{(--)}}}{p} \to S_{2}  \, jj\to \gamma \gamma \, jj$ & \ldelim \} {5}{0.8cm} \multirow{5}{*}{spin-2 resonant production}
\\
\bf 195 & $p \overset{\mbox{\tiny{(--)}}}{p} \to S_{2}  \, jj\to W^{+}W^{-} \, jj\to \ell_{1}^+\nu_{\ell_{1}} \ell_{2}^- 
\bar{\nu}_{\ell_{2}} \,jj$ 
&\\
\bf 196 & $p \overset{\mbox{\tiny{(--)}}}{p} \to S_{2}  \, jj\to ZZ \, jj\to \ell_{1}^+ \ell_{1}^- \ell_{2}^+ 
\ell_{2}^- \,jj$ 
&\\
\bf 197 & $p \overset{\mbox{\tiny{(--)}}}{p} \to S_{2}  \, jj\to  ZZ \, jj\to \ell_{1}^+ \ell_{1}^- \nu_{\ell_{2}}  
\bar{\nu}_{\ell_{2}} \,jj$
&\\
&\\
\hline
\end{tabular}
\caption {\em  Process IDs for a spin-2 particle $S_{2}$+2 jet production 
via vector boson fusion at NLO QCD accuracy.}
\vspace{0.2cm}
\label{tab:spin2}
\end{center}
\end{table}


\subsection{VBF production of two vector bosons and two jets}

The production of four leptons plus two jets in the final state at order
$\mathcal{O}(\alpha^6)$ is dominated by VBF contributions. In \textsc{Vbfnlo},
all resonant and non-resonant $t$-channel exchange contributions (including
contributions from Higgs bosons) giving rise to a specific leptonic final state
are considered. For simplicity, we refer to these reactions as ``VBF diboson
production''. Finite width effects of the weak bosons and spin correlations of
the decay leptons are fully retained.

The available processes and corresponding process IDs are listed in
Table~\ref{tab:prc4}.  Anomalous gauge boson couplings, input via {\tt anomV.dat}, are implemented for
all processes in this class, as well as a simplified model with two Higgs resonances. 
This process class can also be run in the Kaluza-Klein and spin-2 models implemented in \textsc{Vbfnlo}, with the exception of
$W\gamma$, $Z\gamma$, $\gamma\gamma$ and same-sign $WW$ production. 
Details of the calculations can be found in
Refs.~\cite{Jager:2006zc,Jager:2006cp,Bozzi:2007ur,Jager:2009xx,frank,Campanario:2013eta,kaiser,Campanario:2017ffz,Campanario:2020xaf}.

Furthermore, semileptonic decay modes of the vector bosons are implemented
for the opposite-sign $WW$, same-sign $WW$, $WZ$ and $ZZ$ production processes in VBF~\cite{semilep}. 
In this case only anomalous couplings are available as
BSM options.
\begin{table}[htb!]
\newcommand{\lstrut}{{$\strut\atop\strut$}}
\begin{center}
\small
\begin{tabular}{c|l|l}
\hline
&\\
\textsc{ProcId} & \textsc{Process} & \textsc{BSM}\\
&\\
\hline
&\\
\bf 200 & $p \overset{\mbox{\tiny{(--)}}}{p} \to W^{+}W^{-} \, jj \to \ell_{1}^{+} \nu_{\ell_{1}} \ell_{2}^{-}
\bar{\nu}_{\ell_{2}} \, jj$ &  \ldelim \} {6}{0.8cm} \multirow{6}{*}{\begin{parbox}{3.65cm}{anomalous couplings, two-Higgs model, Kaluza-Klein models, spin-2 models}\end{parbox}}\\
\bf 210 & $p \overset{\mbox{\tiny{(--)}}}{p} \to ZZ  \, jj\to \ell_{1}^{+} \ell_{1}^{-} \ell_{2}^{+} \ell_{2}^{-} \, jj$ &\\
\bf 211 & $p \overset{\mbox{\tiny{(--)}}}{p} \to ZZ  \, jj\to \ell_{1}^{+} \ell_{1}^{-} \nu_{\ell_{2}} \bar{\nu}_{\ell_{2}} \, jj$ & \\
\bf 220 & $p \overset{\mbox{\tiny{(--)}}}{p} \to W^{+}Z \,  jj\to \ell_{1}^{+} \nu_{\ell_{1}} \ell_{2}^{+} \ell_{2}^{-} \, jj$ & \\
\bf 230 & $p \overset{\mbox{\tiny{(--)}}}{p} \to W^{-}Z \, jj\to \ell_{1}^{-} \bar{\nu}_{\ell _{1}} \ell_{2}^{+} \ell_{2}^{-} \, jj$ & \\
\bf 240 & $p \overset{\mbox{\tiny{(--)}}}{p} \to \gamma \gamma \, jj$ & \ldelim \} {1}{0.8cm} \multirow{1}{*}{anomalous couplings}\\
\bf 250 & $p \overset{\mbox{\tiny{(--)}}}{p} \to W^{+}W^{+} \,  jj\to \ell_{1}^{+} \nu_{\ell_{1}} \ell_{2}^{+} \nu_{\ell_{2}} \, jj$ & \ldelim \} {2}{0.8cm} \multirow{2}{*}{anomalous couplings, two-Higgs model}\\
\bf 260 & $p \overset{\mbox{\tiny{(--)}}}{p} \to W^{-}W^{-} \,  jj\to \ell_{1}^{-} \bar{\nu}_{\ell_{1}} \ell_{2}^{-} \bar{\nu}_{\ell_{2}} \, jj$ & \\
\bf 270 & $p \overset{\mbox{\tiny{(--)}}}{p} \to W^{+}\gamma \, jj\to \ell^{+} \nu_{\ell} \gamma \, jj$ & \ldelim \} {5}{0.8cm} \multirow{5}{*}{anomalous couplings} \\
\bf 280 & $p \overset{\mbox{\tiny{(--)}}}{p} \to W^{-}\gamma \, jj\to \ell^{-} \bar{\nu}_{\ell} \gamma \, jj$ & \\
\bf 290 & $p \overset{\mbox{\tiny{(--)}}}{p} \to Z \gamma \, jj\to \ell^{+} \ell^{-} \gamma \, jj$ & \\ 
\bf 291 & $p \overset{\mbox{\tiny{(--)}}}{p} \to Z\gamma \, jj\to
\nu_{\ell}\bar{\nu}_{\ell} \gamma \, jj$ & \\
&\\
\hline
&\\
\bf 201$^*$ & $p \overset{\mbox{\tiny{(--)}}}{p} \to W^{+}W^{-} \, jj \to q \bar{q} \, \ell^{-}
\bar{\nu}_{\ell} \, jj$ &  \ldelim \} {11}{0.8cm} \multirow{11}{*}{\begin{parbox}{3.65cm}{anomalous couplings, two-Higgs model}\end{parbox}}\\
\bf 202$^*$ & $p \overset{\mbox{\tiny{(--)}}}{p} \to W^{+}W^{-}  \, jj\to \ell^{+} \nu_\ell \, q \bar{q} \, jj$ &\\
\bf 212$^*$ & $p \overset{\mbox{\tiny{(--)}}}{p} \to ZZ  \, jj\to q \bar{q} \, \ell^{+} \ell^{-} \, jj$ &\\
\bf 221$^*$ & $p \overset{\mbox{\tiny{(--)}}}{p} \to W^{+}Z \,  jj\to q \bar{q} \, \ell^{+} \ell^{-} \, jj$ & \\
\bf 222$^*$ & $p \overset{\mbox{\tiny{(--)}}}{p} \to W^{+}Z \,  jj\to \ell^{+} \nu_{\ell} \, q \bar{q} \, jj$ & \\
\bf 231$^*$ & $p \overset{\mbox{\tiny{(--)}}}{p} \to W^{-}Z \, jj\to q \bar{q} \, \ell^{+} \ell^{-} \, jj$ & \\
\bf 232$^*$ & $p \overset{\mbox{\tiny{(--)}}}{p} \to W^{-}Z \, jj\to \ell^{-} \bar{\nu}_{\ell} \, q \bar{q} \, jj$ & \\
\bf 251$^*$ & $p \overset{\mbox{\tiny{(--)}}}{p} \to W^{+}W^{+} \,  jj\to q \bar{q} \, \ell^{+} \nu_{\ell} \, jj$ & \\
\bf 261$^*$ & $p \overset{\mbox{\tiny{(--)}}}{p} \to W^{-}W^{-} \,  jj\to q \bar{q} \, \ell^{-} \bar{\nu}_{\ell} \, jj$ & \\
&\\
\hline
\end{tabular}
\caption {\em  Process IDs for diboson + 2 jet
production via vector boson fusion at NLO QCD accuracy.}
\vspace{0.2cm}
\label{tab:prc4}
\end{center}
\end{table}


\subsection{VBF production of two Higgs bosons and two jets}

Higgs pair production plus two jets via VBF mainly proceeds via
electroweak quark-quark scattering processes such as $qq'\to qq'HH$ and
crossing-related reactions. In \textsc{Vbfnlo}, tree level cross
sections and NLO QCD corrections to the $t$-channel production process
are provided in the framework of the Standard Model. Two decay modes
for the subsequent decays of the Higgs bosons in the narrow-width
approximation are implemented as well. The processes and corresponding
process IDs are listed in Table~\ref{tab:hhjj}.  More details of the
calculations can be found in Refs.~\cite{Figy:2008zd,Baglio:2012np},
in particular about the generation of the amplitudes of the on-shell
production.

\begin{table}[htb!]
\newcommand{\lstrut}{{$\strut\atop\strut$}}
\begin{center}
\small
\begin{tabular}{c|l}
\hline
&\\
\textsc{ProcId} & \textsc{Process}\\
&\\
\hline
&\\
\bf 160 & $p \overset{\mbox{\tiny{(--)}}}{p} \to HH \,  jj$ \\
\bf 161 & $p \overset{\mbox{\tiny{(--)}}}{p} \to HH \,  jj \to b\bar{b}\tau^+\tau^- \, jj$ \\
\bf 162 & $p \overset{\mbox{\tiny{(--)}}}{p} \to HH \,  jj \to b\bar{b}\gamma\gamma \, jj$ \\
&\\
\hline
\end{tabular}
\caption {\em  Process IDs for Higgs pair + 2 jet
production via vector boson fusion at NLO QCD accuracy.}
\vspace{0.2cm}
\label{tab:hhjj}
\end{center}
\end{table}


\subsection{$W$ production with up to one jet}
\label{sec:wj}

The production of a $W$ boson with up to one jet is implemented in \textsc{Vbfnlo} at NLO
QCD. 
The $W$ boson decays leptonically and full off-shell effects and spin correlations of the final state leptons are included.
The processes are listed in Table~\ref{tab:wj}.
Details of the calculation can be found in Ref.~\cite{robin}.

\begin{table}[t!]
\newcommand{\lstrut}{{$\strut\atop\strut$}}
\begin{center}
\small
\begin{tabular}{c|l}
\hline
&\\
\textsc{ProcId} & \textsc{Process}  \\
&\\
\hline
&\\
\bf 1330 & $p \overset{\mbox{\tiny{(--)}}}{p} \to W^+ \to \ell^+\nu_{\ell} $ \\
\bf 1340 & $p \overset{\mbox{\tiny{(--)}}}{p} \to W^- \to \ell^- \bar{\nu}_{\ell} $ \\
\bf 1630 & $p \overset{\mbox{\tiny{(--)}}}{p} \to W^+ \, j \to \ell^+\nu_{\ell} \, j $ \\
\bf 1640 & $p \overset{\mbox{\tiny{(--)}}}{p} \to W^- \, j\to \ell^- \bar{\nu}_{\ell} \, j $ \\
&\\
\hline
\end{tabular}
\caption {\em  Process IDs for the $W$ production processes with up to one jet at NLO QCD accuracy.}
\vspace{0.2cm}
\label{tab:wj}
\end{center}
\end{table}

\subsection{Double vector boson production}
\label{ssec:diboson}
The production of four-lepton final states mainly proceeds via double vector
boson production with subsequent decays. Additionally, there are processes where
one or more decaying boson is replaced by an on-shell photon, giving rise to
lepton production in association with a photon and double photon production. In
\textsc{Vbfnlo}, the processes listed in Table~\ref{tab:dib} are implemented to
NLO QCD accuracy, including full off-shell effects and spin correlations of the
final state leptons and photons. Anomalous vector boson couplings, input via
{\tt anomV.dat}, are implemented for $WW$, $W^{\pm}Z$ and $W^{\pm}\gamma$
production. The processes with a neutral final state ($WW$, $ZZ$, $Z\gamma$ and
$\gamma\gamma$) also include the gluon-induced fermionic loop diagrams by
default at NLO -- both continuum production via box diagrams and the $s$-channel
Higgs boson contributions are included, and anomalous $HVV$ couplings (input via
{\tt anom\_HVV.dat}) can be used. 
Details of the calculations can be found in Ref.~\cite{johannes}.

Furthermore, semileptonic decay modes of the vector bosons are implemented
for the $WW$, $ZZ$ and $WZ$ production processes~\cite{semilep}. 
Both anomalous $VVV$ and $HVV$ couplings are available as
BSM options.
\begin{table}[t!]
\newcommand{\lstrut}{{$\strut\atop\strut$}}
\begin{center}
\small
\begin{tabular}{c|l|l}
\hline
&\\
\textsc{ProcId} & \textsc{Process} & \textsc{BSM}  \\
&\\
\hline
&\\
\bf 300 & $p \overset{\mbox{\tiny{(--)}}}{p} \to W^{+}W^{-} \to \ell_{1}^{+} \nu_{\ell_{1}} \ell_{2}^{-}\bar{\nu}_{\ell_{2}} $ & anomalous $HVV$ and $VVV$ couplings\\
\bf 310 & $p \overset{\mbox{\tiny{(--)}}}{p} \to W^{+}Z \to  \ell_{1}^{+} \nu_{\ell_1}  \ell_{2}^{+} \ell_{2}^{-} $ & \ldelim \} {2}{0.8cm} \multirow{2}{*}{anomalous $VVV$ couplings}\\
\bf 320 & $p \overset{\mbox{\tiny{(--)}}}{p} \to W^{-}Z \to \ell_{1}^{-} \bar{\nu}_{\ell_{1}}  \ell_{2}^{+} \ell_{2}^{-} $ & \\
\bf 330 & $p \overset{\mbox{\tiny{(--)}}}{p} \to ZZ \to \ell_{1}^{-} \ell_{1}^{+}  \ell_{2}^{-} \ell_{2}^{+} $ & anomalous $HVV$ couplings\\
\bf 340 & $p \overset{\mbox{\tiny{(--)}}}{p} \to W^{+}\gamma \to \ell_{1}^{+} \nu_{\ell_1} \gamma $ & \ldelim \} {2}{0.8cm} \multirow{2}{*}{anomalous $VVV$ couplings}\\
\bf 350 & $p \overset{\mbox{\tiny{(--)}}}{p} \to W^{-}\gamma \to \ell_{1}^{-} \bar{\nu}_{\ell_{1}} \gamma $ & \\
\bf 360 & $p \overset{\mbox{\tiny{(--)}}}{p} \to Z\gamma \to \ell_{1}^{-} \ell_{1}^{+}  \gamma $ & \ldelim \} {2}{0.8cm} \multirow{2}{*}{anomalous $HVV$ couplings}\\
\bf 370 & $p \overset{\mbox{\tiny{(--)}}}{p} \to \gamma\gamma $ & \\
&\\
\hline
&\\
\bf 301$^*$ & $p \overset{\mbox{\tiny{(--)}}}{p} \to W^{+}W^{-} \to q \bar{q} \, \ell^{-}\bar{\nu}_{\ell} $ &  \ldelim \} {2}{0.8cm} \multirow{2}{*}{anomalous $VVV$ and $HVV$ couplings}\\
\bf 302$^*$ & $p \overset{\mbox{\tiny{(--)}}}{p} \to W^{+}W^{-} \to \ell^{+} \nu_{\ell} \, q \bar{q} $ \\
\bf 312$^*$ & $p \overset{\mbox{\tiny{(--)}}}{p} \to W^{+}Z \to  q \bar{q} \, \ell^{+} \ell^{-} $ &  \ldelim \} {5}{0.8cm} \multirow{5}{*}{anomalous $VVV$ couplings}\\
\bf 313$^*$ & $p \overset{\mbox{\tiny{(--)}}}{p} \to W^{+}Z \to  \ell^{+} \nu_{\ell} \, q \bar{q} $ \\
\bf 322$^*$ & $p \overset{\mbox{\tiny{(--)}}}{p} \to W^{-}Z \to q \bar{q} \, \ell^{+} \ell^{-} $  \\
\bf 323$^*$ & $p \overset{\mbox{\tiny{(--)}}}{p} \to W^{-}Z \to \ell^{-} \bar{\nu}_{\ell} \, q \bar{q} $  \\
\bf 331$^*$ & $p \overset{\mbox{\tiny{(--)}}}{p} \to ZZ \to q \bar{q} \, \ell^{-} \ell^{+} $ &  {anomalous $HVV$ couplings}\\
&\\
\hline
\end{tabular}
\caption {\em  Process IDs for the diboson production processes at NLO
  QCD accuracy.}
\vspace{0.2cm}
\label{tab:dib}
\end{center}
\end{table}

\subsection{Triple vector boson production}
\label{ssec:triboson}
The production of six-lepton final states mainly proceeds via triple vector
boson production with subsequent decays. Additionally, there are processes where
one or more decaying boson is replaced by an on-shell photon, giving rise to
lepton production in association with photon(s) and triple photon production. In
\textsc{Vbfnlo}, the processes listed in Tables~\ref{tab:prc5} and~\ref{tab:prc5a} are implemented to
NLO QCD accuracy, including full off-shell effects and spin correlations of the
final state leptons and photons. For processes with three massive gauge bosons
the Higgs boson contributions are included. Anomalous
vector boson couplings\footnote{Some anomalous $VVV$ ($f_{W}$ and $f_{B}$ or,
equivalently, $\Delta \kappa_{\gamma}$ and $\Delta g_{1}^{Z}$) couplings also
imply anomalous $HVV$ couplings -- these are automatically taken into account.}
are implemented for all triboson processes, with the anomalous parameters input via {\tt anomV.dat}. 
The processes $WWZ$, $ZZW^{\pm}$ and $W^{\pm}W^{+}W^{-}$ can also be run in the implemented higgsless
Kaluza-Klein models.  Details of the calculations can be found in
Refs.~\cite{Hankele:2007sb,Campanario:2008yg,Bozzi:2009ig,Bozzi:2010sj,
Bozzi:2011wwa,Bozzi:2011en}.  

Furthermore, semileptonic decay modes of the vector bosons are implemented
for all triple vector boson production processes with zero or one final
state photon~\cite{semilep}. 
In this case only anomalous couplings are available as
BSM options.

\begin{table}[t!]
\newcommand{\lstrut}{{$\strut\atop\strut$}}
\begin{center}
\small
\begin{tabular}{c|l|l}
\hline
&\\
\textsc{ProcId} & \textsc{Process} & \textsc{BSM}  \\
&\\
\hline
&\\
\bf 400 & $p \overset{\mbox{\tiny{(--)}}}{p} \to W^{+}W^{-}Z \to \ell_{1}^{+}\nu_{\ell_{1}} \ell_{2}^{-} \bar{\nu}_{\ell_{2}} 
\ell_{3}^{+} \ell_{3}^{-} $ & \ldelim \} {6}{0.8cm} \multirow{6}{*}{\begin{parbox}{3.65cm}{anomalous couplings, Kaluza-Klein models}\end{parbox}}\\
\bf 410 & $p \overset{\mbox{\tiny{(--)}}}{p} \to ZZW^{+} \to  \ell_{1}^{+} \ell_{1}^{-}  \ell_{2}^{+} \ell_{2}^{-} 
 \ell_{3}^{+} \nu_{\ell_{3}} $ & \\
\bf 420 & $p \overset{\mbox{\tiny{(--)}}}{p} \to ZZW^{-} \to \ell_{1}^{+} \ell_{1}^{-}  \ell_{2}^{+} \ell_{2}^{-} 
 \ell_{3}^{-}  \bar{\nu}_{\ell_{3}}$ & \\
\bf 430 & $p \overset{\mbox{\tiny{(--)}}}{p} \to W^{+}W^{-}W^{+} \to \ell_{1}^{+}\nu_{\ell_1} \ell_{2}^{-}
\bar{\nu}_{\ell_2} \ell_{3}^{+}\nu_{\ell_{3}}$ & \\
\bf 440 & $p \overset{\mbox{\tiny{(--)}}}{p} \to W^{-}W^{+}W^{-} \to \ell_{1}^{-} \bar{\nu}_{\ell_1}\ell_{2}^{+}\nu_{\ell_2}
\ell_{3}^{-} \bar{\nu}_{\ell_{3}} $ & \\
\bf 450 & $p \overset{\mbox{\tiny{(--)}}}{p} \to ZZZ \to \ell_{1}^{-} \ell_{1}^{+} \ell_{2}^{-}
\ell_{2}^{+} \ell_{3}^{-} \ell_{3}^{+} $ & \ldelim \} {4}{0.8cm} \multirow{4}{*}{anomalous couplings}\\
\bf 460 & $p \overset{\mbox{\tiny{(--)}}}{p} \to W^{-}W^{+} \gamma \to \ell_{1}^{-} \bar{\nu}_{\ell_1}
\ell_{2}^{+}\nu_{\ell_2} \gamma$ & \\
\bf 470 & $p \overset{\mbox{\tiny{(--)}}}{p} \to Z Z \gamma \to \ell_{1}^{-} \ell_{1}^{+} \ell_{2}^{-}
\ell_{2}^{+} \gamma$ & \\
\bf 472 & $p \overset{\mbox{\tiny{(--)}}}{p} \to Z Z \gamma \to \ell_{1}^{-} \ell_{1}^{+} \nu_{\ell_2} \bar{\nu}_{\ell_2} \gamma$ & \\
\bf 480 & $p \overset{\mbox{\tiny{(--)}}}{p} \to W^{+} Z \gamma \to \ell_{1}^{+}\nu_{\ell_1} \ell_{2}^{-}
\ell_{2}^{+} \gamma$ & \ldelim \} {7}{0.8cm} \multirow{7}{*}{anomalous couplings}\\
\bf 490 & $p \overset{\mbox{\tiny{(--)}}}{p} \to W^{-} Z \gamma \to \ell_{1}^{-} \bar{\nu}_{\ell_1} \ell_{2}^{-}
\ell_{2}^{+} \gamma$ & \\
\bf 500 & $p \overset{\mbox{\tiny{(--)}}}{p} \to W^{+} \gamma \gamma \to {\ell}^{+}\nu_{\ell} 
\gamma \gamma$ & \\
\bf 510 & $p \overset{\mbox{\tiny{(--)}}}{p} \to W^{-} \gamma \gamma \to {\ell}^{-} \bar{\nu}_{\ell} 
\gamma \gamma$ & \\
\bf 520 & $p \overset{\mbox{\tiny{(--)}}}{p} \to Z \gamma \gamma \to {\ell}^{-} {\ell}^{+} 
\gamma \gamma$ & \\
\bf 521 & $p \overset{\mbox{\tiny{(--)}}}{p} \to Z \gamma \gamma \to \nu_{\ell} \bar{\nu}_{\ell} 
\gamma \gamma$ & \\
\bf 530 & $p \overset{\mbox{\tiny{(--)}}}{p} \to \gamma \gamma \gamma $ & \\
&\\
\hline
\end{tabular}
\caption {\em  Process IDs for the triboson production processes at NLO QCD accuracy with fully leptonic decays.}
\vspace{0.2cm}
\label{tab:prc5}
\end{center}
\end{table}

\begin{table}[t!]
\newcommand{\lstrut}{{$\strut\atop\strut$}}
\begin{center}
\small
\begin{tabular}{c|l|l}
\hline
&\\
\textsc{ProcId} & \textsc{Process} & \textsc{BSM}  \\
&\\
\hline
&\\
\bf 401$^*$ & $p \overset{\mbox{\tiny{(--)}}}{p} \to W^{+}W^{-}Z \to q \bar{q} \, \ell_{1}^{-} \bar{\nu}_{\ell_{1}} 
\ell_{2}^{+} \ell_{2}^{-} $ & \ldelim \} {24}{0.8cm} \multirow{24}{*}{\begin{parbox}{3.65cm}{anomalous couplings}\end{parbox}}\\
\bf 402$^*$ & $p \overset{\mbox{\tiny{(--)}}}{p} \to W^{+}W^{-}Z \to \ell_{1}^{+}\nu_{\ell_{1}} \, q \bar{q} \,
\ell_{2}^{+} \ell_{2}^{-} $ \\
\bf 403$^*$ & $p \overset{\mbox{\tiny{(--)}}}{p} \to W^{+}W^{-}Z \to \ell_{1}^{+}\nu_{\ell_{1}} \ell_{2}^{-} \bar{\nu}_{\ell_{2}} \,
q \bar{q} $ \\
\bf 411$^*$ & $p \overset{\mbox{\tiny{(--)}}}{p} \to ZZW^{+} \to  \ell_{1}^{+} \ell_{1}^{-}  \ell_{2}^{+} \ell_{2}^{-} 
 \, q \bar{q} $ & \\
\bf 412$^*$ & $p \overset{\mbox{\tiny{(--)}}}{p} \to ZZW^{+} \to  q \bar{q} \, \ell_{1}^{+} \ell_{1}^{-}
 \ell_{2}^{+} \nu_{\ell_{2}} $ & \\
\bf 421$^*$ & $p \overset{\mbox{\tiny{(--)}}}{p} \to ZZW^{-} \to \ell_{1}^{+} \ell_{1}^{-}  \ell_{2}^{+} \ell_{2}^{-} 
 \, q \bar{q} $ & \\
\bf 422$^*$ & $p \overset{\mbox{\tiny{(--)}}}{p} \to ZZW^{-} \to q \bar{q} \, \ell_{1}^{+} \ell_{1}^{-}
 \ell_{2}^{-}  \bar{\nu}_{\ell_{2}}$ & \\
\bf 431$^*$ & $p \overset{\mbox{\tiny{(--)}}}{p} \to W^{+}W^{-}W^{+} \to  q \bar{q} \, \ell_{1}^{-}
\bar{\nu}_{\ell_1} \ell_{2}^{+}\nu_{\ell_{2}}$ & \\
\bf 432$^*$ & $p \overset{\mbox{\tiny{(--)}}}{p} \to W^{+}W^{-}W^{+} \to \ell_{1}^{+}\nu_{\ell_1} \, q \bar{q} \,
 \ell_{2}^{+}\nu_{\ell_{2}}$ & \\
\bf 441$^*$ & $p \overset{\mbox{\tiny{(--)}}}{p} \to W^{-}W^{+}W^{-} \to \ell_{1}^{-} \bar{\nu}_{\ell_1} \, q \bar{q} \,
\ell_{2}^{-} \bar{\nu}_{\ell_{2}} $ & \\
\bf 442$^*$ & $p \overset{\mbox{\tiny{(--)}}}{p} \to W^{-}W^{+}W^{-} \to  q \bar{q} \, \ell_{1}^{+}\nu_{\ell_1}
\ell_{2}^{-} \bar{\nu}_{\ell_{2}} $ & \\
\bf 451$^*$ & $p \overset{\mbox{\tiny{(--)}}}{p} \to ZZZ \to  q \bar{q} \, \ell_{1}^{-} \ell_{1}^{+} \ell_{2}^{-}
\ell_{2}^{+}  $ & \\
\bf 461$^*$ & $p \overset{\mbox{\tiny{(--)}}}{p} \to W^{+}W^{-} \gamma \to  q \bar{q} \,
\ell^{-}\bar{\nu}_{\ell} \gamma$ & \\
\bf 462$^*$ & $p \overset{\mbox{\tiny{(--)}}}{p} \to W^{+}W^{-} \gamma \to \ell^{+} \nu_{\ell}
\, q \bar{q} \, \gamma$ & \\
\bf 471$^*$ & $p \overset{\mbox{\tiny{(--)}}}{p} \to Z Z \gamma \to \ell^{-} \ell^{+} \, q \bar{q} \, \gamma$ & \\
\bf 481$^*$ & $p \overset{\mbox{\tiny{(--)}}}{p} \to W^{+} Z \gamma \to  q \bar{q} \, \ell^{-}
\ell^{+} \gamma$ & \\
\bf 482$^*$ & $p \overset{\mbox{\tiny{(--)}}}{p} \to W^{+} Z \gamma \to \ell^{+}\nu_{\ell} \, q \bar{q} \, \gamma$ & \\
\bf 491$^*$ & $p \overset{\mbox{\tiny{(--)}}}{p} \to W^{-} Z \gamma \to  q \bar{q} \, \ell^{-}
\ell^{+} \gamma$ & \\
\bf 492$^*$ & $p \overset{\mbox{\tiny{(--)}}}{p} \to W^{-} Z \gamma \to \ell^{-} \bar{\nu}_{\ell} \, q \bar{q} \, \gamma$ & \\
&\\
\hline
\end{tabular}
\caption {\em  Process IDs for the triboson production processes at NLO QCD accuracy with semileptonic decays.}
\vspace{0.2cm}
\label{tab:prc5a}
\end{center}
\end{table}


\subsection{Double vector boson production in association with a hadronic jet}

%
Diboson production in association with a hard hadronic jet are
available in {\sc{Vbfnlo}} at NLO QCD accuracy under the process IDs of
Tables~\ref{tab:prc7} and~\ref{tab:prc7a}. All off-shell and finite width effects
are included. The $WZ$ and $W\gamma$ processes can be run with anomalous $WWZ$ and $WW\gamma$ couplings, input via {\tt anomV.dat}.
Details can be found in
Refs.~\cite{Campanario:2010hv,Campanario:2010xn,Campanario:2010hp,
Campanario:2009um,Campanario:2012bh,Campanario:2013wta,Campanario:2015nha}.  

Furthermore, semileptonic decay modes of the vector bosons are implemented
for the $W^+W^-$ final state in the SM.
\begin{table}[t!]
\begin{center}
\small
\begin{tabular}{c|l|l}
\hline
&\\
\textsc{ProcId} & \textsc{Process} & \textsc{Bsm} \\
&\\
\hline
&\\
\bf 600 & $p \overset{\mbox{\tiny{(--)}}}{p}  \to W^{+} W^{-} j \to \ell_{1}^{+}\nu_{\ell_1} \ell_{2}^{-} \bar{\nu}_{\ell_2} j $ & \multirow{1}{*}{anomalous $HVV$ couplings} \\
\bf 610 & $p \overset{\mbox{\tiny{(--)}}}{p}  \to W^{-} \gamma j \to \ell^{-} \bar \nu_{\ell} \gamma j $ & \ldelim \} {5}{0.8cm} \multirow{5}{*}{anomalous couplings}\\
\bf 620 & $p \overset{\mbox{\tiny{(--)}}}{p}  \to W^{+} \gamma j  \to \ell^{+} \nu_{\ell} \gamma j $ &\\
\bf 630 & $p \overset{\mbox{\tiny{(--)}}}{p}  \to W^{-} Z j \to \ell_{1}^{-} \bar \nu_{\ell_1} \ell_{2}^{-}
\ell_{2}^{+} j$ & \\
\bf 640 & $p \overset{\mbox{\tiny{(--)}}}{p}  \to W^{+} Z j \to  \ell_{1}^{+}\nu_{\ell_1} \ell_{2}^{-}
\ell_{2}^{+}j $ & \\
\bf 650 & $p \overset{\mbox{\tiny{(--)}}}{p}  \to Z Z j \to \ell_{1}^{+} \ell_{1}^{-} \ell_{2}^{+} \ell_{2}^{-} j $ &  \multirow{1}{*}{anomalous $HVV$ couplings} \\
& \\
\hline
\end{tabular}
\caption{ \em  Process IDs for diboson production in association with a hadronic jet at
NLO QCD with fully leptonic decays.}
\vspace{0.2cm}
\label{tab:prc7}
\end{center}
\end{table}
\begin{table}[t!]
\begin{center}
\small
\begin{tabular}{c|l|l}
\hline
&\\
\textsc{ProcId} & \textsc{Process} & \textsc{Bsm} \\
&\\
\hline
&\\
\bf 601$^*$ & $p \overset{\mbox{\tiny{(--)}}}{p}  \to W^{+} W^{-} j \to q \bar{q} \ell^{-} \bar{\nu}_{\ell} j $ & \\
\bf 602$^*$ & $p \overset{\mbox{\tiny{(--)}}}{p}  \to W^{+} W^{-} j \to \ell^{+}\nu_{\ell} q \bar{q} j $ & \\
&\\
\hline
\end{tabular}
\caption{ \em  Process IDs for diboson production in association with a hadronic jet at
NLO QCD with semileptonic decays.}
\vspace{0.2cm}
\label{tab:prc7a}
\end{center}
\end{table}


\subsection{Triple vector boson production in association with a hadronic jet}

%
$W\gamma\gamma$ production in association with a hard hadronic jet is available
in {\sc{Vbfnlo}} at NLO QCD accuracy under the process IDs of
Table~\ref{tab:tribPj}. All off-shell and finite width effects
are included. Anomalous vector boson couplings are implemented, 
with the anomalous parameters input via {\tt anomV.dat}. Details can be found in
Ref.~\cite{Campanario:2011ud}.  Note that this class of processes is not enabled
by default and so, in order to run these processes, they must be enabled at
compilation, using the {\tt configure} option {\tt -{}-enable-processes=all} or
{\tt -{}-enable-processes=tribosonjet}.
\begin{table}[t!]
\begin{center}
\small
\begin{tabular}{c|l|l}
\hline
&\\
\textsc{ProcId} & \textsc{Process} & \textsc{Bsm}  \\
&\\
\hline
&\\
\bf 800 & $p \overset{\mbox{\tiny{(--)}}}{p}  \to W^{+} \gamma \gamma j  \to \ell^{+} \nu_{\ell} \gamma \gamma j $ & \ldelim \} {2}{0.8cm} \multirow{2}{*}{anomalous couplings}\\
\bf 810 & $p \overset{\mbox{\tiny{(--)}}}{p}  \to W^{-} \gamma \gamma j \to \ell^{-} \bar \nu_{\ell} \gamma \gamma j $ & \\
& \\
\hline
\end{tabular}
\caption{ \em  Process IDs for triboson production in association with a hadronic jet at
NLO QCD.}
\vspace{0.2cm}
\label{tab:tribPj}
\end{center}
\end{table}

\subsection{Higgs production in association with a $W$}
\label{sec:wh}

The production of a Higgs boson in association with a $W$ boson proceeds via the production of
a single off-shell $W$ boson which radiates the Higgs boson.
The $W$ boson decays leptonically, while the Higgs boson can decay in several channels.

In \textsc{Vbfnlo}, the processes listed in Table~\ref{tab:wh} are implemented to
NLO QCD accuracy, including full off-shell effects and spin correlations of the
final state leptons. 
Anomalous couplings can be input via {\tt anomV.dat} and are used both in the production
process as well as the decay of the Higgs boson.
Details of the calculation can be found in Ref.~\cite{robin}.

\begin{table}[t!]
\newcommand{\lstrut}{{$\strut\atop\strut$}}
\begin{center}
\small
\begin{tabular}{c|l|l}
\hline
&\\
\textsc{ProcId} & \textsc{Process} & \textsc{BSM}  \\
&\\
\hline
&\\
\bf 1300 & $p \overset{\mbox{\tiny{(--)}}}{p} \to W^+ H \to \ell^+\nu_{\ell} H $ & \ldelim \} {10}{0.8cm} \multirow{10}{*}{anomalous gauge couplings} \\
\bf 1301 & $p \overset{\mbox{\tiny{(--)}}}{p} \to W^+ H \to \ell^+\nu_{\ell} \gamma\gamma $ & \\
\bf 1302 & $p \overset{\mbox{\tiny{(--)}}}{p} \to W^+ H \to \ell^+\nu_{\ell} \mu^+\mu^- $ & \\
\bf 1303 & $p \overset{\mbox{\tiny{(--)}}}{p} \to W^+ H \to \ell^+\nu_{\ell} \tau^+\tau^- $ & \\
\bf 1304 & $p \overset{\mbox{\tiny{(--)}}}{p} \to W^+ H \to \ell^+\nu_{\ell} b\bar{b} $ & \\
\bf 1305 & $p \overset{\mbox{\tiny{(--)}}}{p} \to W^+ H \to W^+ W^{+}W^{-} \to \ell_{1}^+\nu_{\ell_{1}} \ell_{2}^+\nu_{\ell_{2}} \ell_{3}^- \bar{\nu}_{\ell_{3}}$ &  \\
\bf 1306 & $p \overset{\mbox{\tiny{(--)}}}{p} \to W^+ H \to W^+ ZZ \to \ell_{1}^+\nu_{\ell_{1}} \ell_{2}^+ \ell_{2}^- \ell_{3}^+ \ell_{3}^-$ & \\
\bf 1307 & $p \overset{\mbox{\tiny{(--)}}}{p} \to W^+ H \to W^+ ZZ \to \ell_{1}^+\nu_{\ell_{1}} \ell_{2}^+ \ell_{2}^- \nu_{\ell_{3}}  \bar{\nu}_{\ell_{3}}$ & \\
&\\
\hline
&\\
\bf 1310 & $p \overset{\mbox{\tiny{(--)}}}{p} \to W^- H \to \ell^- \bar{\nu}_{\ell} H $ & \ldelim \} {10}{0.8cm} \multirow{10}{*}{anomalous gauge couplings} \\
\bf 1311 & $p \overset{\mbox{\tiny{(--)}}}{p} \to W^- H \to \ell^- \bar{\nu}_{\ell} \gamma\gamma $ & \\
\bf 1312 & $p \overset{\mbox{\tiny{(--)}}}{p} \to W^- H \to \ell^- \bar{\nu}_{\ell} \mu^+\mu^- $ & \\
\bf 1313 & $p \overset{\mbox{\tiny{(--)}}}{p} \to W^- H \to \ell^- \bar{\nu}_{\ell} \tau^+\tau^- $ & \\
\bf 1314 & $p \overset{\mbox{\tiny{(--)}}}{p} \to W^- H \to \ell^- \bar{\nu}_{\ell} b\bar{b} $ & \\
\bf 1315 & $p \overset{\mbox{\tiny{(--)}}}{p} \to W^- H \to W^- W^{+}W^{-} \to \ell_{1}^-\bar{\nu}_{\ell_{1}} \ell_{2}^+\nu_{\ell_{2}} \ell_{3}^- \bar{\nu}_{\ell_{3}}$ &  \\
\bf 1316 & $p \overset{\mbox{\tiny{(--)}}}{p} \to W^- H \to W^- ZZ \to \ell_{1}^-\bar{\nu}_{\ell_{1}} \ell_{2}^+ \ell_{2}^- \ell_{3}^+ \ell_{3}^-$ & \\
\bf 1317 & $p \overset{\mbox{\tiny{(--)}}}{p} \to W^- H \to W^- ZZ \to \ell_{1}^-\bar{\nu}_{\ell_{1}} \ell_{2}^+ \ell_{2}^- \nu_{\ell_{3}}  \bar{\nu}_{\ell_{3}}$ & \\
&\\
\hline
\end{tabular}
\caption {\em  Process IDs for the $WH$ production processes at NLO QCD accuracy.}
\vspace{0.2cm}
\label{tab:wh}
\end{center}
\end{table}

\subsection{Higgs production in association with a $W$ and a hadronic jet}
\label{sec:whj}

$WH$ production in association with a hadronic jet is available at NLO QCD accuracy in
{\sc{Vbfnlo}} under the process IDs of Table~\ref{tab:whj}.

Full off-shell effects and spin correlations of the final state leptons are included. 
Anomalous couplings can be input via {\tt anomV.dat} and are used both in the production
process as well as the decay of the Higgs boson.

The virtual contributions of the NLO calculation include diagrams where the Higgs is
radiated off a top quark loop attached to a gluon.
They are calculated including the full top mass dependence.
These contributions can be disabled using the flag {\tt dotoploops} in 
{\tt utilities/global.inc}\footnote{Note
that, if this parameter is changed, {\tt make clean} must be run in
the folder {\tt utilities}, and then {\tt make all
install} must be run for the \textsc{Vbfnlo} code.}.
Details of the calculation can be found in Refs.~\cite{robin,Campanario:2014lza}.

\begin{table}[t!]
\newcommand{\lstrut}{{$\strut\atop\strut$}}
\begin{center}
\small
\begin{tabular}{c|l|l}
\hline
&\\
\textsc{ProcId} & \textsc{Process} & \textsc{BSM}  \\
&\\
\hline
&\\
\bf 1600 & $p \overset{\mbox{\tiny{(--)}}}{p} \to W^+ H \, j \to \ell^+\nu_{\ell} H \, j $ & \ldelim \} {10}{0.8cm} \multirow{10}{*}{anomalous couplings} \\
\bf 1601 & $p \overset{\mbox{\tiny{(--)}}}{p} \to W^+ H \, j \to \ell^+\nu_{\ell} \gamma\gamma \, j $ & \\
\bf 1602 & $p \overset{\mbox{\tiny{(--)}}}{p} \to W^+ H \, j \to \ell^+\nu_{\ell} \mu^+\mu^- \, j $ & \\
\bf 1603 & $p \overset{\mbox{\tiny{(--)}}}{p} \to W^+ H \, j \to \ell^+\nu_{\ell} \tau^+\tau^- \, j $ & \\
\bf 1604 & $p \overset{\mbox{\tiny{(--)}}}{p} \to W^+ H \, j \to \ell^+\nu_{\ell} b\bar{b} \, j $ & \\
\bf 1605 & $p \overset{\mbox{\tiny{(--)}}}{p} \to W^+ H \, j \to W^+ W^{+}W^{-} \, j \to \ell_{1}^+\nu_{\ell_{1}} \ell_{2}^+\nu_{\ell_{2}} \ell_{3}^- \bar{\nu}_{\ell_{3}}\, j $ &  \\
\bf 1606 & $p \overset{\mbox{\tiny{(--)}}}{p} \to W^+ H \, j \to W^+ ZZ \, j \to \ell_{1}^+\nu_{\ell_{1}} \ell_{2}^+ \ell_{2}^- \ell_{3}^+ \ell_{3}^-\, j $ & \\
\bf 1607 & $p \overset{\mbox{\tiny{(--)}}}{p} \to W^+ H \, j \to W^+ ZZ \, j \to \ell_{1}^+\nu_{\ell_{1}} \ell_{2}^+ \ell_{2}^- \nu_{\ell_{3}}  \bar{\nu}_{\ell_{3}}\, j $ & \\
&\\
\hline
&\\
\bf 1610 & $p \overset{\mbox{\tiny{(--)}}}{p} \to W^- H \, j \to \ell^- \bar{\nu}_{\ell} H \, j $ & \ldelim \} {10}{0.8cm} \multirow{10}{*}{anomalous couplings} \\
\bf 1611 & $p \overset{\mbox{\tiny{(--)}}}{p} \to W^- H \, j \to \ell^- \bar{\nu}_{\ell} \gamma\gamma \, j $ & \\
\bf 1612 & $p \overset{\mbox{\tiny{(--)}}}{p} \to W^- H \, j \to \ell^- \bar{\nu}_{\ell} \mu^+\mu^- \, j $ & \\
\bf 1613 & $p \overset{\mbox{\tiny{(--)}}}{p} \to W^- H \, j \to \ell^- \bar{\nu}_{\ell} \tau^+\tau^- \, j $ & \\
\bf 1614 & $p \overset{\mbox{\tiny{(--)}}}{p} \to W^- H \, j \to \ell^- \bar{\nu}_{\ell} b\bar{b} \, j $ & \\
\bf 1615 & $p \overset{\mbox{\tiny{(--)}}}{p} \to W^- H \, j \to W^- W^{+}W^{-} \, j \to \ell_{1}^-\bar{\nu}_{\ell_{1}} \ell_{2}^+\nu_{\ell_{2}} \ell_{3}^- \bar{\nu}_{\ell_{3}}\, j $ &  \\
\bf 1616 & $p \overset{\mbox{\tiny{(--)}}}{p} \to W^- H \, j \to W^- ZZ \, j \to \ell_{1}^-\bar{\nu}_{\ell_{1}} \ell_{2}^+ \ell_{2}^- \ell_{3}^+ \ell_{3}^-\, j $ & \\
\bf 1617 & $p \overset{\mbox{\tiny{(--)}}}{p} \to W^- H \, j \to W^- ZZ \, j \to \ell_{1}^-\bar{\nu}_{\ell_{1}} \ell_{2}^+ \ell_{2}^- \nu_{\ell_{3}}  \bar{\nu}_{\ell_{3}}\, j $ & \\
&\\
\hline
\end{tabular}
\caption {\em  Process IDs for the WH plus jet production processes at NLO QCD accuracy.}
\vspace{0.2cm}
\label{tab:whj}
\end{center}
\end{table}


\subsection{QCD-induced production of a vector boson in association with two jets}

The QCD-induced production of a
vector boson in association with two jets is available
in \textsc{Vbfnlo} at NLO QCD accuracy, including leptonic decays of the vector
boson. The process IDs can be found in Table~\ref{tab:qcdvjj}. These
processes are not included by default and they have to be enabled at
compilation, using the {\tt configure} option {\tt
-{}-enable-processes=X} where {\tt X} is one of {\tt
qcdvjj,qcdvvjj,all}. These processes use quadruple precision to handle
instabilities in the virtual amplitudes. If quadruple precision is not
supported by the compiler it can be disabled by adding the {\tt
configure} option {\tt -{}-disable-quad}. An estimate of the error due
to unstable points is printed out at the end of program execution.

\begin{table}[t!]
\newcommand{\lstrut}{{$\strut\atop\strut$}}
\begin{center}
\small
\begin{tabular}{c|l}
\hline
&\\
\textsc{ProcId} & \textsc{Process}\\
&\\
\hline
&\\
\bf 3120 & $p \overset{\mbox{\tiny{(--)}}}{p} \to Z \,  jj\to \ell^{+} \ell^{-} \, jj$  \\
\bf 3121 & $p \overset{\mbox{\tiny{(--)}}}{p} \to Z \,  jj\to \nu_{\ell} \bar{\nu}_{\ell} \, jj$  \\
\bf 3130 & $p \overset{\mbox{\tiny{(--)}}}{p} \to W^{+} \,  jj\to \ell^{+} \nu_\ell \, jj$ \\
\bf 3140 & $p \overset{\mbox{\tiny{(--)}}}{p} \to W^{-} \, jj\to \ell^{-} \bar{\nu}_\ell  \, jj$\\
\bf 3150 & $p \overset{\mbox{\tiny{(--)}}}{p} \to \gamma \, jj$  \\
&\\
\hline
\end{tabular}
\caption {\em  Process IDs for QCD-induced vector boson + 2 jet production 
at NLO QCD accuracy.}
\vspace{0.2cm}
\label{tab:qcdvjj}
\end{center}
\end{table}


\subsection{QCD-induced diboson production in association with two jets}

%
The QCD-induced production of two vector bosons in association with two
jets is available in \textsc{Vbfnlo} at NLO QCD accuracy for the
processes listed in Table~\ref{tab:qcdvvjj}. All spin correlations and
finite width effects are included. These processes are not included by
default and they have to be enabled at compilation, using the {\tt
configure} option {\tt -{}-enable-processes=qcdvvjj} or {\tt
-{}-enable-processes=all}.  These processes use quadruple precision to
handle instabilities in the virtual amplitudes. If quadruple precision
is not supported by the compiler it can be disabled by adding the {\tt
configure} option {\tt -{}-disable-quad}. An estimate of the error due
to unstable points is printed at the end of program execution. Details
of the calculation can be found in
Refs.~\cite{Campanario:2013qba,Campanario:2013gea,Campanario:2014dpa,
Campanario:2014ioa,Campanario:2014wga,qcdvvjj,Campanario:2020xaf}.

\begin{table}[t!]
\newcommand{\lstrut}{{$\strut\atop\strut$}}
\begin{center}
\small
\begin{tabular}{c|l}
\hline
&\\
\textsc{ProcId} & \textsc{Process}\\
&\\
\hline
&\\
\bf 3210 & $p \overset{\mbox{\tiny{(--)}}}{p} \to ZZ \,  jj\to \ell_{1}^{+} \ell_{1}^{-} \ell_{2}^{+} \ell_{2}^{-} \, jj$  \\
\bf 3211 & $p \overset{\mbox{\tiny{(--)}}}{p} \to ZZ \,  jj\to \ell_{1}^{+} \ell_{1}^{-} \nu_{\ell_{2}} \bar{\nu}_{\ell_{2}} \, jj$  \\
\bf 3220 & $p \overset{\mbox{\tiny{(--)}}}{p} \to W^{+}Z \,  jj\to \ell_{1}^{+} \nu_{\ell_{1}} \ell_{2}^{+} \ell_{2}^{-} \, jj$  \\
\bf 3230 & $p \overset{\mbox{\tiny{(--)}}}{p} \to W^{-}Z \, jj\to \ell_{1}^{-} \bar{\nu}_{\ell _{1}} \ell_{2}^{+} \ell_{2}^{-} \, jj$  \\
\bf 3240 & $p \overset{\mbox{\tiny{(--)}}}{p} \to \gamma\gamma \, jj$  \\
\bf 3250 & $p \overset{\mbox{\tiny{(--)}}}{p} \to W^{+}W^{+} \,  jj\to \ell_{1}^{+} \nu_{\ell_{1}} \ell_{2}^{+} \nu_{\ell_{2}} \, jj$ \\
\bf 3260 & $p \overset{\mbox{\tiny{(--)}}}{p} \to W^{-}W^{-} \,  jj\to \ell_{1}^{-} \bar{\nu}_{\ell_{1}} \ell_{2}^{-} \bar{\nu}_{\ell_{2}} \, jj$ \\
\bf 3270 & $p \overset{\mbox{\tiny{(--)}}}{p} \to W^{+}\gamma \,  jj\to \ell^{+} \nu_{\ell} \gamma \, jj$  \\
\bf 3280 & $p \overset{\mbox{\tiny{(--)}}}{p} \to W^{-}\gamma \, jj\to \ell^{-} \bar{\nu}_{\ell} \gamma \, jj$  \\
\bf 3290 & $p \overset{\mbox{\tiny{(--)}}}{p} \to Z\gamma \,  jj\to \ell^{+} \ell^{-} \gamma \, jj$  \\
\bf 3291 & $p \overset{\mbox{\tiny{(--)}}}{p} \to Z\gamma \,  jj\to \nu_{\ell} \bar{\nu}_{\ell} \gamma\, jj$  \\
&\\
\hline
\end{tabular}
\caption {\em  Process IDs for QCD-induced diboson + 2 jet
production at NLO QCD accuracy.}
\vspace{0.2cm}
\label{tab:qcdvvjj}
\end{center}
\end{table}

%

\subsection{Higgs boson production in gluon fusion with two jets}

\label{GFHjj}
$\cal{CP}$-even and $\cal{CP}$-odd Higgs boson production in gluon
fusion, associated with two additional jets, are processes that first
appear at the one-loop level which, therefore, are counted as leading
order in the strong coupling constant. These processes are simulated
with the full mass dependence of the top- and
bottom-quark\footnote{The value $m_{b}(M_{H})$ in the \msbar scheme is
used for both the mass and the Yukawa coupling of the bottom quark,
while for the top quark we use its on-shell value.} running in the
loop in the Standard Model, in the (complex) MSSM and in a generic
two-Higgs-doublet model. Anomalous $HVV$ couplings (for decays into
$W$ or $Z$ bosons) can be input via {\tt anom\_HVV.dat} and the input
file {\tt ggflo.dat} can be used to define additional settings for
these processes. The file {\tt susy.dat}, Sec.~\ref{sec:susydat}, can
be used to specify the supersymmetric parameters when working in the
MSSM. The processes are not included by default and they have to be
enabled at compilation, using the {\tt configure} option {\tt
  -{}-enable-processes=ggf} or {\tt -{}-enable-processes=all}.
\begin{table}[t!]
\newcommand{\lstrut}{{$\strut\atop\strut$}}
\begin{center}
\small
\begin{tabular}{c|l|l}
\hline
&\\
\textsc{ProcId} & \textsc{Process} & \textsc{Bsm}\\
&\\
\hline
&\\
\bf 4100 & $p \overset{\mbox{\tiny{(--)}}}{p} \to H \, jj $ & MSSM, general 2HDM\\
\bf 4101 & $p \overset{\mbox{\tiny{(--)}}}{p} \to H \, jj\to \gamma\gamma \, jj$ & \ldelim \} {5}{0.8cm} \multirow{5}{*}{MSSM}\\
\bf 4102 & $p \overset{\mbox{\tiny{(--)}}}{p} \to H \, jj\to \mu^+\mu^- \, jj$ &\\
\bf 4103 & $p \overset{\mbox{\tiny{(--)}}}{p} \to H \, jj\to \tau^+\tau^- \, jj$ &\\
\bf 4104 & $p \overset{\mbox{\tiny{(--)}}}{p} \to H \, jj\to b\bar{b} \, jj$ & \\
\bf 4105 & $p \overset{\mbox{\tiny{(--)}}}{p} \to H \, jj\to W^{+}W^{-} \, jj\to \ell_{1}^+\nu_{\ell_{1}} \ell_{2}^- 
\bar{\nu}_{\ell_{2}} \,jj$ &  \ldelim \} {4}{0.8cm} \multirow{3}{*}{\begin{parbox}{3.85cm}{MSSM, general 2HDM, anomalous $HVV$}\end{parbox}}\\
\bf 4106 & $p \overset{\mbox{\tiny{(--)}}}{p} \to H \, jj\to ZZ \, jj\to \ell_{1}^+ \ell_{1}^- \ell_{2}^+ 
\ell_{2}^- \,jj$ & \\
\bf 4107 & $p \overset{\mbox{\tiny{(--)}}}{p} \to H \, jj\to ZZ \, jj\to \ell_{1}^+ \ell_{1}^- \nu_{\ell_{2}}  
\bar{\nu}_{\ell_{2}} \,jj$ & \\

&\\
\hline
\end{tabular}
\caption {\em  Process IDs for LO Higgs boson plus 2 jet
  production via gluon fusion.}
\vspace{0.2cm}
\label{tab:prc6}
\end{center}
\end{table}
The relevant process IDs are given in Table~\ref{tab:prc6}. 
Details of the calculations can be found in
 Refs.~\cite{DelDuca:2001eu,DelDuca:2001fn,DelDuca:2006hk,Klamke:2007cu,GF,Campanario:2010mi}.
%

\subsection{Higgs boson production in gluon fusion with three jets}

%
Adding an extra parton to the Higgs production processes of
Sec.\ref{GFHjj} gives rise Hjjj production via gluon fusion. The
corresponding cross sections are implemented at LO QCD accuracy. A
list of all available modes and corresponding process ID is given in
Table~\ref{tab:prcGFHjjj}. These processes are not included by default
and they have to be enabled at compilation, using the {\tt configure}
option {\tt -{}-enable-processes=ggf} or {\tt
  -{}-enable-processes=all}. These processes use quadruple precision
to handle instabilities in the virtual amplitudes. If quadruple
precision is not supported by the compiler it can be disabled by
adding the {\tt configure} option {\tt -{}-disable-quad}. An estimate
of the error due to unstable points is printed at the end of program
execution.  Details of the calculation can be found in
Refs.~\cite{Campanario:2013mga,Campanario:2014oua}.
\begin{table}[t!]
\newcommand{\lstrut}{{$\strut\atop\strut$}}
\begin{center}
\small
\begin{tabular}{c|l|l}
\hline
&\\
\textsc{ProcId} & \textsc{Process} & \textsc{Bsm}\\
&\\
\hline
&\\
\bf 4200 & $p \overset{\mbox{\tiny{(--)}}}{p} \to H \, jjj $ & MSSM, general 2HDM\\
&\\
\hline
\end{tabular}
\caption {\em  Process ID for LO Higgs boson plus 3 jet
  production via gluon fusion.}
\vspace{0.2cm}
\label{tab:prcGFHjjj}
\end{center}
\end{table}
%

\subsection{Gluon-induced diboson production}

%
Gluon-induced diboson production can be studied separately in \textsc{Vbfnlo} at
LO, using the process IDs in
Table~\ref{tab:gludib}\footnote{These are also included by default as higher
order corrections to diboson production.}.  Continuum production via box
diagrams as well as production via an $s$-channel Higgs boson resonance are included, with
interference effects fully taken into account.  Anomalous $HVV$ couplings can be
included, using {\tt anom\_HVV.dat}.  In the loop diagrams, first and second generation quarks are taken to
be massless and third generation quark masses\footnote{Again, $m_{b}(M_{H})$ in the \msbar scheme is used.} are included.

Furthermore, semileptonic decay modes of the vector bosons are implemented
for the $WW$ and $ZZ$ production processes~\cite{semilep}. 
Anomalous $HVV$ couplings are available.

\begin{table}[t!]
\newcommand{\lstrut}{{$\strut\atop\strut$}}
\begin{center}
\small
\begin{tabular}{c|l|l}
\hline
&\\
\textsc{ProcId} & \textsc{Process} & \textsc{BSM}  \\
&\\
\hline
&\\
\bf 4300 & $p \overset{\mbox{\tiny{(--)}}}{p} \to W^{+}W^{-} \to \ell_{1}^{+} \nu_{\ell_{1}} \ell_{2}^{-}\bar{\nu}_{\ell_{2}} $ & \ldelim \} {4}{0.8cm} \multirow{4}{*}{anomalous $HVV$ couplings}\\
\bf 4330 & $p \overset{\mbox{\tiny{(--)}}}{p} \to ZZ \to \ell_{1}^{-} \ell_{1}^{+}  \ell_{2}^{-} \ell_{2}^{+} $ & \\
\bf 4360 & $p \overset{\mbox{\tiny{(--)}}}{p} \to Z\gamma \to \ell_{1}^{-} \ell_{1}^{+}  \gamma $ & \\
\bf 4370 & $p \overset{\mbox{\tiny{(--)}}}{p} \to \gamma\gamma $ & \\
&\\
\hline
&\\
\bf 4301 & $p \overset{\mbox{\tiny{(--)}}}{p} \to W^{+}W^{-} \to q \bar{q} \, \ell^{-}\bar{\nu}_{\ell} $ & \ldelim \} {3}{0.8cm} \multirow{3}{*}{anomalous $HVV$ couplings}\\
\bf 4302 & $p \overset{\mbox{\tiny{(--)}}}{p} \to W^{+}W^{-} \to \ell^{+} \nu_{\ell} \, q \bar{q} $ & \\
\bf 4331 & $p \overset{\mbox{\tiny{(--)}}}{p} \to ZZ \to q \bar{q} \, \ell^{-} \ell^{+} $ & \\
&\\
\hline
\end{tabular}
\caption {\em  Process IDs for the gluon-induced diboson production at LO (one-loop) QCD.}
\vspace{0.2cm}
\label{tab:gludib}
\end{center}
\end{table}

%

\subsection{Gluon-induced diboson production in association with a hadronic jet}

%
Gluon-induced diboson production in association with a hadronic jet can be studied separately in \textsc{Vbfnlo} at
LO, using the process IDs in
Table~\ref{tab:gludibj}\footnote{These are also included by default as higher
order corrections to diboson plus jet production.}.  Continuum production via box and pentagon
diagrams as well as production via an $s$-channel Higgs boson resonance are included, with
interference effects fully taken into account.  Anomalous $HVV$ couplings can be
included, using {\tt anom\_HVV.dat}. In the loop diagrams, first and second generation quarks are taken to
be massless and third generation quark masses\footnote{Again, $m_{b}(M_{H})$ in the \msbar scheme is used.} are included.
Details of the calculation can be found in Ref.~\cite{Campanario:2012bh}.

\begin{table}[t!]
\newcommand{\lstrut}{{$\strut\atop\strut$}}
\begin{center}
\small
\begin{tabular}{c|l|l}
\hline
&\\
\textsc{ProcId} & \textsc{Process} & \textsc{BSM}  \\
&\\
\hline
&\\
\bf 4600 & $p \overset{\mbox{\tiny{(--)}}}{p} \to W^{+}W^{-} j \to \ell_{1}^{+} \nu_{\ell_{1}} \ell_{2}^{-}\bar{\nu}_{\ell_{2}} j $ & \multirow{1}{*}{anomalous $HVV$ couplings}\\
\bf 4601 & $p \overset{\mbox{\tiny{(--)}}}{p} \to W^{+}W^{-} j \to q \bar{q} \ell^{-}\bar{\nu}_{\ell} j $ & \\
\bf 4602 & $p \overset{\mbox{\tiny{(--)}}}{p} \to W^{+}W^{-} j \to \ell^{+} \nu_{\ell} q \bar{q} j $ & \\
\bf 4650 & $p \overset{\mbox{\tiny{(--)}}}{p} \to ZZ j \to \ell_{1}^{-} \ell_{1}^{+}  \ell_{2}^{-} \ell_{2}^{+} j $ & \multirow{1}{*}{anomalous $HVV$ couplings}\\
&\\
\hline
\end{tabular}
\caption {\em  Process IDs for the gluon-induced diboson production in association with a hadronic jet at LO (one-loop) QCD.  }
\vspace{0.2cm}
\label{tab:gludibj}
\end{center}
\end{table}


\section{Interfacing and Matching to Herwig 7}
\label{sec:herwig}

\textsc{Vbfnlo}'s processes have been interfaced via the BLHA accord
to the \textsc{Herwig}~7 event generator
\cite{Bellm:2015jjp,Bellm:2019zci}. The \textsc{Matchbox} module of
\textsc{Herwig}~7 \cite{Platzer:2011bc} is able to perform the NLO
matching using the subtractive (MC\@ NLO-type) and multiplicative
(Powheg-type) matching to NLO.

Further to this, all processes can also be used for the (N)LO multi-jet
merging as outlined in
\cite{Platzer:2012bs,Bellm:2017ktr}. \textsc{Herwig}~7 can also be
used to perform fixed-order calculations with the VBFNLO matrix
elements, if this eases comparison to the matched or merged
simulations. Matching and merging systematics can then be addressed as
{\it e.g.} performed in \cite{Rauch:2016upa,Jager:2020hkz}.  All VBF
and VBS processes including leptonic decays, but not the semileptonic
modes, have been included.

Note that all the VBF processes are computed in the VBF approximation
only, i.e.\ $s$-channel contributions and interference between $t$-
and $u$-channel diagrams, which both are tiny in the VBF region, are
neglected. BLHAv1~\cite{Binoth:2010xt} stipulates that all Born matrix
elements are calculated by the Monte Carlo program. Therefore, to
prevent inconsistencies, care must be taken that the VBF approximation
is also present in the Born matrix elements. In version 2 of the
interface~\cite{Alioli:2013nda} all matrix elements are provided by
\textsc{Vbfnlo}, so in that case this is ensured automatically. 

We further note that the parameters such as couplings, masses and
widths are supplied as set in the \textsc{Herwig} input files and that
cuts are applied by the \textsc{Matchbox} module, as well. Wihtout
additional adjustments these different setup modes might result in
small differences in between the results obtained from \textsc{Vbfnlo}
standalone as compared to those produced from \textsc{Herwig}. An
integrator module resembling \textsc{Vbfnlo}'s \textsc{Monaco}
algorithm is also available for \textsc{Matchbox}.

Explicit tests of the interface, using a `tuned' comparison for which
all parameters have been made equal for the \textsc{Vbfnlo} and
\textsc{Herwig} runs have been performed with
\textsc{Herwig}~7\footnote{\tt
\url{https://herwig.hepforge.org/}}~\cite{Bellm:2015jjp}.  Example
input files are provided in the directory {\tt regress/runs/BLHA} of
the source code. Using the BLHA interface and a corresponding Monte
Carlo program also allows one to write out \textsc{Vbfnlo} events at
NLO. The corresponding setup for Herwig is also shown in the example
input files mentioned above, and further examples are provided online
within the \textsc{Herwig} tutorials.

\subsection{Usage}
\label{sec:herwigUsage}

If {\sc Herwig 7} is installed using the {\tt herwig-bootstrap}
\footnote{\tt \url{https://herwig.hepforge.org/tutorials/installation/bootstrap.html}},
\textsc{Vbfnlo} will be directly available.
To use a local build of \textsc{Vbfnlo} instead,
the flag {\tt -{}-with-VBFNLO=PATH} can be passed to the {\tt herwig-bootstrap}
or during the configuration step in the {\sc Herwig} compilation.

{\sc Herwig} 7.3 expects \textsc{Vbfnlo} 3.0 to have a custom 
{\sc BLHA} interface where \_VBFNLO have been appended to the standard 
{\sc BLHA} functions. For this reason, if you compile {\sc Vbfnlo} from
source and expect to use it with {\sc Herwig} 7.3 through the 
{\sc BLHA} interface, the -{}-enable-custom-blha flag has to be added to 
the \textsc{Vbfnlo} configuration. Trying to compile {\sc Herwig} 7.3 
with {\sc Vbfnlo} 3.0 without this flag will result in a compilation
error.

To use \textsc{Vbfnlo} as an amplitude provider,
the line {\tt read Matchbox/VBFNLO.in} has to be included in the
{\sc Herwig} input file. The cuts for the process need to be set
in {\sc Herwig} and are automatically passed to \textsc{Vbfnlo}.
If a Frixione Isolation cut has to be set, the line
{\tt set PhotonIsolationCut:CutType VBFNLO} will need to be provided.
The VBF processes inside \textsc{Vbfnlo} are calculated in the
VBF approximation. Telling {\sc Herwig} to not include s-channel
contributions, which are ignored in the VBF approximation, 
can be done with {\tt read Matchbox/VBFDiagramsOnly.in}.

The {\sc BLHA} states that the MC has to drive the process generation
while the OLP has to give the amplitudes asked. In the interface
between \textsc{Vbfnlo} and {\sc Herwig}, an extension allowing the use
of the \textsc{Vbfnlo} phasespace has been implemented. To use the
\textsc{Vbfnlo} phasespace generator, which improves the convergence,
we have to add {\tt read Matchbox/VBFNLOPhasespace.in} to the
{\sc Herwig} input card.

\begin{table}
  \centering
  \begin{tabular}{c | c}
    \hline
    &\\
    \textsc{Line} & \textsc{description} \\
    &\\
    \hline
    \multirow{3}{*}{{\tt read Matchbox/VBFNLO.in}} \\
    & {\sc Herwig} will use \\ 
    & \textsc{Vbfnlo} as an OLP \\
    \\
    \multirow{3}{*}{{\tt set PhotonIsolationCut:CutType VBFNLO}} \\
    & Using the Frixione Isolation \\
    & cut convention in \textsc{Vbfnlo} \\
    \\
    \multirow{4}{*}{{\tt read Matchbox/VBFDiagramsOnly.in}}\\
    & Lets {\sc Herwig} know that only \\
    & t-channel type diagrams contribute \\
    & to the process of interest \\
    \\
    \multirow{1}{*}{{\tt read Matchbox/VBFNLOPhasespace.in}} & Uses the \textsc{Vbfnlo} phasespace \\
    \\
    \hline
  \end{tabular}
  \caption{Summary of relevant lines in the {\sc Herwig} input card
  concerning \textsc{Vbfnlo} with a brief description of their effect.}
  \label{tab:herwigInputLines}
\end{table}
A summary of the relevant lines concerning \textsc{Vbfnlo} 
with a brief description of their effect
may be found in Table~\ref{tab:herwigInputLines}.
A {\sc Herwig} input card, ready to use \textsc{Vbfnlo} as an amplitude
provider, can be found in the {\tt share/VBFNLO} folder.
It is recommended to start from that input card, which can be copied to the
run folder, and make the
appropriate changes for the required process.
For {\sc Herwig} to be able to access the libraries required and
to be able to call the {\tt Herwig} command from any folder,
the pertinent environment variables must be set. If the installation
was done with the {\tt herwig-bootstrap}, it can be done
with {\tt source \$\{INSTALLDIR\}/bin/activate}.

The process is built with the command line {\tt Herwig build \{inputcardname\}},
which creates a {\tt \{runfilename\}} file that is used to generate the events.
Before the generation step, the phasespace grid has to be optimized
with {\tt Herwig integrate \{runfilename\}}.
This step can be parallelized using the flag {\tt -{}-maxjobs=\{numberofjobs\}}
which will create integration jobs that can be integrated separately
with {\tt Herwig integrate -{}-jobid=n \{runfilename\}},
where {\tt n} is the identification number assigned to the
integration job, and
merged afterwards with {\tt Herwig mergegrids}. Then, the event generation
is started with {\tt Herwig run \{runfilename\}}.

As an example, if our input card is named {\tt LHC-Matchbox.in} and
creates a runfile named {\tt LHC.run}, the following commands
\newline
\\
{\tt Herwig build -{}-maxjobs=10 LHC-Matchbox.in} \newline
{\tt for i in \$(seq 0 9)\; do Herwig integrate -{}-jobid=\$i LHC.run \& done} \newline
{\tt Herwig mergegrids} \newline
{\tt Herwig run LHC.run} \newline
\\
will create 10 different integration jobs, merge the grids after
the integration has finished and
make an event generation run.

The comparison with \textsc{Vbfnlo} standalone can be performed by
turning off the parton shower, hadronization and MPI in {\sc Herwig},
and using the same cuts and parameters in both programs.
To help with the possible difficulties with the last step,
curated input cards for \textsc{Vbfnlo} and {\sc Herwig}
can be found inside
{\tt VBFNLO/regress/runs} and {\tt VBFNLO/regress/runs/BLHA}
respectively.


\newpage
\section{\textsc{Input files and parameters}}
\label{sec:input} 
\textsc{Vbfnlo} is steered through the following input files:
\begin{itemize}
\item {\tt vbfnlo.dat:} General parameters for a run.
\item {\tt cuts.dat:} Values for kinematic cuts.
\item {\tt ggflo.dat:} Additional parameters for gluon-fusion induced processes.
\item {\tt susy.dat:} Parameters describing the MSSM scenario.
\item {\tt anom\_HVV.dat:} Parameters for anomalous Higgs boson couplings.
\item {\tt anomV.dat:} Parameters for anomalous gauge boson couplings.
\item {\tt kk\_input.dat:} Settings for the Warped Higgsless and
Three-Site Higgsless Models.
\item {\tt kk\_coupl\_inp.dat:} Numerical values if externally
calculated Kaluza-Klein couplings and masses should be used.
\item {\tt spin2coupl.dat:} Settings for the spin-two models.
\item {\tt histograms.dat:} Histogram options.
\item {\tt random.dat:} Seed for the random number generator.
\end{itemize}

The following subsections give a detailed description of all available parameters.


\subsection{{\tt vbfnlo.dat}}
\label{sec:vbfnlodat} 

{\tt vbfnlo.dat} is the main input file for \textsc{Vbfnlo}.

\subsubsection{{\tt vbfnlo.dat} $-$ general parameters}
\label{sec:general}
\begin{itemize}
\item {\tt PROCESS:} Process ID as described in Sec.~\ref{sec:proc}.  Default is
100: Higgs boson production via VBF.

\item {\tt LOPROCESS\_PLUS\_JET:} If set to {\tt true}, the leading order
  process with one additional jet is generated, i.e.\ only the real radiation
	contribution is generated.  This option is available for all but gluon-induced processes.
	Default is {\tt false}.
\item{\tt LEPTONS:} Choice of the final state leptons (decay products of $W$ and
$Z$ bosons) according to the MC particle numbering scheme~\cite{Amsler:2008zz}:
\begin{eqnarray}
11 &:& e^{-} \nonumber \\
12 &:& \nu_{e} \nonumber \\ 
13 &:& \mu^{-} \nonumber \\
14 &:& \nu_{\mu} \nonumber \\ 
15 &:& \tau^{-} \nonumber \\
16 &:& \nu_{\tau} \nonumber \\
98 &:& \mbox{leptons are either generation 1 or generation 2} \nonumber \\
99 &:& \mbox{any lepton} \nonumber
\end{eqnarray}
The ordering of particles is not important. 
If the selected configuration is not available, the appropriate first-generation
leptons are used as default values.  If {\tt 99(98)} is set, results are output
summed over all possible lepton combinations (all possible lepton combinations
of generation 1 or 2) -- for individual events (in the cuts, histograms and Les
Houches event output) the specific leptons for that particular event are
generated randomly.  
\item{\tt DECAY\_QUARKS:} Choice of the final state quarks from hadronic vector boson decays
(decay products of $W$ and $Z$ bosons) according to the MC particle numbering scheme~\cite{Amsler:2008zz}:
\begin{eqnarray}
1 &:& d \nonumber \\
2 &:& u \nonumber \\ 
3 &:& s \nonumber \\
4 &:& c \nonumber \\ 
5 &:& b \nonumber \\
93 &:& \mbox{all possible combinations of $u, \bar{u}, d, \bar{d}, s, \bar{s}, c, \bar{c}$} \nonumber \\
94 &:& \mbox{all possible combinations of $u, \bar{u}, d, \bar{d}, s, \bar{s}, c, \bar{c}, b, \bar{b}$} \nonumber
\end{eqnarray}
If {\tt 93} is set, results are output summed over all possible quark / anti-quark 
combinations of the first and second generation.
For {\tt DECAY\_QUARKS = 94} additionally $b$ and $\bar{b}$ are considered.
For individual events in the Les Houches event output, the specific quarks for
that particular event are generated randomly, weighted by the contribution
of that specific quark combination.
If the selected configuration is not available, all possible combinations
of first and second generation quarks are used as default values
({\tt DECAY\_QUARKS = 93}).
\item {\tt LO\_ITERATIONS:} Sets the number of iterations for the integration
of LO cross sections. Usually more than one iteration is used in
order to adapt the integration grid and thus improve the efficiency of the
MC integration algorithm\footnote{For all NLO calculations the virtual
  contributions are calculated using the already optimized leading order grid.}. 
For an adapted grid file (see {\tt LO\_GRID})
this parameter can be set to 1. Default is 4.
\item {\tt NLO\_ITERATIONS:} Analogous to  {\tt LO\_ITERATIONS}, but for the real
emission part of an NLO calculation. Since the corresponding phase-space 
is different from the LO configuration, a second independent
MC integration has to be performed. Default is 4.
\item {\tt LO\_POINTS:} Determines the number of phase-space points that are
generated in each iteration. In the last iteration there are $2^N$ points,
where $N$ = {\tt LO\_POINTS}. In each previous iteration, the number of points
is half the value of the following one. Example: For 4 iterations ({\tt
LO\_ITERATIONS = 4}) and {\tt LO\_POINTS = 20}, there are $2^{17}$ generated
points in the first, $2^{18}$ in the second , $2^{19}$ in the third and
$2^{20}\approx 10^6$ in the last iteration\footnote{The virtual contributions
  are calculated for $2^{N}$ points only.}. 
Default is $N={20}$.
\item {\tt NLO\_POINTS:} Similar to {\tt LO\_POINTS}, but for the real emission
part of an NLO calculation.  Default is 20.
\item {\tt LO\_GRID:} Sets the name of the grid files that are generated at the
end of each iteration. Choosing {\tt name} as the input parameter, in each
iteration $X$ a grid file {\tt name.out.X} will be produced in the working
directory. If a grid file {\tt name} is already present in the input directory
(specified by {\tt -{}-input=INPUT}), the program reads in this file when
executed.  Note that some optimised grids for several processes (using the standard cuts
given in the {\tt regress} files) are provided on the \textsc{Vbfnlo} webpage.
\item{\tt FLOOP\_GRID:} Similar to {\tt LO\_GRID}, but for the gluon-induced fermion
loop contributions (the kinematics of which can differ significantly from the LO
kinematics).  The number of iterations used is given by {\tt LO\_ITERATIONS} and the number of points used is {\tt LO\_POINTS--8}.
\item {\tt NLO\_GRID:} Similar to {\tt LO\_GRID}, but for the real emission part
of an NLO calculation.
\item {\tt PHTN\_GRID:} Similar to {\tt NLO\_GRID}, but for the real photon
emission part of an NLO electroweak calculation.
\item {\tt NLO\_SWITCH:} Switch for the NLO part of a process, if available.
If set to {\tt true}, cross sections and histograms are calculated to NLO
QCD  accuracy. Default is set to {\tt false}.
\item{\tt EWCOR\_SWITCH:} Switch for the electroweak corrections (note that this
is only available for VBF Higgs boson production). If set to {\tt true}, cross
sections and histograms are calculated to NLO electroweak  accuracy. This option
can only be used if \textsc{LoopTools} was enabled at compilation.  Default is
set to {\tt false}.  
\item{\tt FERMIONLOOP:} Flag for the gluon-induced fermionic loop processes,
such as $gg \rightarrow WW$ (currently only available for neutral diboson
processes in the ranges 300-370 and 4300-4370). The options are: 
  \begin{itemize}
  \item {\tt 0} switches off these processes
  \item {\tt 1} includes only the box contribution 
  \item {\tt 2} includes only diagrams via an $s$-channel Higgs resonance
  \item {\tt 3} includes both contributions including interference effects. 
  \end{itemize}
The default value is {\tt 3} (all contributions included).  
\item{\tt NLO\_SEMILEP\_DECAY:} This flag defines how the hadronic decay in processes
  with semileptonic vector boson decay should be calculated: 
  \begin{itemize}
  \item {\tt 0} the hadronic decay $V\to q\bar{q}$ is calculated at leading order QCD
  \item {\tt 1} the hadronic decay $V\to q\bar{q}$ is calculated including approximate next-to
                leading order QCD effects:
                A factor of~\cite{Albert:1979ix} $$1+\frac{\alpha_s(Q^2)}{\pi}$$ is applied to the full cross section
                (including off-shell and singly or non-resonant contributions),
                which reproduces the NLO corrections for the on-shell $V\to q\bar{q}$
                decay. The real emission of a gluon is not resolved.
  \end{itemize}
The default value is {\tt 0} (leading order calculation for the decay).
\item {\tt ECM:} The center-of-mass energy, $\sqrt{s}$, of the collider,
measured in GeV. Default is 13000~GeV.
\item {\tt BEAM1, BEAM2:} Define the type of particle of each beam. Possible
options are $+1$ for proton beams and $-1$ for anti-proton beams. Default
is proton-proton collisions, ($+1$, $+1$).
\item {\tt ID\_MUF:} Choice of the factorization scale. 
See Table~\ref{tab:fscales} for a list of available options. Default is 0.\\
For the QCD-induced processes, the following definitions are used (the momentum of the gauge boson system is 
reconstructed from the momenta of the corresponding fermions in the final state):
\begin{align}
H_T &= \sum_{i \in \text{partons}} p_{T,i} + E_T(V_1) + E_T(V_2),\label{eq:HT} \\
H_T^{\prime} &= \sum_{i \in \text{jets}}  p_{T,i} e^{| y_i - y_{12} |} + E_T(V_1) + E_T(V_2),\label{eq:HTprime} 
\end{align}
with
$$ E_T = \sqrt{p_T^2+m^2}, \quad \text{and} \quad y_{12} = \frac{1}{2}(y_1 + y_2).$$
\item {\tt ID\_MUR:} Choice of the renormalization scale. 
See Table~\ref{tab:rscales} for a list of available options. Default is 0.\\
For the QCD-induced processes the definitions of Eqs.~\eqref{eq:HT} and \eqref{eq:HTprime} are used.
\begin{table}[t!]
\newcommand{\lstrut}{{$\strut\atop\strut$}}
\begin{center}
\begin{tabular}{c|p{7.5cm}|c}
\hline
&&\\
{\tt ID\_MUF } & \textsc{Factorization Scale} & \textsc{Process class} \\
&&\\
\hline
&&\\
\multirow{2}{*}{\bf 0} & user defined constant scale set & \multirow{2}{*}{\tt all}\\
 & by {\tt MUF\_USER} & \\
\multirow{2}{*}{\bf 1} & momentum transfer of exchanged &  \multirow{2}{*}{{\tt vbf} (except {\tt Hjjj})} \\
 & gauge boson &  \\
\multirow{2}{*}{\bf 2} & \multirow{2}{*}{$\min(\pt(j_{i})$)} & \multirow{2}{*}{\tt vbf, tribosonjet} \\
 & & \\
\multirow{2}{*}{\bf 3 or 4} & invariant mass of the electroweak system & {\tt diboson, dibosonjet,} \\ 
 & ($V$ / $VV$ / $VH$ / $VVV$) & {\tt triboson, tribosonjet} \\ 
\multirow{2}{*}{\bf 5} &  \multirow{2}{*}{$\sqrt{ \pt(j_1) \times \pt(j_2)}$} & \multirow{2}{*}{\tt ggf}\\ 
 & & \\
\multirow{2}{*}{\bf 6} & \multirow{2}{*}{constant scale = Higgs boson mass} & \multirow{2}{*}{\tt all}\\ 
 & & \\
\multirow{3}{*}{\bf 7} & \multirow{3}{7cm}{Minimum transverse energy of the ``primary'' bosons} & {\tt diboson} (except {\tt W}), \\
 & & {\tt dibosonjet, triboson,} \\
 & & {\tt tribosonjet} \\
\multirow{2}{*}{\bf 8} & \multirow{2}{*}{$\frac{1}{2}H_T$ as defined in Eq.~\eqref{eq:HT}} & \multirow{2}{*}{\tt vbf,qcdvjj,qcdvvjj}  \\
 & & \\
\multirow{2}{*}{\bf 9} & \multirow{2}{*}{$\frac{1}{2}H_T^\prime$ as defined in Eq.~\eqref{eq:HTprime}} & \multirow{2}{*}{\tt vbf,qcdvjj,qcdvvjj}  \\
 & & \\
\multirow{2}{*}{\bf 10} & \multirow{2}{*}{$\frac{1}{2} \left( E_T(j_1j_2) + E_T(V/VV) \right)$} & \multirow{2}{*}{\tt vbf,qcdvjj,qcdvvjj}  \\
 & & \\
\multirow{2}{*}{\bf 11} & \multirow{2}{*}{$E_T(V/VV)$} & \multirow{2}{*}{\tt vbf,qcdvjj,qcdvvjj}  \\
 & & \\
\multirow{2}{*}{\bf 12} & \multirow{2}{*}{$\sqrt{\pt(j_1) \times \pt(j_2)}$} & \multirow{2}{*}{\tt vbf,qcdvjj,qcdvvjj}  \\
 & & \\
&&\\
\hline
\end{tabular}
\caption {\em Factorization scale options. The definition of the process classes can be found in Section~\ref{compilation}.}
\vspace{0.2cm}
\label{tab:fscales}
\end{center}
\end{table}
\begin{table}[t!]
\newcommand{\lstrut}{{$\strut\atop\strut$}}
\begin{center}
\begin{tabular}{c|p{7.5cm}|c}
\hline
&&\\
{\tt ID\_MUR } & \textsc{Renormalization Scale} & \textsc{Process class} \\
&&\\
\hline
&&\\
\multirow{2}{*}{\bf 0} & user defined constant scale set & \multirow{2}{*}{\tt all}\\
 & by {\tt MUR\_USER} & \\
\multirow{2}{*}{\bf 1} & momentum transfer of exchanged &  \multirow{2}{*}{{\tt vbf} (except {\tt Hjjj})} \\
 & gauge boson &  \\
\multirow{2}{*}{\bf 2} & \multirow{2}{*}{$\min(\pt(j_{i})$)} & \multirow{2}{*}{\tt vbf, tribosonjet} \\
 & & \\
\multirow{2}{*}{\bf 3 or 4} & invariant mass of the electroweak system & {\tt diboson, dibosonjet,} \\ 
 & ($V$ / $VV$ / $VH$ / $VVV$) & {\tt triboson, tribosonjet} \\ 
\multirow{2}{*}{\bf 5} & \multirow{2}{*}{$\alpha^{4}_s=\alpha_s(\pt(j_{1})) \times \alpha_s(\pt(j_{2})) \times \alpha^{2}_s(m_{H})$} & \multirow{2}{*}{\tt ggf}\\ 
 & & \\
\multirow{2}{*}{\bf 6} & \multirow{2}{*}{constant scale = Higgs boson mass} & \multirow{2}{*}{\tt all}\\ 
 & & \\
\multirow{3}{*}{\bf 7} & \multirow{3}{7cm}{Minimum transverse energy of the ``primary'' bosons} & {\tt diboson} (except {\tt W}), \\
 & & {\tt dibosonjet, triboson,} \\
 & & {\tt tribosonjet} \\
\multirow{2}{*}{\bf 8} & \multirow{2}{*}{$\frac{1}{2}H_T$ as defined in Eq.~\eqref{eq:HT}} & \multirow{2}{*}{\tt vbf,qcdvjj,qcdvvjj}  \\
 & & \\
\multirow{2}{*}{\bf 9} & \multirow{2}{*}{$\frac{1}{2}H_T^\prime$ as defined in Eq.~\eqref{eq:HTprime}} & \multirow{2}{*}{\tt vbf,qcdvjj,qcdvvjj}  \\
 & & \\
\multirow{2}{*}{\bf 10} & \multirow{2}{*}{$\frac{1}{2} \left( E_T(j_1j_2) + E_T(V/VV) \right)$} & \multirow{2}{*}{\tt vbf,qcdvjj,qcdvvjj}  \\
 & & \\
\multirow{2}{*}{\bf 11} & \multirow{2}{*}{$E_T(V/VV)$} & \multirow{2}{*}{\tt vbf,qcdvjj,qcdvvjj}  \\
 & & \\
\multirow{2}{*}{\bf 12} & \multirow{2}{*}{$\sqrt{\pt(j_1) \times \pt(j_2)}$} & \multirow{2}{*}{\tt vbf,qcdvjj,qcdvvjj}  \\
 & & \\
&&\\
\hline
\end{tabular}
\caption {\em Renormalization scale options. The definition of the process classes can be found in Section~\ref{compilation}.}
\vspace{0.2cm}
\label{tab:rscales}
\end{center}
\end{table}
\item {\tt MUF\_USER:} If {\tt ID\_MUF} is set to 0, this parameter sets the
user defined constant factorization scale measured in GeV. Default is 100~GeV.
\item {\tt MUR\_USER:} If {\tt ID\_MUR} is set to 0, this parameter sets the
user defined constant renormalization scale measured in GeV. Default is 100~GeV.
\item {\tt XIF:} Factor by which the factorization scale is multiplied. May be
used to analyze the scale dependence of differential cross sections. Default is
$1$.
\item {\tt XIR:} Factor by which the renormalization scale is multiplied. May be
used to analyze the scale dependence of differential cross sections. Default is
$1$.
\end{itemize}
Note that alternative scale choices can be implemented in the file {\tt utilities/scales.F90}.


\subsubsection{{\tt vbfnlo.dat} $-$ physics parameters}
\label{sec:physics}

\begin{itemize}
\item {\tt HMASS:} Standard Model Higgs boson mass in GeV. Default value is 126~GeV. 
\item {\tt HTYPE:} Type of Higgs boson produced:
  \begin{itemize}
   \item {\tt HTYPE = 0} : SM Higgs boson, with mass {\tt HMASS}
   \item {\tt HTYPE = 1} : Light $\mathcal{CP}$-even MSSM Higgs boson
   \item {\tt HTYPE = 2} : Heavy $\mathcal{CP}$-even MSSM Higgs boson
   \item {\tt HTYPE = 3} : $\mathcal{CP}$-odd MSSM Higgs boson (Note: this is not produced at LO
unless we are working in the MSSM with complex parameters and include
Higgs-propagator effects at LO.)
  \end{itemize}
  Note that for {\tt HTYPE = 1-3}, the input {\tt HMASS} is not used.  Default
value is 0 (SM Higgs boson).
\item {\tt HSCHEME:} Distribution of Higgs boson produced (VBF-H only, process IDs 100-117,1010).
  \begin{itemize}
   \item {\tt HSCHEME = 0} : on-shell
   \item {\tt HSCHEME = 1} : Breit-Wigner distribution
   \item {\tt HSCHEME = 2} : Passarino CPS scheme~\cite{Passarino:2010qk,Goria:2011wa}
  \end{itemize}
  Default value is 0 (on-shell).
\item{\tt HWIDTH:} Although \textsc{Vbfnlo} can calculate the Higgs total and
partial widths, it is also possible to set the Higgs boson width with this input
parameter.  Default is -999~GeV, which means that the internally calculated
value of the width is used.  A value of -998~GeV uses the value by the LHC HXSWG~\cite{Dittmaier:2011ti}. If a SLHA file is being used, the SLHA value will
be taken rather than the input {\tt HWIDTH}.
\item {\tt MODEL:} This flag determines whether we are working in the SM
(1), the MSSM (2) or the Two-Higgs model (3). Default is SM (1).  Note
that if {\tt HTYPE = 1-3} is chosen with {\tt MODEL = 1}, the code will
run in the Standard Model, but with a Higgs boson mass equal to that
given by the specified MSSM parameters.
\item{\tt H2MASS:} Mass of the second Higgs boson for {\tt MODEL = 3}.
Default value is 126~GeV.
\item{\tt H2WIDTH:} Width of the second Higgs boson for {\tt MODEL = 3}.
Default is -999~GeV, which means that an internally calculated value of
the width is used.
\item{\tt SIN2BA:} Multiplicative factor for the squared $HVV$-coupling of
the first Higgs boson for {\tt MODEL = 3}. Default value is 1.
\item{\tt COS2BA:} Multiplicative factor for the squared $HVV$-coupling of
the second Higgs boson for {\tt MODEL = 3}. Note that unitarity requires that
${\tt SIN2BA}+{\tt COS2BA}=1$. Default value is -999, which means that the
value which fulfills the unitarity relation is chosen.
\item {\tt TOPMASS:} Top-quark mass in GeV. Default value is 172.4~GeV.  If a
SLHA file is being used, the SLHA value will be taken rather than the input {\tt
TOPMASS}.
\item {\tt BOTTOMMASS:} Bottom-quark pole mass in GeV, used in the calculation
of the Higgs boson width and branching ratios.  In the gluon fusion processes
and gluon-induced contributions to diboson production, $m_{b}(M_{H})$ in the \msbar scheme is used
(which is calculated internally from the input pole mass). Default value is
4.855~GeV, which corresponds to $m_{b}^{\overline{\text{MS}}}(m_{b}) = 4.204$~GeV.  The
explicit formula used is given on the \textsc{Vbfnlo} webpage.  If a SLHA file
is being used, the SLHA value will be taken rather than the input {\tt
BOTTOMMASS}.
\item {\tt CHARMMASS:} Charm-quark pole mass in GeV used in the calculation of
the Higgs boson width and branching ratios. Default value is 1.65~GeV,
corresponding to $m_{c}^{\overline{\text{MS}}}(m_{c}) = 1.273$~GeV.  If a SLHA file is
being used, the SLHA value will be taken rather than the input {\tt CHARMMASS}.
\item {\tt TAU\_MASS:} Tau mass in GeV used in the calculation of the Higgs boson
width and branching ratios. Default value is 1.77684~GeV.  If a SLHA file is
being used, the SLHA value will be taken rather than the input {\tt TAU\_MASS}.
\item {\tt EWSCHEME:} Sets the scheme for the calculation of electroweak
parameters. A summary of the six available options is given in
Table~\ref{tab:Schemes}. Note that if {\tt EWSCHEME = 4} is
chosen, all variables in Table~\ref{tab:Schemes} are taken as inputs.  As the
parameters are not independent, this can lead to problems if the input values
are not consistent.  In this scheme, all photon couplings are set according to
the input variable {\tt INVALFA} and all other couplings are set according to
{\tt FERMI\_CONST}.  Note also that the choice of {\tt EWSCHEME} can
have a large effect on the relative size of the electroweak corrections, as the
charge renormalization depends on the way in which the electromagnetic coupling 
in the LO cross section is parametrized. Full details of all
changes and their effects, together with the explicit formulae used, are
available on the \textsc{Vbfnlo} webpage. Default value is 3.
\begin{table}[t!]
\newcommand{\lstrut}{{$\strut\atop\strut$}}
\begin{center}
\begin{tabular}{c|c|l|c}
\hline
&&&\\
{\tt EWSCHEME} & \textsc{Parameter} & \textsc{Default Value} & \textsc{Input/Calculated} \\
&&&\\
\hline
&&&\\
 & {\tt FERMI\_CONST} & $1.16637\times 10^{-5} \ \mathrm{GeV}^{-2}$ & \textsc{Input}\\
 & {\tt INVALFA} & $128.944341122$ & \textsc{Input}\\
\bf 1 & {\tt SIN2W} & $0.230990$ & \textsc{Calculated} \\
 & {\tt WMASS} & $79.9654 \ \mathrm{GeV}$ & \textsc{Calculated}\\
 & {\tt ZMASS} & $91.1876 \ \mathrm{GeV}$ & \textsc{Input}\\
&&&\\
\hline
&&&\\
 & {\tt FERMI\_CONST} & $1.16637 \times 10^{-5} \ \mathrm{GeV}^{-2}$ & \textsc{Input}\\
 & {\tt INVALFA} & $132.340643024$ & \textsc{Calculated}\\
\bf 2 & {\tt SIN2W} & $0.222646$ & \textsc{Input}\\
 & {\tt WMASS} & $ 80.3980 \ \mathrm{GeV}$ & \textsc{Calculated}\\
 & {\tt ZMASS} & $91.1876 \ \mathrm{GeV}$ & \textsc{Input}\\
&&&\\
\hline
&&&\\
 &{\tt FERMI\_CONST} & $1.16637 \times 10^{-5} \ \mathrm{GeV}^{-2}$ & \textsc{Input}\\
 &{\tt INVALFA} & $132.340705199$ & \textsc{Calculated}\\
\bf 3 &{\tt SIN2W} & $0.222646$ & \textsc{Calculated}\\
 &{\tt WMASS} & $80.3980 \ \mathrm{GeV}$ & \textsc{Input}\\
 & {\tt ZMASS} & $91.1876 \ \mathrm{GeV}$ & \textsc{Input}\\
&&&\\
\hline
&&&\\
 &{\tt  FERMI\_CONST} & $1.16637 \times 10^{-5} \ \mathrm{GeV}^{-2}$ & \textsc{Input}\\
 &{\tt  INVALFA} & $137.035999679$ & \textsc{Input}\\
\bf 4 &{\tt  SIN2W} & $0.222646$ & \textsc{Input}\\
 &{\tt  WMASS} & $80.3980 \ \mathrm{GeV}$ & \textsc{Input}\\
 &{\tt  ZMASS} & $91.1876 \ \mathrm{GeV}$ & \textsc{Input}\\
&&&\\
\hline
&&&\\
 &{\tt  INVALFA(ZMASS)} & $ 128.944341122$ & \textsc{Input}\\
\bf 5 &{\tt  SIN2W} & $0.222646$ & \textsc{Calculated}\\
 &{\tt  WMASS} & $80.3980 \ \mathrm{GeV}$ & \textsc{Input}\\
 &{\tt  ZMASS} & $91.1876 \ \mathrm{GeV}$ & \textsc{Input}\\
&&&\\
\hline
&&&\\
 &{\tt  INVALFA(0)} & $137.035999679$ & \textsc{Input}\\
\bf 6 &{\tt  SIN2W} & $0.222646$ & \textsc{Calculated}\\
 &{\tt  WMASS} & $80.398 \ \mathrm{GeV}$ & \textsc{Input}\\
 &{\tt  ZMASS} & $91.1876 \ \mathrm{GeV}$ & \textsc{Input}\\
&&&\\
\end{tabular}
\caption {\em  Electroweak input parameter schemes.}
\vspace{0.2cm}
\label{tab:Schemes}
\end{center}
\end{table}
\item {\tt FERMI\_CONST:} Fermi constant, used as input for the calculation of
electroweak parameters in {\tt EWSCHEME} = 1-4. Default value is $1.16637 \times
10^{-5} \ \mathrm{GeV}^{-2}$.  If a SLHA file is being used, the SLHA value will
be taken rather than the input {\tt FERMI\_CONST}. 
\item {\tt INVALFA:} One over the fine structure constant, used as input for
{\tt EWSCHEME} $=$ 1, 4, 5 and 6.  Within the other schemes this parameter is
calculated. The default value depends on the choice of {\tt EWSCHEME}, as given
in Table~\ref{tab:Schemes}.  If {\tt EWSCHEME = 5} is chosen, the value of
$\alpha$ should be $\alpha(M_{Z})$, whereas if {\tt EWSCHEME = 6} is chosen, the
value of $\alpha$ should be $\alpha(0)$.  In order to ensure backwards
compatibility with previous versions of \textsc{Vbfnlo}, as an alternative {\tt
ALFA}, the fine structure constant, can be used as an input in {\tt vbfnlo.dat},
which is read and used only if {\tt INVALFA} is not present. If a SLHA file is
being used, the SLHA value will be taken rather than the input {\tt INVALFA} or
{\tt ALFA}. 
\item {\tt DEL\_ALFA:} Value of $\Delta \alpha$, where
\begin{equation}
 \alpha(M_{Z}) = \frac{\alpha(0)}{1 - \Delta \alpha}
\end{equation}
This is used as input for {\tt EWSCHEME} $=$ 6.  Default value is $0.059047686$.
 Note that this is only used for the electroweak corrections during the
calculation of the charge renormalization constant.
\item {\tt SIN2W:} Sinus squared of the weak mixing angle. Used as input for
  {\tt EWSCHEME} $= 2$ and 4. Within the other schemes this parameter is
  calculated. Default input value is $0.222646$.  If a SLHA file is being used, the SLHA value will be taken rather than the input {\tt SIN2W}.
\item {\tt WMASS:} $W$ boson mass in GeV. This parameter is calculated in {\tt
EWSCHEME = 1} and {\tt 2}.  Default input value is 80.398~GeV.  If a SLHA file is being used, the SLHA value will be taken rather than the input {\tt WMASS}.
\item {\tt ZMASS:} $Z$ boson mass in GeV. Default value is 91.1876~GeV.  If a SLHA file is being used, the SLHA value will be taken rather than the input {\tt ZMASS}.
\item {\tt ANOM\_CPL:} If set to {\tt true}, anomalous Higgs boson or gauge
boson couplings are used if available for the selected process. Anomalous
coupling parameters are set via the files {\tt anom\_HVV.dat} and {\tt
anomV.dat}.  These are available for   
  \begin{itemize}
    \item $Hjj$, single and double vector production processes in VBF 
    \item diboson production processes, including $WH$
    \item diboson + jet production processes, including $WHj$
    \item triple vector boson production processes
    \item triboson + jet production processes
    \item $Hjj \rightarrow VV jj$ production via gluon fusion
    \item gluon-induced (contributions to) diboson production
  \end{itemize}
  Default is set to {\tt false}.
\item {\tt KK\_MOD:} Option for the Warped Higgsless Model and Three-Site
Higgsless Model. This is available for all $VVjj$ production modes in VBF except $W\gamma$ and
same sign $W^\pm W^\pm$ production and for the triboson processes $W^\pm W^+W^-$,
$W^+W^-Z$ and $W^\pm ZZ$. Default is set to {\tt false}.  Note that this needs
to be enabled at compilation, using the {\tt configure} option {\tt
-{}-enable-kk}.  Kaluza-Klein parameters are specified via the files {\tt
kk\_input.dat} or {\tt kk\_coupl\_inp.dat}.
\item {\tt SPIN2:} Option for the spin-2 models. This is available for the
$VVjj$ production modes in VBF except same sign $W^\pm W^\pm$ production.  Default is set
to {\tt false}. 
Spin-2 parameters are set via
the file {\tt spin2coupl.dat}.
\item {\tt EW\_APPROX:} Option controlling the electroweak corrections in
\textit{Hjj} production via VBF.
  \begin{itemize}
   \item {\tt EW\_APPROX} = 0 : No approximations involved.  This option is not
available when working in the MSSM.
   \item {\tt EW\_APPROX} = 1 : Only top/bottom (and stop/sbottom in the MSSM) loops
are calculated. 
   \item {\tt EW\_APPROX} = 2 : All fermion (and sfermion in the MSSM) loops are
calculated.
   \item {\tt EW\_APPROX} = 3 : MSSM option -- all SM-type (i.e.\ fermions,
gauge and Higgs bosons) and sfermion loops are calculated.
   \item {\tt EW\_APPROX} = 4 : MSSM option -- all MSSM corrections to the Higgs boson
vertex are calculated, together with all SM-type and sfermion corrections
elsewhere.
   \item {\tt EW\_APPROX} = 5 : MSSM option -- all MSSM corrections to the Higgs boson
vertex, the quark vertex and the vector boson self energies are calculated,
together with all SM-type boxes and pentagons (i.e.\ only chargino and
neutralino box and pentagon diagrams are neglected).
  \end{itemize}
  When working in the SM, {\tt EW\_APPROX} options 3-5 are equivalent to {\tt
EW\_APPROX} = 0.  Default is 5: full corrections in the SM, and the most
complete available corrections in the MSSM.
\end{itemize}

\subsubsection{{\tt vbfnlo.dat} $-$ parameters for event output}
\label{sec:lha}
\textsc{Vbfnlo} generates parton level events according to the most
recent Les Houches Accord~(LHA) format~\cite{Alwall:2006yp} and in the
\textsc{HepMC} format~\cite{Dobbs:2001ck} for all processes at leading order,
except for $W\gamma j$, $WZj$ and $W\gamma\gamma j$ production.
For $WH(j)$ production the event output is only available in the case of
a non-decaying Higgs boson.
\begin{itemize}
\item {\tt LHA\_SWITCH:} Switch on or off output of LHA event files. Default is
set to {\tt false}.   Note that LHA event file output is not yet available for
diboson plus jet or triboson plus jet processes.
\item {\tt LHA\_FILE:} Name of output LHA event file.  Default is {\tt ``event.lhe''}.
\item {\tt HEPMC\_SWITCH:} Switch on or off output of \textsc{HepMC} event files. Default is
set to {\tt false}.   Note that \textsc{HepMC} event file output is not yet available for
diboson plus jet or triboson plus jet processes.
\item {\tt HEPMC\_FILE:} Name of output \textsc{HepMC} event file.  Default is
{\tt ``event.hepmc''}.
\item {\tt UNWEIGHTING\_SWITCH:} Option for event weights. If set to {\tt true}, 
  events are unweighted
  (event weight $=$ +1). If set to {\tt false}, events are weighted.  Default is set 
  to {\tt true}.
\item {\tt DESIRED\_EVENT\_COUNT:} With {\tt UNWEIGHTING\_SWITCH = true}, the number of
  unweighted events written out to disk can be specified. For multi-channel processes 
  (for example diboson / triboson production with a final-state photon) this can be
  time-consuming.\\
  {\tt DESIRED\_EVENT\_COUNT = 0} stands for collecting all events generated during the
  normal integration (behaviour of \textsc{Vbfnlo} up to version 2.7 beta 2).\\
  Default is {\tt DESIRED\_EVENT\_COUNT = 0}.
\item {\tt PARTIAL\_UNWEIGHTING:} For some processes or parameter settings it may not be possible
  to get the desired number of unweighted events: Single events with large weights in some
  problematic phase space regions can spoil the unweighting efficiency. For this case
  {\tt PARTIAL\_UNWEIGHTING = true} will give the desired number of events with weight 1
  and additionally a few events with weight > 1 which occured during the event generation.\\
  Default is {\tt PARTIAL\_UNWEIGHTING = false}.
\item {\tt TAUMASS:} Option to include the mass of $\tau$ leptons in the event files. 
Default is set to {\tt false}.\\
All calculations (except for branching ratios) are done with massless $\tau$ leptons.
However, as subsequent programs need the non-zero mass for the $\tau$ decay, \textsc{Vbfnlo}
can include the $\tau$ mass in the output and rescale the momentum consistently.
The $\tau$ leptons in the events have to fulfill $m_{\tau\tau} > 2 \cdot m_\tau$ and 
$m_{\tau\nu_\tau} > m_\tau$, otherwise the inclusion of $m_\tau$ is not possible. 
All events where this requirement is not fulfilled will be dismissed. 
Usually this criterium should be always met when using reasonable cuts.
To ensure that no events are thrown away please use a cut on the invariant lepton pair mass
of at least $m_{\ell\ell} > 2 \, m_\tau$.
\end{itemize}

\noindent {\bf Important note for processes with more than one phase space (usually processes involving
a final state photon):} \nopagebreak \\
\textsc{Vbfnlo} writes the events of different phase spaces block-wise into the event file.
Therefore the event file should always be used completely, otherwise some parts of phase
space are underrepresented. Using only parts of the event file gives only correct results
if the events are taken randomly from the whole file.
This note is put out at the end of the \textsc{Vbfnlo} run for the relevant processes
if event output is requested.

\subsubsection{{\tt vbfnlo.dat} $-$ PDF  parameters}
\label{sec:pdf}
\textsc{Vbfnlo} can use either built-in parton distribution functions (PDFs)
or the \textsc{Lhapdf}~\cite{Whalley:2005nh} library, either in version 5 or 6.  
\begin{itemize}
\item {\tt PDF\_SWITCH:} Flag to choose which PDFs are used. The options are
  \begin{itemize}
    \item {\tt 0} : built-in PDFs CT18~\cite{Hou:2019efy}
    \item {\tt 1} : an interface to \textsc{Lhapdf} is provided
    \item {\tt 2} : built-in PDFs MRST2004qed~\cite{Martin:2004dh} are used at LO and NLO (if this
option is chosen, photon-induced processes can be included when calculating the
electroweak corrections to \textit{Hjj} production via VBF)
    \item {\tt 3} : the built-in PDFs MSTW2008~\cite{Martin:2009iq} are used
   \end{itemize}
 Default is 0.
\end{itemize}

The following options are used only if \textsc{Lhapdf} has been selected (i.e.\ {\tt PDF\_SWITCH~=~1}).
\begin{itemize}
\item {\tt LO\_PDFNAME:} Name of the LO PDF set. 
\item {\tt NLO\_PDFNAME:} Name of the NLO PDF set. 
\item {\tt LO\_PDFMEMBER:} Member PDF of the LO PDF set. Default is 0.
\item {\tt NLO\_PDFMEMBER:} Member PDF of the NLO PDF set. Default is 0.
\end{itemize}
The names of the PDF sets can be found on the \textsc{Lhapdf} 
website\footnote{\label{footnote:pdfsets}\url{https://lhapdf.hepforge.org/pdfsets}}.
In case of {\sc Lhapdf}~6 the file ending ({\tt .LHgrid/.LHpdf}) is no longer needed, but using the {\sc Lhapdf}~5 naming
scheme is still supported.
The list of PDF sets which are already available in {\sc Lhapdf}~6 can be obtained from the {\sc Lhapdf} website as
well\footnote{\url{https://www.hepforge.org/archive/lhapdf/pdfsets/6.0/}}.
The PDF name can also be obtained from the file {\tt PDFsets.index}, which is located in the \textsc{Lhapdf}~5 subfolder 
{\tt share/lhapdf}\footnote{\label{footnote:pdfsetsindex}When using \textsc{Lhapdf}~6 this file is located at
{\tt <LHAPDF-PATH>/share/LHAPDF/pdfsets.index} .}.

For compatibility with earlier versions of \textsc{Vbfnlo}, the following
two variables are also supported for setting the PDF sets when using \textsc{Lhapdf}. If both are
present, {\tt LO\_PDFNAME} and {\tt NLO\_PDFNAME} take precedence over
{\tt LO\_PDFSET} and {\tt NLO\_PDFSET}, respectively.
\begin{itemize}
\item {\tt LO\_PDFSET:} LHAGLUE number for the LO PDF set.
\item {\tt NLO\_PDFSET:} LHAGLUE number for the NLO PDF set.
\end{itemize}
The LHAGLUE numbers can be found on the \textsc{Lhapdf} 
website\textsuperscript{\ref{footnote:pdfsets}}
or in the file {\tt PDFsets.index}, which is located in the \textsc{Lhapdf}~5 subfolder 
{\tt share/lhapdf}\textsuperscript{\ref{footnote:pdfsetsindex}}.


\subsubsection{{\tt vbfnlo.dat} $-$  parameters for output and histograms}
\label{sec:hist}

\textsc{Vbfnlo} provides output for histograms in the following formats:
\textsc{Gnuplot}\footnote{\tt \url{http://www.gnuplot.info/}},
\textsc{Root}\footnote{\tt \url{https://root.cern.ch/}} and \textsc{Topdrawer}\footnote{\tt
\url{https://www.pa.msu.edu/reference/topdrawer-docs/}}, 
as well as raw data tables.  Options controlling the histogram ranges and smearing are defined in
{\tt histograms.dat}. Additional histograms can be defined by the user in the
file {\tt utilities/histograms.F}.
\begin{itemize}
\item {\tt XSECFILE:} Name of output file containing LO and NLO cross sections
with the associated errors.  Default is {\tt xsection}.
\item {\tt ROOT:} Enable output of histograms in \textsc{Root} format. Default
is set to {\tt false}. Additionally, custom \textsc{Root}
histograms can be defined by the user with the file {\tt utilities/rootuserhists.cpp}, which are filled with the
weighted events from Monte Carlo integration.  This option needs to be enabled
when building \textsc{Vbfnlo} using the option {\tt -{}-with-root}.
\item {\tt TOP:} Enable output of histograms  in 
  \textsc{Topdrawer} format. Default is set to {\tt false}.
\item {\tt GNU:} Enable output of histograms in \textsc{Gnuplot}
  format. Default is set to {\tt true}.
\item {\tt DATA:} Enable output of raw data in a directory hierarchy. Additionally, a \textsc{Gnuplot}
                  command file is written into the data output directory.
  Default is set to {\tt true}.
\item {\tt REPLACE:} Switch to overwrite existing histogram output files. 
Default is set to {\tt true}. 
\item {\tt ROOTFILE:} Name of the \textsc{Root} output file. Default is 
{\tt histograms}.
\item {\tt TOPFILE:}  Name of the \textsc{Topdrawer} output file. Default is 
{\tt histograms}.
\item {\tt GNUFILE:}  Name of the \textsc{Gnuplot} output file. Default is 
{\tt histograms}.
\item {\tt DATAFILE:}  Name of the data output directory. Default is 
{\tt histograms}.
\end{itemize}


\subsection{{\tt cuts.dat} $-$ parameters for kinematic cuts}
\label{sec:cuts}
The following general set of cuts has been implemented in \textsc{Vbfnlo}.  Alternative cuts can be added in the file {\tt utilities/cuts.F}. 

A brief documentation on how to implement additional cuts into the file {\tt cuts.F}
can be found on the \VBFNLO{} webpage,
\url{https://ific.uv.es/vbfnlo/} .

\subsubsection{{\tt cuts.dat} $-$ jet-specific cuts}

\begin{itemize}
\item {\tt RJJ\_MIN:} Minimum separation of two identified jets, $\Delta R_{jj}=
\sqrt{\Delta y^{2}_{jj}+\Delta \phi^{2}_{jj}}$, used by the generalised
$k_\perp$ jet finding algorithm~\cite{Seymour:1997kj} that combines all partons.
 Default is 0.8.
\item {\tt Y\_P\_MAX:} Maximum allowed (pseudo)rapidity\footnote{As all quarks and leptons are considered massless in \textsc{Vbfnlo},
       pseudorapidity and rapidity coincide for almost all final state particles, with one exception: Jets which consist of more than one parton
       have a non-vanishing mass and therefore rapidity $\neq$ pseudorapidity.} for observation of final state 
partons (detector edge).  Default is 5.0.
\item {\tt PGENKTJET:} Exponent of the generalised $k_\perp$ algorithm.
  This yields the $k_\perp$ algorithm when setting
  the variable to 1, the Cambridge/Aachen
  algorithm~\cite{Dokshitzer:1997in} for 0 and the anti-$k_\perp$
  algorithm~\cite{Cacciari:2008gp} when setting it to $-1$. 
  Default value of the floating-point number is 1.0.
\item {\tt PT\_JET\_MIN:} List of minimum transverse momenta for identified jets
in descending order. The later values can be omitted if they are the same as
previous values. Default is 20~GeV.
\item {\tt Y\_JET\_MAX:} Maximum allowed rapidity for identified jets. 
Default is 4.5.
\end{itemize}

\subsubsection{{\tt cuts.dat} $-$ lepton specific cuts}

\begin{itemize}
\item {\tt Y\_L\_MAX:} Maximum (pseudo)rapidity for charged leptons. Default is 2.5.
\item {\tt PT\_L\_MIN:} List of minimum transverse momenta for charged leptons
in descending order. The later values can be omitted if they are the same as
previous values. Default is 10~GeV.
\item {\tt MLL\_MIN:} Minimum invariant mass for pairs of charged
  leptons. Whether this applies only to opposite-sign leptons or to all
  pairs is chosen by the variable {\tt MLL\_OSONLY}.
  Default is 15~GeV.
\item {\tt MLL\_MAX:} Maximum invariant mass for pairs of
  charged leptons. Whether this applies only to opposite-sign leptons or to all
  pairs is chosen by the variable {\tt MLL\_OSONLY}.
  Default is $10^{20}$~GeV.
\item {\tt MLL\_OSONLY:} If set to {\tt true}, the variables 
  {\tt MLL\_MIN} and {\tt MLL\_MAX} only apply to pairs of oppositely
  signed leptons, otherwise to all pairs of charged leptons. 
  Default is {\tt true}.
\item {\tt RLL\_MIN:} Minimum separation of charged lepton pairs, $\Delta R_{\ell\ell}$. 
Default is 0.4.
\item {\tt RLL\_MAX:} Maximum separation of  charged lepton pairs, $\Delta R_{\ell\ell}$. 
Default is 50.
\end{itemize}

\subsubsection{{\tt cuts.dat} $-$ photon specific cuts}
\begin{itemize} 
\item {\tt Y\_G\_MAX:} Maximum (pseudo)rapidity for photons. Default is 1.5.
\item {\tt PT\_G\_MIN:} List of minimum transverse momenta for photons
in descending order. The later values can be omitted if they are the same as
previous values. Default is 20~GeV.
\item {\tt RGG\_MIN:} Minimum separation of photon pairs, $\Delta R_{\gamma\gamma}$. 
Default is 0.6.
\item {\tt RGG\_MAX:} Maximum separation of photon pairs, $\Delta R_{\gamma\gamma}$. 
Default is 50.
\end{itemize}
\VBFNLO{} has a photon isolation cut implemented as defined in Ref.~\cite{Frixione:1998jh},
\begin{equation}
 \sum_{i} E_{T_{i}} \theta(\delta - R_{i\gamma}) \le \epsilon \mbox{ } p_{T_{\gamma}} \left(\frac{1 - \cos \delta}{1 - \cos \delta_{0}}\right)^n \mbox{   for all } \delta < \delta_{0} \,,
\end{equation}
where $i$ is a parton with transverse energy $E_{T_{i}}$ and a separation $R_{i\gamma}$ with a photon of transverse momentum $p_{T\gamma}$.
The parameters $\delta_{0}$ and $\epsilon$ can be adjusted:
\begin{itemize} 
\item {\tt PHISOLCUT:} Photon isolation $\delta_{0}$.  Default is 0.7.
\item {\tt EFISOLCUT:} Efficiency $\epsilon$ of photon isolation cut. Default is 1.
\item {\tt EXISOLCUT:} Exponent $n$ of the photon isolation cut. Default is 1.
\end{itemize}

\subsubsection{{\tt cuts.dat} $-$ additional cuts}
\begin{itemize}
\item {\tt RJL\_MIN:} Minimum separation of an identified jet and a charged
lepton, $\Delta R_{j\ell}$. Default is 0.6. 
\item {\tt RJG\_MIN:}  Minimum separation of an identified jet and a photon,
$\Delta R_{j\gamma}$. Default is 0.6. 
\item {\tt RLG\_MIN:}  Minimum separation of a charged lepton and a photon,
$\Delta R_{\ell\gamma}$. Default is 0.6. 
\item {\tt MLG\_MIN:} Minimum invariant mass for any combination of a charged lepton and a photon. Default is 0~GeV.
\item {\tt MLG\_MAX:} Maximum invariant mass for any combination of a charged lepton and a photon. Default is $14$~TeV.
\item {\tt PTMISS\_MIN:} Minimum missing transverse momentum of the event $${p}_T^{miss}=-\sum_i p_{T,i} \;,$$ summing over
      all visible jets, leptons and photons. Default is 0~GeV.
\item {\tt MJ\_WINDOWV:} To exclude jet/dijet/trijet masses with $|m_\text{jets} - (M_W+M_Z)/2|$ smaller than this cut value. 
Default is 0~GeV. $M_W$ and $M_Z$ are the physical masses here. 
\end{itemize}

\subsubsection{{\tt cuts.dat} $-$ jet veto}
\begin{itemize}
\item {\tt JVETO:} If set to {\tt true}, a jet veto is applied. For processes with fully leptonic decays of
the vector bosons the following criteria apply: 
\begin{itemize}
\item For {\tt vbf} and {\tt ggf} processes it is applied to central jets beyond the two tagging
jets, where the central region is bounded by the rapidities of the two tagging jets.
\item For all other processes the jet veto is applied to additional jets beyond
the leading-order number, ordered by decreasing transverse momentum.
\end{itemize}
For processes with one hadronically decaying vector boson, one or two additional jets are allowed,
depending on the value of {\tt SINGLE\_DECAYJET} (one jet for {\tt SINGLE\_DECAYJET = 2}, else
two jets).\\
Default is {\tt false}.
\item {\tt YMAX\_VETO:} Maximum rapidity of the additional jet. Default is 4.5.
\item {\tt PTMIN\_VETO:} Minimum transverse momentum of the additional jet. Default is 50~GeV.
\item {\tt DELY\_JVETO:} Minimum rapidity separation of a central jet from the two
tagging jets for {\tt vbf} and {\tt ggf} processes. Default is 0.
\end{itemize}

\subsubsection{{\tt cuts.dat} $-$ VBF specific cuts}

These cuts apply only to {\tt vbf}, {\tt qcdv(v)jj} and {\tt ggf} processes.
Furthermore, they apply to the processes with semileptonic decays if {\tt VBFCUTS\_ALWAYS = true}.
\begin{itemize}
\item {\tt ETAJJ\_MIN:} Minimum required rapidity gap, $\Delta \eta_{jj}$,
between  two tagging jets, which are the two leading jets in a $\pt$ ordering for fully leptonically
decaying vector bosons. For tagging jet definition in semileptonic decay processes see
{\tt DEF\_TAGJET}. Default is 0.
\item {\tt YSIGN:} If set to {\tt true}, the two tagging jets are required to
be found in opposite detector hemispheres. Default is {\tt false}.
\item {\tt LRAPIDGAP:} If set to {\tt true} all charged leptons are
required to lie between the two tagging jets in rapidity. Default is {\tt false}.
\item {\tt DELY\_JL:} Minimum rapidity distance of the charged leptons from the tagging jets,
  if {\tt LRAPIDGAP} is set to {\tt true}.  Default is 0.
\item {\tt GRAPIDGAP:} If set to {\tt true} all photons are
required to lie between the two tagging jets in rapidity. Default is {\tt false}.
\item {\tt DELY\_JG:} Minimum rapidity distance of photons from tagging jets,
  if {\tt GRAPIDGAP} is set to {\tt true}.  Default is 0.
\item {\tt MDIJ\_MIN:} Minimum dijet invariant mass of two tagging
  jets. Default is 0~GeV.
\item {\tt MDIJ\_MAX:} Maximum dijet invariant mass of
two tagging jets. Default is $10^{20}$~GeV. 
\end{itemize}

\subsubsection{{\tt cuts.dat} $-$ special cuts for semileptonic decays}

These cuts apply only to processes with one hadronically decaying vector boson.
\begin{itemize}
\item {\tt DEF\_TAGJET:} Switch for different tagging jet definitions.
\begin{itemize}
 \item {\tt DEF\_TAGJET = 1}: two jets with largest $p_T$
 \item {\tt DEF\_TAGJET = 2}: two jets with largest $p_T$ and $|y| > \texttt{ETA\_CENTRAL}$. Furthermore both jets
       have to lie in opposite detector hemispheres.
 \item {\tt DEF\_TAGJET = 3}: two jets with largest separation in rapidity
 \item {\tt DEF\_TAGJET = 4}: two jets with largest $p_T$, except for the two jets (or one jet if {\tt SINGLE\_DECAYJET > 0}
       and the mass of a single jet is closer to $m_V$ than any two-jet combination)
       with an invariant mass closest to $m_V$.
\end{itemize}
      Default is 1.
\item {\tt ETA\_CENTRAL:} Defines the outer limit of the ``central region'' if {\tt DEF\_TAGJET = 2}. For other values
      of {\tt DEF\_TAGJET} the ``central region'' is defined by the rapidity values of the two tagging jets.
      Default is 2.
\item {\tt PTMIN\_TAG\_1:} Minimum transverse momentum for the harder tagging jet. Default is 20 GeV.
\item {\tt PTMIN\_TAG\_2:} Minimum transverse momentum for the softer tagging jet. Default is 20 GeV.
\item {\tt HARD\_CENTRAL:} If set to {\tt true} an additional jet is required within the central region. Default is {\tt false}.
\item {\tt PTMIN\_CENTRAL:} Minimum transverse momentum for the additional hard jet in the central region if 
      {\tt PTMIN\_CENTRAL = true}. Default is 20~GeV. 
\item {\tt VBFCUTS\_ALWAYS:} Usually the VBF cuts are only applied for VBF processes.
      With {\tt VBFCUTS\_ALWAYS = true}, the VBF cuts are applied to the semileptonic 
      diboson and triboson production processes as well, as they supplement
      the $s$-channel part of the electroweak $V(V)\, jj$ production to the VBF processes.\\
      Default is {\tt false}.
\item {\tt RECONST\_HAD\_V:} Choose between different options for a cut on the reconstructed invariant mass of
      the hadronically decaying vector boson. All possible jet combinations are taken into account, the one
      which gives an invariant mass closest to the vector boson mass is used to apply the cut.
\begin{itemize}
 \item {\tt RECONST\_HAD\_V = 0}: No cut on the reconstructed invariant vector boson mass.
 \item {\tt RECONST\_HAD\_V = 1}: Use two-jet invariant mass, tagging jets are not considered.
 \item {\tt RECONST\_HAD\_V = 2}: Use invariant mass of a single jet, tagging jets are not considered (``subjet-analysis'',
       only needed when {\tt SINGLE\_DECAYJET > 0}).
 \item {\tt RECONST\_HAD\_V = 3}: Use one- or two-jet invariant mass, tagging jets are not considered.
\end{itemize}
       Default is 0.
\item {\tt V\_MASS\_RANGE:} Mass range for the cut on the reconstructed invariant vector boson mass:
      {\tt |m\_V - m\_reconst| < V\_MASS\_RANGE}. Default is 20 GeV.
\item {\tt SINGLE\_DECAYJET:} This switch controls the number of additional jets which is required
      for processes with semileptonic decay of the vector bosons with respect to the fully leptonic process:
\begin{itemize}
 \item {\tt SINGLE\_DECAYJET = 0}: two jets are required from the hadronic decay (plus two tagging jets
       for the VBF processes).
 \item {\tt SINGLE\_DECAYJET = 1}: a single jet is allowed if both quarks from the hadronic decay are collimated into one jet.
 \item {\tt SINGLE\_DECAYJET = 2}: a single jet is allowed for all phase space points.
\end{itemize}
      {\tt SINGLE\_DECAYJET = 2} can only be used with {\tt LOPROCESS\_PLUS\_JET} 
      set to {\tt false} in {\tt vbfnlo.dat}.\\
      Default is {\tt 0}.
\item {\tt QSQAMIN\_ZDEC:} Minimal photon virtuality of hadronically decaying $Z/\gamma^*$. \\
      Default is 0~GeV$^2$. However, any values below the $Q^2$ threshold (see Section~\ref{lowvirt}) 
      for $u/d$-quarks of $(0.373\,\text{GeV})^2$ have no effect.\\
      As the Higgs boson does not couple directly to photons, no special treatment or cut is needed for
      process 1010. Therefore {\tt QSQAMIN\_ZDEC} is not evaluated there.\\
\end{itemize}

\subsubsection{Treatment of hadronic decays of low-virtuality photons}
\label{lowvirt}

      As the hadronic decay of a $Z$ boson also includes the hadronic decay of photons,
      a divergence in $\gamma^*\to q\bar{q}$ occurs, when the final state quarks are treated massless
      and no cuts are applied.
      In the case of the semileptonic decays within \textsc{Vbfnlo}, this can lead to problems in two cases:
      \begin{itemize}
       \item In the real emission case with one additional visible jet from QCD radiation both decay quarks
             are allowed to form a single jet. 
       \item This can also happen already at leading order, when {\tt SINGLE\_DECAYJET} is set to {\tt true}.
      \end{itemize}
      Therefore we apply a cut on the minimal photon virtuality, which is estimated from $e^+e^-\to \text{hadrons}$
      individually for each final state quark flavour. The threshold for each quark is chosen such
      that the NLO approximation for $\sigma(e^+e^-\to \text{hadrons})$ gives the same contribution as the experimental
      continuum data plus the contribution from the sharp resonances of the respective quark flavours \cite{Beringer:1900zz}.\\
      This procedure approximates the correct rates from low-$q^2$ photons. The kinematics of the quarks
      in this region are not modeled correctly, but this is of minor importance as the low-$q^2$ region
      is only relevant for semileptonic decays when the decay products form a single jet.\\
      Unfortunately, this procedure can cause problems when interfacing \textsc{Vbfnlo} with
      a parton shower Monte Carlo program by using the Les Houches event output without jet cuts or
      with {\tt SINGLE\_DECAYJET = true}.
      Events which involve $q\bar{q}$ pairs with an invariant mass below 1 GeV or below $2 \cdot m_q$ may
      be dismissed by the parton shower Monte Carlo. This can be avoided by using the option
      {\tt QSQAMIN\_ZDEC}.


\subsection{{\tt ggflo.dat} $-$ general parameters for gluon fusion processes}

In \textsc{Vbfnlo} the double real-emission corrections to $gg\to \phi$,  which
lead to $\phi$ + 2~jet events at order $\alpha_s^{4}$, are
included\cite{Campanario:2010mi}. Here, $\phi$ can be set to be a Standard Model
Higgs boson or any of the three neutral MSSM Higgs bosons (including mixing
between $h$, $H$ and $A$ in the real or complex MSSM) by using the variables
{\tt HTYPE} and {\tt MODEL} in {\tt vbfnlo.dat}, or a mixture of scalar and
pseudoscalar Higgs bosons as in a generic two-Higgs-doublet model (2HDM) of type
II.  Contributions contain top- and bottom-quark triangles, boxes and pentagon
diagrams, i.e.\ the full mass dependence of the loop induced
production\footnote{The bottom quark mass is taken as $m_{b}(M_{H})$ in the \msbar scheme.}.
Interference effects between loops with bottom- and top-quarks, as well as
between ${\cal{CP}}$-even and ${\cal{CP}}$-odd couplings of the heavy quarks,
are fully taken into account. An option to use the large top mass approximation,
which works well for intermediate Higgs boson masses (provided that the
transverse momenta of the final state partons are smaller than the top-quark
mass), is also implemented.

If Higgs boson plus two jet production via gluon fusion is selected in {\tt
vbfnlo.dat}, i.e.\ {\tt PROCESS = 4100-4107}, the following additional
parameters can be adjusted in the {\tt ggflo.dat} file:

\begin{itemize}
\item {\tt Q\_LOOP:} Input that sets how the Higgs boson coupling is determined. 
  \begin{itemize}
   \item {\tt Q\_LOOP = 0:} Effective theory in the large top-quark mass limit
($m_{t} \rightarrow \infty$).
   \item {\tt Q\_LOOP = 1:} Coupling derived from top-quark loop.
   \item {\tt Q\_LOOP = 2:} Coupling derived from bottom-quark loop.
   \item {\tt Q\_LOOP = 3:} Coupling derived from both top- and bottom-quark loops.
  \end{itemize}
 Default is set to  {\tt 3}.
\item {\tt SUBPRQQ:} Switch that determines whether the subprocesses with a quark-quark initial
  state are included. Default is set to {\tt true}.  
\item {\tt SUBPRQG:} Switch that determines whether the subprocesses with a quark-gluon initial
  state are included. Default is set to {\tt true}.  
\item {\tt SUBPRGG:} Switch that determines whether the subprocesses with a gluon-gluon initial
  state are included. Default is set to {\tt true}.  

\item {\tt HIGGS\_MIX:} Switch for Higgs mixing.  If set to 0, there is no Higgs 
mixing, which is the default.  If set to 1, the mixing is determined via the user-input variables:
   \begin{itemize}
    \item {\tt CP\_EVEN\_MOD:} Changes the strength of the $\cal{CP}$-even
            coupling. Default is set to  1d0.
    \item {\tt CP\_ODD\_MOD:} Changes the strength of the $\cal{CP}$-odd
            coupling. Default is set to 1d0.
   \end{itemize}
  using the Lagrangian
   \begin{eqnarray}
     {\cal L}_{\rm Yukawa}=\overline{q} \, (C_{\rm even} \, y_q + \ii \,C_{\rm odd}\, \gamma_5 \,\tilde{y}_q)\, q\,  \Phi
   \end{eqnarray} 
 with $C_{\rm even}$ = {\tt CP\_EVEN\_MOD} and $C_{\rm odd}$ = {\tt CP\_ODD\_MOD} and
\begin{eqnarray}
 y_{b}  &=& \tilde{y_{b}} = \frac{1}{v} m_{b}\tan\beta  \nonumber \\
 y_{t}  &=& \tilde{y_{t}} = \frac{1}{v} m_{t}\cot\beta  \, .
\end{eqnarray}
When working in the Standard Model, $\tan \beta$ is set to 1.  Note that this
option is not yet available when decays of the Higgs boson to fermions or photons are included. The
option {\tt HIGGS\_MIX~=~2} is only used when working in the MSSM, and
incorporates mixing between all three neutral Higgs bosons according to the Z
propagator matrix, as follows
 \begin{eqnarray}
   h_{i} &=& Z_{ih} \, h_{\rm MSSM} + Z_{iH}\, H_{\rm MSSM} + Z_{iA}\, A_{\rm MSSM} \,,
 \end{eqnarray}
 where $i$ = {\tt HTYPE}.

\end{itemize}


\subsection{{\tt susy.dat} $-$ parameters in the MSSM}
\label{sec:susydat}
The file {\tt susy.dat} is used to specify the supersymmetric parameters when
working in the MSSM.  This file is used either if {\tt MODEL = 2} in {\tt
vbfnlo.dat}, or if {\tt HTYPE = 1-3} in {\tt vbfnlo.dat}.  As described earlier,
if {\tt MODEL = 2} the code will run in the MSSM, which means that the Higgs
boson masses, widths and couplings will all be set to the MSSM values.  If {\tt
MODEL = 1} but {\tt HTYPE = 1-3}, the code will run in the Standard Model, and
only the Higgs boson mass will be affected.  Consequently, this file affects
processes 100-107 (Higgs boson plus 2 jet production via VBF), processes 110-117
(Higgs boson plus 3 jet production via VBF), processes 2100-2107 (production of
a Higgs boson in association with a photon and 2 jets via VBF) and gluon fusion
processes 4100-4107.  In particular, the electroweak corrections to Higgs boson
production via VBF are affected by the inputs of {\tt susy.dat}.  As stated
earlier, the Higgs boson contributions to the production of massive gauge bosons
will be affected by {\tt susy.dat} by fixing the Higgs boson mass and couplings,
but the code will only provide a reasonable approximation to the full MSSM
result in the decoupling region (i.e.\ when the Higgs is SM-like).

\begin{itemize}
 \item {\tt FEYNH\_SWITCH:} Determines whether \textsc{FeynHiggs} is used to
calculate the MSSM Higgs boson sector.  Default is {\tt false}.  When working in the
MSSM, especially when including electroweak corrections, it is recommended that
\textsc{FeynHiggs} is used, as the Higgs boson mixing and the renormalization scheme used
in \textsc{Vbfnlo} are consistent with those used by \textsc{FeynHiggs}.  If a
SLHA file is used instead of \textsc{FeynHiggs}, inconsistencies may occur in
the calculation of parameters.
 \item {\tt SLHA\_SWITCH:} Determines whether the MSSM parameters are taken from
a SLHA file.  Default is {\tt true}.  If set to {\tt true} the values in the
SLHA input file are used instead of internal settings or calculations\footnote{Note that a SLHA file can also be used as the input for \textsc{FeynHiggs} by setting both {\tt FEYNH\_SWITCH} and {\tt SLHA\_SWITCH} to {\tt true}.}.
 \item {\tt SLHA\_FILE:} Name of the SLHA input file.  An example file --
{\tt sps1a.slha} -- is provided, which is the default.
 \item {\tt BENCH:}  Various benchmark scenarios are pre-programmed in the code,
as an alternative to using a SLHA file.  These are:
   \begin{itemize}
    \item {\tt 0:} User-input scenario  (see below for further inputs)
    \item {\tt 1:} $M_{h}^{\max}$ scenario ~\cite{Carena:2002qg}
    \item {\tt 2:} No-mixing scenario ~\cite{Carena:2002qg}
    \item {\tt 3:} Gluophobic scenario ~\cite{Carena:2002qg}
    \item {\tt 4:} Small $\alpha_{\rm eff}$ scenario ~\cite{Carena:2002qg}
    \item {\tt 5:} CPX scenario ~\cite{Schael:2006cr}
    \item {\tt 10:} SPS1a\footnote{The SPS points are defined with low-energy
parameters, as given at {\tt \href{https://www.ippp.dur.ac.uk/~georg/sps/}{https://www.ippp.dur.ac.uk/$\sim$georg/sps/}}).}
~\cite{Allanach:2002nj}
    \item {\tt 11:} SPS1b
    \item {\tt 12-19:} SPS 2 -- 9
   \end{itemize}
  Default is {\tt 1} -- the $M_{h}^{\max}$ scenario. 
 \item {\tt PROPLEVEL:}  Level at which Higgs propagator corrections are included:
   \begin{itemize}
    \item {\tt 0:} Effective Higgs-mixing angle used.
    \item {\tt 1:} Propagator factors included at leading order.
    \item {\tt 2:} Propagator factors included at leading order and loop level.
    \item {\tt 3:} Propagator factors included as an additional loop correction.
   \end{itemize}
  Default is {\tt 1}.  These options are discussed in more detail on the \textsc{Vbfnlo} webpage, as well as in~\cite{Figy:2010ct}.
  \item {\tt DELMB\_SWITCH:} Switch determining whether or not to correct the bottom-quark
Yukawa coupling.  Default is {\tt false}.
  \item {\tt MH\_LOOPS:} Flag determining the value of the internal Higgs masses used in the electroweak loops:
   \begin{itemize}
    \item {\tt 0:} Tree level Higgs masses used in loops.
    \item {\tt 1:} Corrected Higgs masses used in loops.
   \end{itemize}
  \item {\tt LOOPSQR\_SWITCH:} Flag determining whether the squared contributions of the electroweak corrections from the (s)fermion sector will be included -- i.e.\ if set to {\tt true} the amplitude is given by:
     \begin{equation}
       |\mathcal{M}_{\text{Born}}|^{2} + 2 \Re\left[\mathcal{M}^{*}_{\text{Born}} \mathcal{M}_{loop} \right] + |\mathcal{M}_{\text{(s)fermion loop}}|^{2}
     \end{equation}
    Note that the loop squared component is only added if $|\mathcal{M}_{\text{(s)fermion loop}}|$ is greater than 10\% of $|\mathcal{M}_{\text{Born}}|$.  Default is set to {\tt true}.
\end{itemize}
If a SLHA file is not being used, and {\tt BENCH = 0-5}, the following inputs are also needed.
\begin{itemize}
  \item {\tt TANB:} Value of $\tan \beta$.  Default value is 10.  Note that for
the SPS benchmarks ({\tt BENCH = 10-19}), this value of $\tan \beta$ is not
used.
  \item {\tt MASSA0:} Mass of $\cal{CP}$-odd Higgs boson $M_{A}$.  This should be used
when working in the MSSM with real parameters. Default value is 400~GeV.  Note
that for the SPS benchmarks ({\tt BENCH=10-19}), this value of $M_{A}$ is not
used.
  \item {\tt MASSHP:} Mass of charged Higgs boson $M_{H^{\pm}}$.  This should
be used when working in the MSSM with complex parameters.  Default value is -1
GeV. 
\end{itemize}
 Note that it is standard to use as input the mass of the $\cal{CP}$-odd Higgs boson,
$M_{A}$, when working in the MSSM with real parameters and the mass of the
charged Higgs boson, $M_{H^{\pm}}$, when working in the MSSM with complex
parameters\footnote{This is because, when working with complex parameters, there
is mixing between all three neutral Higgs bosons, and thus all neutral Higgs boson
masses receive loop corrections.}.  The mass that is not being used as input should
be set to -1~GeV.

If desired, the SUSY breaking parameters that define the scenario can be input
by the user, by selecting {\tt BENCH = 0}.  Default values are those for SPS1a. 
These parameters are (in the notation used by \textsc{FeynHiggs}):
\begin{itemize}
 \item {\tt M3SQ} etc.: The soft SUSY breaking parameters for the sfermion section.
 \item {\tt AT} etc.: Trilinear coupling parameters.
 \item {\tt MUE:} Higgs boson mixing parameter.
 \item {\tt M\_1} etc.: Gaugino mass parameters.
\end{itemize}
By default, lower generation parameters are set to the corresponding higher generation parameters -- e.g.\ {\tt Ac = At}.


\subsection{Parameters for anomalous couplings}

\textsc{Vbfnlo} supports anomalous $HVV$ couplings, where $V=W,Z,\gamma$, in
both the production and the decay of a Higgs boson in VBF-type reactions,  i.e.
for process IDs 100-107. Anomalous $HVV$ couplings are also included in gluon-induced diboson (plus jet) production (processes 430x, 433x, 4360, 4370, 4600, 4650 and processes 30x,
33x, 360, 370, 600 and 650), as well as in gluon fusion processes $Hjj \rightarrow VVjj$
(IDs 4105-7). The anomalous $HVV$ couplings can be specified in the {\tt
anom\_HVV.dat} input file.

Anomalous triple and quartic gauge boson couplings are available for single
and double vector boson production via VBF (process IDs 120-150, 200-291)~\cite{anomWW,anomVBF},
all triple vector boson production processes (process IDs 400-530, 800, 810)~\cite{anomVVV,anomVBF}, diboson
production $WW$, $W^{\pm}\gamma$, $W^{\pm}Z$ and $WH$ (process IDs 300-323, 340-350, 1300-1317)~\cite{johannes,robin}
and for diboson plus jet processes $W^{\pm}\gamma j$~\cite{Campanario:2010xn}, $W^{\pm}Zj$~\cite{Campanario:2010hv}
and $WH j$~\cite{robin} (IDs 610-640, 1600-1617). The respective
parameters are set in the input file {\tt anomV.dat}.  Note, however, that not all
parameters in {\tt anomV.dat} affect all processes and no neutral triple gauge boson couplings are included.
By altering the triple gauge boson couplings (in {\tt anomV.dat}), the couplings between a Higgs and a
pair of gauge bosons are also affected -- these changes are taken into account
automatically for internal and external Higgs bosons occurring in these processes (and during the Higgs width
calculations). 

\subsubsection{{\tt anom\_HVV.dat} $-$ anomalous $HVV$ couplings}
\label{sec:HVV}
The file {\tt anom\_HVV.dat} controls the anomalous Higgs boson coupling parameters. 
It is used if the input {\tt ANOM\_CPL} in {\tt vbfnlo.dat} is set to {\tt true}. 
Among the anomalous coupling input parameters, the user can choose between three
different parametrizations.

The anomalous $HVV$ couplings can be described by the following effective Lagrangian involving the dimension five operators
\begin{eqnarray}
\label{eq:efflag}
 \mathcal{L} &=& \frac{g_{5e}^{HZZ}}{2 \Lambda_{5}} H Z_{\mu \nu} Z^{\mu \nu} + \frac{g_{5o}^{HZZ}}{2 \Lambda_{5}} H \widetilde{Z}_{\mu \nu} Z^{\mu \nu} + \frac{g_{5e}^{HWW}}{\Lambda_{5}} H W^{+}_{\mu \nu} W_{-}^{\mu \nu} + \frac{g_{5o}^{HWW}}{\Lambda_{5}} H \widetilde{W}^{+}_{\mu \nu} W_{-}^{\mu \nu} + \nonumber \\
&& \frac{g_{5e}^{HZ\gamma}}{\Lambda_{5}} H Z_{\mu \nu} A^{\mu \nu} + \frac{g_{5o}^{HZ\gamma}}{\Lambda_{5}} H \widetilde{Z}_{\mu \nu} A^{\mu \nu} + \frac{g_{5e}^{H\gamma \gamma}}{2 \Lambda_{5}} H A_{\mu \nu} A^{\mu \nu} + \frac{g_{5o}^{H\gamma \gamma}}{2 \Lambda_{5}} H \widetilde{A}_{\mu \nu} A^{\mu \nu} \nonumber \\
\end{eqnarray}
where the subscript {\em e} or {\em o} refers to the $\cal{CP}$-even
or $\cal{CP}$-odd nature of the individual operators~\cite{Figy:2004pt}, $V^{\mu\nu}$ is the field strength tensor of the gauge boson $V$ and $\widetilde{V}^{\mu\nu}$ is the dual field strength.

An alternative approach is to write the effective Lagrangian in terms of the dimension-6 operators ${\cal O}_{WW}$,  ${\cal O}_{BB}$, ${\cal O}_W$ 
and ${\cal O}_B$ and their corresponding $\cal{CP}$-odd 
operators according to Refs.~\cite{Hagiwara:1993qt,Hagiwara:1993ck}:
\begin{eqnarray}
\label{eq:HVV_F}
 \mathcal{L}_{\rm eff} &=& \frac{f_{W}}{\Lambda^{2}} \mathcal{O}_{W}+ \frac{f_{B}}{\Lambda^{2}} \mathcal{O}_{B} + \frac{f_{WW}}{\Lambda^{2}} \mathcal{O}_{WW} + \frac{f_{BB}}{\Lambda^{2}} \mathcal{O}_{BB} + \mbox{ CP-odd part}
\end{eqnarray}
The explicit form of the operators is
\begin{eqnarray}
\label{eq:operators}
\mathcal{O}_{W} &=& (D_{\mu} \phi^{\dagger}) \widehat{W}^{\mu \nu} (D_{\nu} \phi) \nonumber \\
\mathcal{O}_{B} &=& (D_{\mu} \phi^{\dagger}) \widehat{B}^{\mu \nu} (D_{\nu} \phi) \nonumber \\
\mathcal{O}_{WW} &=& \phi^{\dagger} \widehat{W}_{\mu\nu} \widehat{W}^{\mu\nu} \phi \nonumber \\
\mathcal{O}_{BB} &=& \phi^{\dagger} \widehat{B}_{\mu\nu} \widehat{B}^{\mu\nu} \phi \,,
\end{eqnarray}
with
\begin{eqnarray}
 \widehat{W}_{\mu\nu} &=& i g T^{a} W^{a}_{\mu\nu} \nonumber \\
 \widehat{B}_{\mu\nu} &=& i g' Y B_{\mu \nu},
\end{eqnarray}
where $g$ and $g'$ are the SU(2) and U(1) gauge couplings, and $T^{a}$ are the
SU(2) generators.  The $\cal{CP}$-odd part of the Lagrangian has the same form, although
only three parameters (they are denoted with a tilde, see Appendix~\ref{listofoperators}) are needed. 

The different parametrizations, and the relationships between them, are discussed in more detail on the \textsc{Vbfnlo} webpage, where the explicit forms of the $HVV$ couplings are given.

\begin{enumerate}
\item  A parametrization in terms of couplings in the effective Lagrangian 
  approach given by Eq.~\eqref{eq:efflag}. 
  \begin{itemize} 
   \item {\tt PARAMETR1:} Parameter that switches on the effective Lagrangian
     parameterization of Eq.~(\ref{eq:efflag}). The default value is {\tt true}.
   \item {\tt LAMBDA5:} Mass scale $\Lambda_{5}$ in units of GeV with 480~GeV chosen as default.
   \item {\tt G5E\_HWW, G5E\_HZZ, G5E\_HGG, G5E\_HGZ:} Parameters that determine
     the couplings $g_{5e}^{HVV}$ of the $\cal{CP}$-even dimension five operators.
     Their default values are set to 0.
   \item {\tt G5O\_HWW, G5O\_HZZ, G5O\_HGG, G5O\_HGZ:} Parameters that determine
     the couplings $g_{5o}^{HVV}$ of the $\cal{CP}$-odd dimension five operators.
     Their default values are set to 0.
  \end{itemize}
\item The parameterization of the anomalous couplings by the L3 Collaboration as
given in Ref.~\cite{Achard:2004kn}. The parameters are $d$, $d_B$, $\Delta
g_1^Z$ and $\Delta \kappa_\gamma$, which are related to the coefficients $f_{i}/\Lambda^{2}$ of the effective Lagrangian of Eq.~\eqref{eq:HVV_F} in the following manner~\cite{Hankele:2006ma}: 
\begin{align} \label{eq:param2}
\begin{split}
&d = -\frac{m_W^2}{\Lambda^2}\ f_{WW}\,, \hspace{3.9cm} \widetilde{d} =
-\frac{m_W^2}{\Lambda^2}\ f_{\widetilde{W}W}\,,\\
&d_B = - \frac{m_W^2}{\Lambda^2}\
\frac{\sin^2{\theta_w}}{\cos^2{\theta_w}}\ f_{BB}\,, \hspace{2.3cm}
\widetilde{d}_B = -\frac{m_W^2}{\Lambda^2}\
\frac{\sin^2{\theta_w}}{\cos^2{\theta_w}}\ f_{\widetilde{B}B}\,,\\
&\Delta \kappa_\gamma = \kappa_\gamma -1 = \frac{m_W^2}{2 \Lambda^2} \
(f_B + f_W)\,, \hspace{1cm} \widetilde{\kappa}_\gamma = \frac{m_W^2}{2
  \Lambda^2}\ f_{\widetilde{B}}\,,\\
&\Delta g_1^Z = g_1^Z -1 = \frac{m_Z^2}{\Lambda^2}\, \ \frac{f_W}{2}\,.
\end{split}
\end{align}
\begin{itemize}
\item {\tt PARAMETR2:} Parameter that switches on the above mentioned parameterization of Eqs.~\eqref{eq:param2}.
 The default  value is  {\tt false}.
\item {\tt D\_EVEN, DB\_EVEN, DKGAM\_EVEN, DG1Z\_EVEN:} Parameters that are 
  the $\cal{CP}$-even couplings in this parameterization.  Default values are set to zero.
\item {\tt D\_ODD, DB\_ODD, KGAM\_ODD:} Parameters that are  
the $\cal{CP}$-odd couplings in this parameterization with default values
equal to 0.
\item  {\tt HVV1:} Parameter that determines which anomalous $HVV$
  couplings are used for the run. For {\tt HVV1} $=$ 0, only the $HZ\gamma$ coupling,
  for {\tt HVV1} $=$ 1, only the $H\gamma\gamma$ coupling, for {\tt HVV1}
  $=$ 2, only the $HZZ$ coupling and for {\tt HVV1} $=$ 3, only the
  $HWW$ coupling is used. If {\tt HVV1} is set to 4, all possible
  anomalous couplings are used. This is also the default value.
\end{itemize}
\item  The parametrization of the anomalous couplings in terms of coefficients 
$f_i/\Lambda^2$ of the effective Lagrangian in Eq.~\eqref{eq:HVV_F}.
\begin{itemize}
\item {\tt PARAMETR3:} Parameter that switches on the parametrization stated above.
  The default value is {\tt false}.
\item {\tt FWW\_EVEN, FBB\_EVEN, FW\_EVEN, FB\_EVEN:} Parameters that represent
the coefficients of the $\cal{CP}$-even operators -- i.e.\ $f_{i}/\Lambda^{2}$
-- with default values set to zero.
\item {\tt FWW\_ODD, FBB\_ODD, FB\_ODD:} Parameters that are the coefficients of
the $\cal{CP}$-odd operators -- i.e.\ $f_{i}/\Lambda^{2}$ -- with default values
0 GeV$^{-2}$.
\item {\tt HVV2:} Parameter that allows the user to choose which anomalous HVV
  couplings are used. For {\tt HVV2} = 0, only the $HZ\gamma$ coupling,
  for {\tt HVV2} = 1, only the $H\gamma\gamma$ coupling, for {\tt HVV2}
  = 2, only the $HZZ$ coupling and for {\tt HVV2} = 3, only the
  $HWW$ coupling is used. If set to 4 all possible
  anomalous couplings are used. The default value is 4.
\end{itemize}
\end{enumerate}
Moreover, for all parametrizations, two different form factors can be chosen
as described in Refs.~\cite{Figy:2004pt,Hankele:2006ma}. They model effective, 
momentum dependent $HVV$ vertices, motivated by new physics
entering with a large scale $\Lambda$ at the loop level.
\begin{eqnarray}
\label{eq:ff1}
F_1 &=& \frac{\Lambda^2}{q_1^2 - \Lambda^2}\ \frac{\Lambda^2}{q_2^2 -
  \Lambda^2}\,, 
\\
\label{eq:ff2}
F_2 &=& -2 \,\Lambda^2 \, C_0\!\left(q_1^2, q_2^2, (q_1+q_2)^2,
\Lambda^2\right). 
\end{eqnarray}
Here the $q_i$ are the momenta of the vector bosons and 
$C_{0}$ is the scalar one-loop three point function 
in the notation of Ref.\cite{Passarino:1978jh}.
\begin{itemize}
\item {\tt FORMFACTOR:} Flag that switches on the above 
 form factor. The default value is set to {\tt false}.
\item {\tt MASS\_SCALE:} Characteristic mass scale, $\Lambda$, of new physics 
 in units of GeV. The default value is 200~GeV.
\item {\tt FFAC:} Parameter that is used to select one particular form factor 
 out of Eqs.~(\ref{eq:ff1}) and~(\ref{eq:ff2}). If {\tt FFAC}~=~1,
  the form factor $F_{1}$ is used for the parametrization.
  {\tt FFAC} $=$ 2 selects $F_{2}$, which is also the default value.
\end{itemize}
Finally, the following parameters can be used to rescale the SM $HVV$ couplings.
\begin{itemize}
\item {\tt TREEFACW:} Parameter that multiplies the $HWW$
  tensor present in the SM Lagrangian. Default is 1.
\item {\tt TREEFACZ:} Parameter that multiplies the $HZZ$
  tensor present in the SM Lagrangian. Default is 1.  
\item {\tt LOOPFAC:} Parameter that multiplies the $HZ\gamma$ and $H\gamma\gamma$
  vertices induced by SM loops. The default is chosen to be 1.
\end{itemize}
In order to comply with previous versions of \textsc{Vbfnlo}, an input {\tt
TREEFAC} can be used, and both {\tt TREEFACW} and {\tt TREEFACZ} are set to this
input.

Note that, when working in the SM, the loop-induced couplings $HZ\gamma$ and
$H\gamma\gamma$ are used only in the calculation of the Higgs width and decays,
not in the production amplitude.  For the case of anomalous Higgs couplings ({\tt
ANOM\_CPL = .true.} and one of the processes 10x, (4)300, (4)330, (4)360, (4)370
has been selected), these contributions are included in the production as well
as in the decay of the Higgs boson.


\subsubsection{{\tt anomV.dat} $-$ anomalous triple and quartic gauge
  boson couplings}
\label{sec:TGC}

The anomalous triple and quartic gauge boson couplings can be set in the file
{\tt anomV.dat}.  They are parameterized using an effective Lagrangian, as
described in Refs.~\cite{Buchmuller:1985jz, Hagiwara:1993ck, Eboli:2006wa, Degrande:2013rea}
\begin{equation}
 \mathcal{L}_{\rm eff} = \frac{f_{i}}{\Lambda^{n}} \mathcal{O}_{i}^{n+4},
\end{equation}
where $n+4$ signifies the dimension of the operator $\mathcal{O}_{i}$. 
\textsc{Vbfnlo} defines the anomalous gauge couplings in terms of the
coefficients $f_{i}/\Lambda^{n}$ of the dimension-6 and dimension-8 
operators. The full list of implemented operators can be found in 
Appendix~\ref{listofoperators}.

A common alternative parameterization (which \textsc{Vbfnlo} can also use as
input) of the trilinear couplings $WW\gamma$ and $WWZ$ uses the following
effective Lagrangians:
\begin{equation}
\begin{split}
\label{anovertex}
\mathcal{L}_{{{WW}}\gamma} =& -ie \left[ W_{\mu\nu}^\dagger W^\mu
A^\nu- W_\mu^\dagger A_\nu W^{\mu\nu} +\kappa_\gamma W_\mu^\dagger
W_\nu F^{\mu\nu} +{\lambda_\gamma\over m_W^2} W_{\sigma\mu}^\dagger
W^\mu_\nu F^{\nu\sigma}\right]\,
\end{split}
\end{equation}
for the anomalous $WW\gamma$ vertex, and
\begin{equation}
\begin{split}
\label{anovertexZ}
\mathcal{L}_{{{WWZ}}} = & -ie \cot \theta_w \,\left[ g_1^Z\left(
  W_{\mu\nu}^\dagger W^\mu Z^\nu- W_\mu^\dagger Z_\nu
  W^{\mu\nu}\right) +\kappa_Z W_\mu^\dagger W_\nu Z^{\mu\nu}
+{\lambda_Z\over m_W^2} W_{\sigma\mu}^\dagger W^\mu_\nu
Z^{\nu\sigma}\right]\,
\end{split}
\end{equation}
for the anomalous $WWZ$ vertex. It is customary to rephrase the
electroweak modifications around the SM Lagrangian in terms of new
quantities,
\begin{equation}
(\Delta g_1^Z,\Delta \kappa_Z, \Delta\kappa_\gamma) = ( g_1^Z,\kappa_Z,\kappa_\gamma) - 1 .
\end{equation}

These quantities are related to the coefficients of the dimension-6 operators
$\mathcal{O}_{W}$, $\mathcal{O}_{B}$ and $\mathcal{O}_{WWW}$ as shown in 
Eqs.~\eqref{eq:param2} for $\Delta\kappa_\gamma$ and $\Delta g_1^Z$.
The corresponding relations for $\Delta\kappa_Z$ and $\Delta \lambda_\gamma, \, \Delta\lambda_Z$ are
\begin{align} \label{eq:param2part2}
\begin{split}
&\Delta \kappa_Z = \frac{m_Z^2}{2 \Lambda^2} \
(f_W \cos^2 \theta_W - f_B \sin^2 \theta_W)\,,\\
&\Delta \lambda_\gamma = \Delta\lambda_Z = \frac{3 g^2 m_W^2}{2 \Lambda^2}\, \ f_{WWW} \,.
\end{split}
\end{align}

In order to include anomalous vector boson couplings, the parameter {\tt
ANOM\_CPL} must be switched to {\tt true} in {\tt vbfnlo.dat}.  The parameters
described above are input via the file {\tt anomV.dat}:

\begin{itemize}
 \item {\tt TRIANOM:} Switches between parameterizations of the anomalous
$WW\gamma$ and $WWZ$ couplings.  {\tt TRIANOM = 1} uses the coefficients of the
dimension-6 operators as input: 
   \begin{itemize}
     \item {\tt FWWW:} Coefficient of the operator $\mathcal{O}_{WWW}$, i.e.\
$f_{WWW}$/$\Lambda^{2}$.  Default is set to $0$~GeV$^{-2}$.
     \item {\tt FW:} Coefficient of the operator $\mathcal{O}_{W}$, i.e.\
$f_{W}$/$\Lambda^{2}$.  Default is set to $0$~GeV$^{-2}$.
     \item {\tt FB:} Coefficient of the operator $\mathcal{O}_{B}$, i.e.\
$f_{B}$/$\Lambda^{2}$.  Default is set to $0$~GeV$^{-2}$.
   \end{itemize}
  {\tt TRIANOM = 2} uses the alternative parameterization of Eqs.~\eqref{anovertex}~and~\eqref{anovertexZ} as input:
   \begin{itemize}
    \item {\tt LAMBDA0:} The quantity $\lambda_{\gamma} (= \lambda_{Z})$. 
Default is set to 0.0.
    \item {\tt ZDELTAKAPPA0:} The quantity $\Delta \kappa_{Z}$.  Default is set
to 0.0.
    \item {\tt ZDELTAG0:} The quantity $\Delta g_{1}^{Z}$.  Default is set to 0.0.
    \item {\tt ADELTAKAPPA0:} The quantity $\Delta \kappa_{\gamma}$.  Default
is set to 0.0.
    \item  Note that, as can be seen from Eqs.~\eqref{eq:param2} and~\eqref{eq:param2part2}, the quantities $\Delta
\kappa_{Z}$, $\Delta \kappa_{\gamma}$ and $\Delta g_{1}^{Z}$ are not
independent, but obey the relation
 \begin{equation}
  \label{eq:anomVconsistency}
  \Delta \kappa_{Z} = \Delta g_{1}^{Z} - \frac{\sin^{2} \theta_{W}}{\cos^{2} \theta_{W}} \Delta \kappa_{\gamma}.
 \end{equation}
 If one of these quantities is zero, it will be set by \textsc{Vbfnlo} to be
consistent with the other values.  If the input values are inconsistent, $\Delta
\kappa_{\gamma}$ will be reset to give the correct relation.
   \end{itemize}

  Default is {\tt TRIANOM = 1}. The LEP limits on those couplings can be found in \cite{Schael:2013ita}.
Note that $Vjj$ and $VVj$ processes (process IDs 120-150 and 610-640) only take account of the above anomalous  coupling parameters and not the following parameters.
 \item {\tt FWW, FBB:} The coefficients of the remaining $\cal{CP}$-even
dimension-6 operators, i.e.\ $f_{i}$/$\Lambda^{2}$.  Note that these are not
implemented for $Vjj$ production via VBF (process IDs 120-150) and $VVj$
processes (IDs 610-640).  Default value is 0 GeV$^{-2}$.
 \item {\tt FWWWt, FWt, FBt, FBWt, FDWt, FWWt, FBBt:} The coefficients of the
$\cal{CP}$-odd di\-mension-6 operators, i.e.\ $f_{i}$/$\Lambda^{2}$.  Note that
these are
not implemented for $Vjj$ production via VBF (process IDs 120-150) and $VVj$
processes (IDs 610-640).  Default value is 0  GeV$^{-2}$.
 \item {\tt  FS0 -- FS2, FM0 -- FM7, FM5hc, FT0 -- FT9:} Parameters that give
the values of the coefficients of the dimension-8 operators, i.e.\
$f_{i}$/$\Lambda^{4}$. The default values for these parameters are~0 GeV$^{-4}$.  Note that
these are only relevant for triboson production and $VVjj$ production via VBF.
\end{itemize}

In addition, a form factor can be applied. Several choices are available. One
option is to use a dipole form factor, which takes the form
\begin{equation}
 F = \left(1 + \frac{s}{\Lambda^{2}} \right)^{-p},
  \label{eq:dipoleformfac}
\end{equation}
for all processes except $\gamma jj$ production in VBF (process ID 150),
where $\Lambda$ is the characteristic scale where the form factor effect becomes relevant.  $s$ is a universal scale (the
invariant mass squared of the produced bosons) for each phase-space point.
A complex version of the dipole form factor~\cite{Kmatrix},
\begin{equation}
  F = \left(1 + i \left( \frac{s}{\Lambda^{2}} \right)^{p} \right)^{-1},
  \label{eq:cdipoleformfac}
\end{equation}
is available as well.
The second possibility is to use a step function of the form
\begin{equation}
  F = \Theta\left(\Lambda^{2} - s\right)
  \label{eq:stepformfac}
\end{equation}
such that the effect of the anomalous coupling is switched off beyond the characteristic scale $\Lambda$.
The third possibility uses a kink form factor~\cite{Rauch:2016pai},
\begin{equation}
  F =
  \begin{cases}
    1 & \text{for $s\le\Lambda^{2}$,} \\
    \left(\frac{\Lambda^{2}}{s} \right)^{p} & \text{for $s>\Lambda^{2}$,}
  \end{cases}
  \label{eq:kinkformfac}
\end{equation}
where the effect of the damping sets in exactly at the characteristic scale. \\
Moreover, for the two operators $\mathcal{O}_{S,1}$ and the isospin-conserving combination $\mathcal{O}_{S,0}\equiv\mathcal{O}_{S,2}$ the
so called K-matrix unitarisation procedure is available. For this unitarisation
method no additional input parameters need to be set. 
\begin{table}[t!]
  \begin{center}
    \begin{tabular}{c|l|c}
      \hline
      &&\\
      {\tt FORMFACTYPE} & \textsc{Form Factor Type} & \textsc{Definition} \\
      &&\\
      \hline
      &&\\
      {\tt 1} & dipole form factor & Eqs.~\eqref{eq:dipoleformfac},~\eqref{eq:dipoleformfac2}\\
      {\tt 2} & complex dipole form factor & Eqs.~\eqref{eq:cdipoleformfac}\\
      {\tt 3} & step form factor & Eq.~\eqref{eq:stepformfac}\\
      {\tt 4} & kink form factor & Eq.~\eqref{eq:kinkformfac}\\
    \end{tabular}
    \caption {\em  Form factor types.}
    \vspace{0.2cm}
    \label{tab:formfactor}
  \end{center}
\end{table}
\begin{itemize}
  \item {\tt FORMFAC: }  Switch determining whether the above form factor $F$ is
    included.  Default is set to {\tt false}.
  \item {\tt FORMFACTYPE: } Parameter that is used to select one particular type
    of form factor. Possible choices are denoted in Table~\ref{tab:formfactor}.
    Default is set to 1 (dipole formfactor).
 \item {\tt FFMASSSCALE: } Mass scale $\Lambda$.  Default is set to 2000~GeV.
 \item {\tt FFEXP:} The exponent $p$.  Default is set to 2.
 \item {\tt KMATRIX: }  Switch determining whether to use K-matrix unitarisation
       for $\mathcal{O}_{S,1}$ and $\mathcal{O}_{S,0}\equiv\mathcal{O}_{S,2}$
       as described in section~\ref{sec:kmatrix}. Default is set to
       {\tt false}.
\item {\tt TMATRIX: } Switch determining whether to use the $T_u$-model unitarisation procedure for the dimension-8 operators as described in section~\ref{sec:tmatrix}. Default is set to {\tt false}.   
\end{itemize}
Individual form factor mass scales and exponents can be set for the trilinear
$WWZ$ and $WW\gamma$ couplings, using either of the parametrizations above.  If
chosen, these values overwrite the ``universal'' mass scale and exponent ({\tt
FFMASSSCALE} and {\tt FFEXP}) set above for the selected parameters.
\begin{itemize}
 \item {\tt FORMFAC\_IND: } Switch determining whether individual or universal
form factors are used.  Default is set to {\tt false} -- universal form factors are used.
\end{itemize}
If {\tt TRIANOM = 1} then
\begin{itemize}
 \item {\tt MASS\_SCALE\_FWWW: } Mass scale $\Lambda$ for coefficient $f_{WWW}$.
 Default is 2000~GeV.
 \item {\tt FFEXP\_FWWW: } Exponent $p$ for coefficient $f_{WWW}$.  Default is 2.
 \item {\tt MASS\_SCALE\_FW: } Mass scale $\Lambda$ for coefficient $f_{W}$. 
Default is 2000~GeV.
 \item {\tt FFEXP\_FW: } Exponent $p$ for coefficient $f_{W}$.  Default is 2.
 \item {\tt MASS\_SCALE\_FB: } Mass scale $\Lambda$ for coefficient $f_{B}$. 
Default is 2000~GeV.
 \item {\tt FFEXP\_FB: } Exponent $p$ for coefficient $f_{B}$.  Default is 2.
\end{itemize}
If {\tt TRIANOM = 2} then
\begin{itemize}
 \item {\tt MASS\_SCALE\_AKAPPA: } Mass scale $\Lambda$ for parameter $\Delta
\kappa_{\gamma}$.  Default is 2000~GeV.
 \item {\tt FFEXP\_AKAPPA: } Exponent $p$ for parameter $\Delta \kappa_{\gamma}$.  Default
is 2.
 \item {\tt MASS\_SCALE\_ZKAPPA: } Mass scale $\Lambda$ for parameter $\Delta
\kappa_{Z}$.  Default is 2000~GeV.
 \item {\tt FFEXP\_ZKAPPA: } Exponent $p$ for parameter $\Delta \kappa_{Z}$.  Default
is 2.
 \item {\tt MASS\_SCALE\_LAMBDA: } Mass scale $\Lambda$ for parameter $\lambda$.
 Default is 2000~GeV.
 \item {\tt FFEXP\_LAMBDA: } Exponent $p$ for parameter $\lambda$.  Default is
2.
 \item {\tt MASS\_SCALE\_G: } Mass scale $\Lambda$ for parameter $\Delta g_{1}^{Z}$. 
Default is 2000~GeV.
 \item {\tt FFEXP\_G: } Exponent $p$ for parameter $\Delta g_{1}^{Z}$.  Default is 2.
 \item As with the anomalous parameters themselves, the formfactors in the
parameterization of {\tt TRIANOM = 2} are not independent (see
Eq.~\eqref{eq:anomVconsistency}).  If one mass scale is set to zero, it will be
set by \textsc{Vbfnlo} to be consistent with the other values.  If the input
values are inconsistent, they will be reset to give the correct relation.
\end{itemize}

For $\gamma jj$ production in VBF (process ID 150), the dipole form factor takes the form
\begin{equation}
 F = \left(1 + \frac{q_1^2}{\Lambda^{2}} + \frac{q_2^2}{\Lambda^{2}} + \frac{q_3^2}{\Lambda^{2}} \right)^{-p},
  \label{eq:dipoleformfac2}
\end{equation}
with $\Lambda$ and $p$ set by {\tt FFMASSSCALE} and {\tt FFEXP}, respectively.
$q_1^2$, $q_2^2$ and $q_3^2$ are the invariant masses squared of the three vector bosons involved in the
$VVV$ vertex.
For this process only the universal form factor can be applied.

\subsubsection{K-Matrix unitarisation}
\label{sec:kmatrix}

The K-matrix unitarisation procedure has been implemented 
for the dimension-8 operators $\mathcal{O}_{S,1}$ and the isospin-conserving
combination $\mathcal{O}_{S,0}\equiv\mathcal{O}_{S,2}$
by using the relations~\cite{Rauch:2016pai} to the operators $\mathcal{O}_4$
and $\mathcal{O}_5$ of the electroweak chiral Lagrangian and the procedure
worked out there~\cite{Alboteanu:2008my}.
This method guarantees the preservation of unitarity when either of these operators is used 
in the study of anomalous quartic gauge couplings. In contrast to using form factors,
no additional input parameters need to be set. The anomalous contributions are automatically suppressed
at the energy scale where unitarity would be violated without unitarisation. After this energy scale is reached,
the anomalous contributions are kept at a finite value, representing the maximally possible anomalous contribution.
See Refs.~\cite{Alboteanu:2008my,Degrande:2013rea,Kmatrix,Kilian:2014zja,Rauch:2016pai}
for details of the K-matrix unitarisation procedure and its implementation.

\subsubsection{$T_u$-model}
\label{sec:tmatrix}
Similarly to the K-matrix, the T$_u$-model guarantees the preservation of unitarity when studying anomalous quartic gauge couplings without needing any additional ad hoc input parameters. Furthermore the T$_u$-model preserves unitarity not only at large energy scales but also for large virtualities of the incoming vector bosons in the vector boson scattering process by suppressing the anomalous contribution sufficiently. Finally the T$_u$-model works for all dimension-8 operators listed in sec. \ref{listofoperators}. It can therefore be seen as an improvement of the K-matrix unitarisation and should be used instead. See Refs.~\cite{PerezRivera2018,Perez:2018kav} for details of the T$_u$-model and its implementation.

\subsubsection{Using {\tt anom\_HVV.dat} and {\tt anomV.dat} simultaneously}
\label{sec:HVVandVVV}

Some of the anomalous coupling parameters affect both the $VVV$ and $HVV$ couplings.  
These parameters ($f_{W}$ and $f_{B}$ or, equivalently, $\Delta g_{1}^{Z}$ and 
$\Delta \kappa_{\gamma}$) are consequently inputs in both {\tt anom\_HVV.dat} 
and {\tt anomV.dat}.  As stated previously, when working with anomalous $VVV$ 
couplings with a process that contains internal Higgs bosons (e.g.\ $WWZ$ triboson production), 
the anomalous $HVV$ couplings resulting from $f_{W}$ and $f_{B}$ (input 
via {\tt anomV.dat}) are calculated and used.

In diboson $WW$ production (process ID 300), however, {\tt anomV.dat} and {\tt anom\_HVV.dat} 
can be used simultaneously.  The anomalous $VVV$ couplings are taken directly from {\tt anomV.dat}.  
The values of $f_{W}$ and $f_{B}$ for the $HVV$ couplings are also taken from {\tt anomV.dat}, 
but the additional parameters ($f_{BB}$, $f_{WW}$ etc.) are taken from {\tt anom\_HVV.dat} 
using the appropriate parametrisation\footnote{{\tt TRIANOM=1} in {\tt anomV.dat} corresponds 
to {\tt PARAMETR3=true} in {\tt anom\_HVV.dat}, and {\tt TRIANOM=2} corresponds to {\tt PARAMETR2=true}.}. 
 Alternatively, the $HVV$ couplings can be specified directly via $g_{5}^{HXY}$ by setting 
{\tt PARAMETR1=true} in {\tt anom\_HVV.dat}.  In this case, the values $g_{5}^{HXY}$ are used if 
they are consistent with the {\tt anomV.dat} inputs -- if they are inconsistent, a warning message 
is printed and $HVV$ parametrisation 2 or 3 is used instead.


\subsection{{\tt kk\_input.dat} $-$ parameters for Higgsless models}

\textsc{Vbfnlo} allows the calculation of the vector boson fusion processes
$WWjj$, $WZjj$ and $ZZjj$ in the Warped Higgsless scenario~\cite{Csaki:2003zu,
Englert:2008wp} at LO and NLO QCD level (see, for example,
Ref.~\cite{Englert:2008tn} for a phenomenological application). This feature is
used if the switch {\tt KK\_MOD} in {\tt vbfnlo.dat} is switched to {\tt true}
and must be enabled at compilation using the {\tt configure} option {\tt
-{}-enable-kk}.

The triple vector-boson processes $WWW$, $WWZ$
and $WZZ$ can be calculated in both this model and a Three-Site Higgsless
Model~\cite{Chivukula:2006cg}. The model parameters (masses and couplings) can be generated by
\textsc{Vbfnlo} via the input file {\tt kk\_input.dat} for a choice
of the relevant five dimensional gauge parameters. The input parameters are: 
\begin{itemize}
\item {\tt VBFCALC\_SWITCH:} Switch that determines whether \textsc{Vbfnlo} should calculate
the  parameters needed by the model. Default is {\tt true}.  Alternatively, the user can input their own model parameters (masses and couplings) as described below.
\item {\tt MHLM\_SWITCH:} Switch that determines whether the Three-Site Higgsless
Model~\cite{Chivukula:2006cg} ({\tt true}) or the Warped Higgsless
Model~\cite{Csaki:2003zu} ({\tt false}) should be used. Default is
{\tt false} (Warped Higgsless Model).
\item The following parameters are used only when working in the Warped Higgsless Model ({\tt MHLM\_SWITCH = false}):
\begin{itemize}
\item  {\tt RDWN:} Location of the UV brane for the generation of the model
parameters in the Warped Higgsless Model. Default is $R=9.75\times 10^{-9}$,
which amounts to the Kaluza-Klein excitations having masses of
$m_{W_1}=700~\mathrm{GeV}$, $m_{Z_1}=695~\mathrm{GeV}$, and
$m_{Z'_1}=718~\mathrm{GeV}$. Smaller values of $R$ result in a heavier
Kaluza-Klein spectrum.
\item {\tt KKMAXW:} The maximum number of Kaluza-Klein $W^\pm_k$ states to be
included on top of the Standard Model $W^\pm$ bosons, which correspond to
$W^\pm_{k=0}$.    All states $k\geq 3$ are phenomenologically irrelevant.
Default is 1.
\item {\tt KKMAXZ:} The maximum number of Kaluza-Klein $Z_k$ states to be
included on top of the Standard Model $Z$ boson, which corresponds to $Z_{k=0}$.
All states $k\geq 3$ are phenomenologically irrelevant. Default is 1.
\item {\tt KKMAXG:} The maximum number of Kaluza-Klein $Z'_k$ bosons that are
excitations of the Standard Model photon $Z'_{k=0}$. States $k\geq 2$ are
phenomenologically irrelevant. Default is 1.
\end{itemize}
\item {\tt WPMASS:} $W_{k=1}$ mass in the Three-Site Higgsless Model ({\tt MHLM\_SWITCH = true}).  Default is 500~GeV.
\end{itemize}

The explicit breaking of higher dimensional gauge invariance is balanced
according to the description of Ref.~\cite{KK}, where more details regarding the model and
its implementation can be found.  Information on the widths and the sum
rules relating the various gauge boson couplings~\cite{Csaki:2003dt} are written
to the file {\tt kkcheck.dat}.

\textsc{Vbfnlo} generates the text file {\tt kk\_coupl\_inp.dat}, which
documents the calculated model parameters, i.e.\ Kaluza-Klein gauge boson
masses and couplings of the specified input parameters. This file can
also be used as an input file for advanced users who want to run the code with
their own set of parameters -- e.g.\ for simulation of more general
technicolor-type scenarios. To that end, select {\tt VBFCALC\_SWITCH = false}
in the file {\tt kk\_input.dat}. \textsc{Vbfnlo} will then calculate the gauge
boson widths on the  basis of these parameters by the decay to the lower lying
states.  The inputs in {\tt kk\_coupl\_inp.dat} are as follows:
\begin{itemize}
 \item {\tt KKMAXW:} The maximum number of Kaluza-Klein $W^\pm_k$ states 
included on top of the Standard Model $W^\pm$ bosons.
 \item {\tt KKMAXZ:} The maximum number of Kaluza-Klein $Z_k$ states 
included on top of the Standard Model $Z$ boson.
 \item {\tt KKMAXG:} The maximum number of Kaluza-Klein $Z'_k$ bosons that are
excitations of the Standard Model photon $Z'_{k=0}$.
 \item Kaluza-Klein masses:
   \begin{itemize}
    \item {\tt KKMASSW:} List of masses of the Kaluza-Klein $W^{\pm}_{k}$ states.
    \item {\tt KKMASSZ:} List of masses of the Kaluza-Klein $Z_{k}$ states.
    \item {\tt KKMASSG:} List of masses of the Kaluza-Klein $Z'_{k}$ states.
   \end{itemize}
 \item Couplings
   \begin{itemize}
    \item {\tt CPL\_W$x$-W$y$-ZX:} List of couplings between Kaluza-Klein
states: $W^{\pm}_{x}-W^{\pm}_{y}-Z_{0}$ to $W^{\pm}_{x}-W^{\pm}_{y}-Z_{\mbox{\tiny{\tt KKMAXZ}}}$.
    \item {\tt CPL\_W$x$-W$y$-GX:} List of couplings between Kaluza-Klein
states: $W^{\pm}_{x}-W^{\pm}_{y}-Z'_{0}$ to
$W^{\pm}_{x}-W^{\pm}_{y}-Z'_{\mbox{\tiny{\tt KKMAXG}}}$.
    \item {\tt CPL\_W$x$-W$y$-W$z$-WX:} List of couplings between Kaluza-Klein
states: $W^{\pm}_{x}-W^{\pm}_{y}-W^{\pm}_{z}-W^{\pm}_{0}$ to
$W^{\pm}_{x}-W^{\pm}_{y}-W^{\pm}_{z}-W^{\pm}_{\mbox{\tiny{\tt KKMAXW}}}$.
    \item {\tt CPL\_W$x$-W$y$-Z$z$-ZX:} List of couplings between Kaluza-Klein
states: $W^{\pm}_{x}-W^{\pm}_{y}-Z_{z}-Z_{0}$ to
$W^{\pm}_{x}-W^{\pm}_{y}-Z_{z}-Z_{\mbox{\tiny{\tt KKMAXZ}}}$.
    \item {\tt CPL\_W$x$-W$y$-G$z$-GX:} List of couplings between Kaluza-Klein
states: $W^{\pm}_{x}-W^{\pm}_{y}-Z'_{z}-Z'_{0}$ to
$W^{\pm}_{x}-W^{\pm}_{y}-Z'_{z}-Z'_{\mbox{\tiny{\tt KKMAXG}}}$.
    \item {\tt CPL\_W$x$-W$y$-G$z$-ZX:} List of couplings between Kaluza-Klein
states: $W^{\pm}_{x}-W^{\pm}_{y}-Z'_{z}-Z_{0}$ to
$W^{\pm}_{x}-W^{\pm}_{y}-Z'_{z}-Z_{\mbox{\tiny{\tt KKMAXZ}}}$.
   \end{itemize}
\end{itemize}
The default values in {\tt kk\_coupl\_inp.dat} are those produced when
\textsc{Vbfnlo} is used to calculate the couplings and masses using the default
values of {\tt kk\_input.dat} as described above.


\subsection{{\tt spin2coupl.dat} $-$ parameters for spin-2 models}

The file {\tt spin2coupl.dat} is used to set the parameters for the spin-2
models.  It is read if the switch {\tt SPIN2} in {\tt vbfnlo.dat} is set to {\tt
true}.

\textsc{Vbfnlo} uses an effective model to describe the interactions of spin-2
particles with electroweak gauge bosons for two cases: an isospin singlet spin-2 state
and a spin-2 triplet in the adjoint representation, as described in
Refs.~\cite{frank,Frank:2012wh,Frank:2013gca}. For the singlet spin-2 field, $T^{\mu\nu}$, the effective
Lagrangian is
\begin{align}
 \mathcal{L}_{\text{singlet}} & = \frac{1}{\Lambda} T_{\mu \nu} \left( f_{1} B^{\alpha \nu} B^{\mu}_{\alpha} + f_{2} W_{i}^{\alpha \nu} W^{i,\mu}_{\alpha} + f_3 \widetilde{B}^{\alpha \nu} B^\mu_{\alpha} 
+f_4\widetilde{W}_i^{\alpha \nu} W^{i, \mu}_{\alpha} \right. \nonumber \\
 & \quad \quad \quad \quad \left. +  2 f_{5} (D^{\mu} \Phi)^{\dagger} (D^{\nu}\Phi) + f_9 \, G^{\alpha \nu}_{a} G^{a, \mu}_{\alpha} \right),
\end{align}

and for the spin-2 triplet field, $T_{j}^{\mu\nu}$, the effective Lagrangian is given by
\begin{equation}
 \mathcal{L}_{\text{triplet}} = \frac{1}{\Lambda} T_{\mu\nu j} \left( f_{6} (D^{\mu}\Phi)^{\dagger} \sigma^{j} (D^{\nu} \Phi) + f_{7} W^{j,\mu}_{\alpha} B^{\alpha\nu} \right),
\end{equation}
where $W$, $B$ and $G$ are the usual field strength tensors of the SM gauge fields,
$\widetilde{W}$ and $\widetilde{B}$ the dual field strength tensors, $\Phi$ is
the Higgs field and $D^{\mu}$ is the covariant derivative.  $f_{i}$ are variable
coupling parameters and $\Lambda$ is the characteristic energy scale of the new
physics.

In order to preserve unitarity, a formfactor is introduced to multiply the
amplitudes.  The formfactor has the form:
\begin{equation}
 f(q_{1}^{2}, q_{2}^{2}, p^{2}_{\text{sp2}}) = \left( \frac{\Lambda_{ff}^{2}} {\left|q_{1}^{2}\right| + \Lambda_{ff}^{2}} \cdot \frac{\Lambda_{ff}^{2}} {\left|q_{2}^{2}\right| + \Lambda_{ff}^{2}} \cdot \frac{\Lambda_{ff}^{2}} {\left|p_{\text{sp2}}^{2}\right| + \Lambda_{ff}^{2}} \right) ^{n_{ff}}.
\end{equation}
Here $p^{2}_{\text{sp2}}$ is the invariant mass of a virtual $s$-channel spin-2 particle
and $q_{1,2}^{2}$ are the invariant masses of the electroweak bosons.  The
energy scale $\Lambda_{ff}$ and the exponent $n_{ff}$ describe the scale of the
cutoff and the suppression power.

The input parameters used by \textsc{Vbfnlo} are:
\begin{itemize}
 \item {\tt F1,F2,F3,F4,F5,F9:} Coupling parameters for the spin-2 singlet field. 
 \item {\tt F6,F7:} Coupling parameters for the spin-2 triplet field. 
 \item {\tt LAMBDA:} Energy scale of the couplings in GeV. 
 \item {\tt LAMBDAFF:} Energy scale of the formfactor in GeV. 
 \item {\tt NFF:} Exponent of the formfactor. 
\end{itemize}
The gluon coupling {\tt F9} only affects the implemented VBF processes via the total 
width of the spin-2 particle. 
Note that the electroweak part of the graviton scenario corresponds 
to {\tt F1=F2=F5=1} and {\tt F3=F4=F6=F7=F9=0}. \\

\textsc{Vbfnlo} also needs the masses and branching ratios of the spin-2
particles into SM gauge bosons. 
\begin{itemize}
 \item {\tt SP2MASS:} Mass of the spin-2 singlet particle in GeV.  
 \item {\tt MSP2TRIPPM:} Mass of charged spin-2 triplet particles in GeV. 
 \item {\tt MSP2TRIPN:} Mass of neutral spin-2 triplet particle in GeV. 
 \item {\tt BRRAT:} Branching ratio for spin-2 singlet particle into SM gauge bosons. 
 Default value is {\tt 1}.
 \item {\tt BRRATTRIPPM:} Branching ratio for charged spin-2 triplet particles into SM gauge bosons.  
Default value is {\tt 1}.
 \item {\tt BRRATTRIPN:} Branching ratio for neutral spin-2 triplet particle into SM gauge bosons.  Default 
value is {\tt 1}.
\end{itemize}

Default values for different processes and scenarios can be found in 
Refs.~\cite{frank,Frank:2012wh,Frank:2013gca}\footnote{Note that in Ref.~\cite{frank}, 
process 191 ($pp \rightarrow \mbox{spin-2} \, jj \rightarrow \gamma \gamma \, jj$) is referred to as process 240.}.


\subsection{{\tt histograms.dat} $-$ parameters for histogram options}
\label{sec:histdat}

\textsc{Vbfnlo} can output histogram data in a variety of different formats
(\textsc{Root}, \textsc{Gnuplot}, \textsc{TopDrawer}, raw data tables), as
described in Sec.\ref{sec:hist}.

The file {\tt histograms.dat} allows the user to set the $x$-axis range,
enable a bin smearing, and enable the calculation of the Monte Carlo
error per bin\footnote{By default 100 bins are used.  This number can
be altered in {\tt utilities/histograms.F}.}.

\noindent\textsc{Vbfnlo} can calculate the Monte Carlo error for each bin and output
this to the root and the raw histogram data output for 1D and 2D histograms.
For the gnuplot histogram output only the 1D histograms can display the error bars.
\begin{itemize}
 \item {\tt CALC\_ERROR\_GNUPLOT:} Enable or disable $y$-error bars in 1D \textsc{Gnuplot} histograms.
                                   Default is {\tt false}.
 \item {\tt CALC\_ERROR\_ROOT:} Enable or disable $y$/$z$-error bars in 1D/2D \textsc{Root} histograms.
                                Please note: only the histograms defined in the \textsc{Fortran}
                                file {\tt histograms.F} can be generated with error bars.
                                Default is {\tt false}.
 \item {\tt CALC\_ERROR\_1D:} Enable or disable $y$-error bars in raw 1D histogram output.
                              Default is {\tt true}.
 \item {\tt CALC\_ERROR\_2D:} Enable or disable $z$-error bars in raw 2D histogram output.
                              Default is {\tt false}.
\end{itemize}
Furthermore, \textsc{Vbfnlo} uses a smearing between adjacent bins to avoid
artefacts at NLO when the real emission kinematics and the corresponding
subtraction term fall into different bins. As this can lead to remnants at the sharp
edges caused by cuts, the smearing can be switched off.
\begin{itemize}
 \item {\tt SMEARING:} Enable or disable smearing.  Default is {\tt true}.
 \item {\tt SMEAR\_VALUE:} Set the bin fraction where the bin smearing is active.
                          The part that is put to the next bin becomes larger when
                          the x-value is closer to a bin border. Default is {\tt 0.1}.
\end{itemize}

\noindent The range of the $x$-axis of the produced histograms is input in the format: {\tt xmin xmax}. 
The following describes those histograms which are already implemented in
\textsc{Vbfnlo}.  By altering the file {\tt utilities/histograms.F}, however, it
should be easy for the user to add new histograms -- \textsc{Vbfnlo} will
automatically read in the range of each created histogram.
\begin{itemize}
 \item {\tt HIST\_ID1:} Range for $p_{T}$ of tagging jets.  Default range is 0
to 250~GeV.
 \item {\tt HIST\_ID2:} Range for $p_{T}$ of tagging jet with higher $p_{T}$.
Default range is 0 to 500~GeV.
 \item {\tt HIST\_ID3:}  Range for $p_{T}$ of tagging jet with lower
$p_{T}$. Default range is 0 to 200~GeV.   
 \item {\tt HIST\_ID4:} Range for rapidity of tagging jets.  Default
range is -5 to 5.
 \item {\tt HIST\_ID5:} Range for rapidity of tagging jet with higher $p_{T}$. 
Default range is -5 to 5.  
 \item {\tt HIST\_ID6:} Range for rapidity of tagging jet with lower $p_{T}$. 
Default range is -5 to 5.
 \item {\tt HIST\_ID7:} Range for $p_{T}^{\max}$ of leptons.  Default range
is 0 to 500~GeV.
 \item {\tt HIST\_ID8:} Range for $p_{T}^{\min}$ of leptons.  Default range
is 0 to 500~GeV. 
 \item {\tt HIST\_ID9:} Range for $\eta_{\max}$ of leptons.  Default range
is 0 to 5.
 \item {\tt HIST\_ID10:} Range for $\eta_{\min}$ of leptons.  Default range
is 0 to 5.
 \item {\tt HIST\_ID11:} Range for azimuthal angle (in degrees) of tagging jets.  
Default range is -180 to 180.
\end{itemize}
\textsc{Vbfnlo} can also produce 2D histograms.  In this case, both the $x$ and
$y$ ranges can be set using {\tt histograms.dat} in the format {\tt xmin xmax
ymin ymax}.
\begin{itemize}
 \item {\tt 2DHIST\_ID1:} Range for 2D histogram of $d^{2}\sigma /(d\eta_{jj}
dm_{jj})$.  Default ranges are 0 to 6 for the $\eta$ ($x$) axis and 0 to 800~GeV
for the $m_{jj}$ ($y$) axis.
\end{itemize}
If the $x$-axis and $y$-axis ranges are not provided in the file {\tt histograms.dat} the
default values, which are set in {\tt utilities/histograms.F}, are used.

A brief documentation on how to implement additional histograms into the file {\tt histograms.F}
can be found on the \VBFNLO{} webpage,
\url{https://ific.uv.es/vbfnlo/}.


\subsection{{\tt random.dat} $-$ parameters for the random number generator}
\label{sec:random}

For the Monte Carlo integration different random number generators can be used by setting
{\tt RTYPE} in {\tt random.dat} to one of the following.
\begin{description}
  \item[0] ranmar based on~\cite{MarsagliaZaman:1990}. This is the default.
  \item[1] Sobol sequences as introduced in~\cite{Sobol:1967} and based on the
    implementation in~\cite{BratleyFox:1988}.
  \item[2] Internal random number generator of the compiler, availability and performance
    depend on the used compiler. Might not work with MPI\@.
  \item[3] xoroshiro128+ as provided at \url{http://vigna.di.unimi.it/xorshift/}. This
    generator is faster than the other choices.
    It also allows quick skipping of random numbers and thus significantly improves the
    performance in combination with MPI\@.
\end{description}

For all those generators (except Sobol) an initialization seed is needed.
This allows reproducible random number generation.

To run with different seeds, set the variable {\tt SEED} 
to a different integer value for each run.


\subsection[Calculating the $s$-channel contributions for the VBF processes]{Calculating the $s$-channel vector boson exchange contributions for the VBF processes}
\label{sec:schan}
The $VVV$ processes with semileptonic decays can be used to calculate
the $s$-channel vector boson exchange contributions which are neglected in the $VVjj$ processes
with fully leptonic decays in VBF. The same holds for semileptonic $VV$ production and $Vjj$
production in VBF.

Using the available processes with semileptonic decays it is possible to calculate the $s$-channel
contributions to the processes $Zjj$, $W^\pm jj$, $W^+W^-jj$, $W^\pm
Zjj$, $W^\pm\gamma jj$, $Z\gamma jj$
and $ZZjj$ in VBF with fully leptonic decays.
In order to achieve this, the corresponding processes with semileptonic decay should be
run with the same settings as the VBF process, except for the scale choice in case a 
dynamical scale has been used in the VBF processes.
Additionally, the following settings should be made in {\tt cuts.dat}:
\begin{itemize}
 \item {\tt DEF\_TAGJET = 1}
 \item {\tt PTMIN\_TAG\_1 = 0} and {\tt PTMIN\_TAG\_2 = 0}
 \item {\tt HARD\_CENTRAL = false}
 \item {\tt VBFCUTS\_ALWAYS = true}
 \item {\tt RECONST\_HAD\_V = 0}
 \item {\tt SINGLE\_DECAYJET = 0}
 \item {\tt RECONST\_HAD\_V = 0}
\end{itemize}
The option {\tt DECAY\_QUARKS} in the file {\tt vbfnlo.dat} has to be set to the value 93 (or 94
if initial and final-state $b$-quarks should be included).
The list of corresponding $s$- and $t$-channel contributions in \textsc{Vbfnlo} can be found in Table~\ref{tab:schannel}.

\begin{table}[t!]
\newcommand{\lstrut}{{$\strut\atop\strut$}}
\begin{center}
\small
\begin{tabular}{l|c|c}
\hline
&\\
\textsc{Process} & \textsc{ProcID} of $t$-channel & \textsc{ProcID}s of $s$-channel \\
&\\
\hline
&\\
$p \overset{\mbox{\tiny{(--)}}}{p} \to Z \, jj \to \ell^{+} \ell^{-} \, jj$            & \bf 120 & \bf 312, 331 \\
$p \overset{\mbox{\tiny{(--)}}}{p} \to W^{+} \, jj \to \ell^{+} \nu_\ell \, jj$        & \bf 130 & \bf 302, 313 \\
$p \overset{\mbox{\tiny{(--)}}}{p} \to W^{-} \, jj \to \ell^{-} \bar{\nu}_\ell  \, jj$ & \bf 140 & \bf 301, 323 \\
&\\
\hline
&\\
$p \overset{\mbox{\tiny{(--)}}}{p} \to W^{+}W^{-} \, jj \to \ell_{1}^{+} \nu_{\ell_{1}} \ell_{2}^{-} \bar{\nu}_{\ell_{2}} \, jj$ 
& \bf 200 & \bf 403, 431, 442 \\
$p \overset{\mbox{\tiny{(--)}}}{p} \to W^{+}Z \,  jj\to \ell_{1}^{+} \nu_{\ell_{1}} \ell_{2}^{+} \ell_{2}^{-} \, jj$
& \bf 220 & \bf 402, 412 \\
$p \overset{\mbox{\tiny{(--)}}}{p} \to W^{-}Z \, jj\to \ell_{1}^{-} \bar{\nu}_{\ell _{1}} \ell_{2}^{+} \ell_{2}^{-} \, jj$
& \bf 230 & \bf 401, 422 \\
$p \overset{\mbox{\tiny{(--)}}}{p} \to ZZ  \, jj\to \ell_{1}^{+} \ell_{1}^{-} \ell_{2}^{+} \ell_{2}^{-} \, jj$ 
& \bf 210 & \bf 411, 421, 451 \\
$p \overset{\mbox{\tiny{(--)}}}{p} \to W^{+} \gamma \,  jj\to \ell^{+} \nu_{\ell} \gamma \, jj$
& \bf 270 & \bf 462, 482 \\
$p \overset{\mbox{\tiny{(--)}}}{p} \to W^{-} \gamma \, jj\to \ell^{-} \bar{\nu}_{\ell} \gamma \, jj$
& \bf 280 & \bf 461, 492 \\
$p \overset{\mbox{\tiny{(--)}}}{p} \to Z \gamma \,  jj\to \ell^{+} \ell^{-} \gamma \, jj$
& \bf 290 & \bf 471, 481, 491 \\
&\\
\hline
\end{tabular}
\caption {\em Corresponding $s$- and $t$-channel vector boson exchange contributions of electroweak $V(V)jj$ production in \textsc{Vbfnlo}.}
\vspace{0.2cm}
\label{tab:schannel}
\end{center}
\end{table}

With this prescription a few simplifications are in place: interferences
between $s$- and $t$-channel contributions are still neglected. Additionally,
the NLO QCD corrections of the hadronic vector boson decay can be included only
in an approximation (see Section~\ref{sec:vbfnlodat}).
\subsection[Implementation of new cuts]
{Implementation of new cuts}
\label{sec:new_cuts}
New cuts can be implemented easily in {\tt VBFNLO} by modifying the file {\tt utilities/cuts.F} (called {\tt cuts.F} for short) 
and related files as follows.
\begin{enumerate}
\item In the subroutine {\tt InitCuts} in {\tt cuts.F} read the value of the cut from the input file {\tt cuts.dat} using
\newline
\\
{\tt call read\_real("CUTNAME",CutValue,DefaultValue)} \newline
\\
where {\tt CUTNAME} is the name of the cut (e.g. {\tt MTW\_MIN}), as written in the input file {\tt cuts.dat}; 
{\tt CutValue} is the internal variable storing the value of the cut (e.g. {\tt MtwMin}); 
{\tt DefaultValue} is the default value of the cut, in case the input value is not specified in the {\tt cuts.dat} (e.g. {\tt 50d0}). 
The new variable {\tt CutValue} must be declared in the file {\tt utilities/cuts.inc} and added to the  
common block {\tt basiccuts} there. 
If the value of the cut is not a real number, the call to {\tt read\_real} should be replaced by {\tt read\_logical} (for logical value) 
or {\tt read\_int} (for integer value) etc. 
Note that this step is not strictly necessary, but it will make setting and changing the value of the cut much simpler. 
It is possible to skip this step and use a hardwired cut value in the {\tt BASIC\_CUTS} function as instructed in the next steps.
\item In the function {\tt BASIC\_CUTS} in the file {\tt cuts.F} calculate the value of the quantity, named {\tt ObsValue}, on which to cut (e.g. {\tt Mtw}). 
The new variable {\tt ObsValue} must be declared locally. 
One then adds the cut condition to the logical value {\tt pass\_cuts} as follows
\newline
\\
{\tt pass\_cuts = pass\_cuts .and. (ObsValue .gt. CutValue)} \newline
\\
Of course, if the cut specifies the maximum value of a quantity, {\tt .gt.} should be replaced by {\tt .lt.} etc. 
The {\tt CutValue} is read in from the {\tt cuts.dat} as explained in the above step. It can be replaced 
by a hardwired value here. 
\item The {\tt utilities} directory must be cleaned by using the {\tt make clean} command before {\tt VBFNLO} is 
remade and installed using the {\tt make} and {\tt make install} commands from the program directory.  
\end{enumerate}
Information about the final state particles is stored in the arrays {\tt jets}, {\tt leptons}, {\tt invisible} (neutrinos) and {\tt photons}. 
The entries are as follows 
\begin{itemize}
\item {\tt jets(i,j)}: information about the jets is stored here, entries are $p_T$-ordered
\begin{itemize}
\item {\tt j} = jet number
\item {\tt i = 0-3} = 4-momentum
\item {\tt i = 4} = mass
\item {\tt i = 5} = transverse momentum $p_T$
\item {\tt i = 6} = rapidity $y$
\item {\tt i = 7} = azimuthal angle $\phi$
\end{itemize}
\item {\tt leptons(i,j)}: information about the charged leptons is stored here
\begin{itemize}
\item {\tt j} = lepton number
\item {\tt i = 0-3} = 4-momentum
\item {\tt i = 4} = mass (all leptons are treated as massless, i.e. this entry is always equal to zero)
\item {\tt i = 5} = transverse momentum $p_T$
\item {\tt i = 6} = rapidity $y$
\item {\tt i = 7} = azimuthal angle $\phi$
\item {\tt i = 8} = lepton ID (i.e. PDG particle number)
\end{itemize}
\item {\tt invisible(i,j)}: information about the invisible particles (i.e. neutrinos) is stored here
\begin{itemize}
\item {\tt j} = neutrino number
\item {\tt i = 0-3} = 4-momentum
\item {\tt i = 4} = mass (all neutrinos are treated as massless, i.e. this entry is always equal to zero)
\item {\tt i = 5} = transverse momentum $p_T$
\item {\tt i = 6} = rapidity $y$
\item {\tt i = 7} = azimuthal angle $\phi$
\item {\tt i = 8} = neutrino ID (i.e. PDG particle number)
\end{itemize}
\item {\tt photons(i,j)}: information about the photons is stored here
\begin{itemize}
\item {\tt j} = photon number
\item {\tt i = 0-3} = 4-momentum
\item {\tt i = 4} = mass (all photons are massless, i.e. this entry is always equal to zero)
\item {\tt i = 5} = transverse momentum $p_T$
\item {\tt i = 6} = rapidity $y$
\item {\tt i = 7} = azimuthal angle $\phi$
\end{itemize}
\end{itemize}
\subsection[Implementation of new histograms]
{Implementation of new histograms}
\label{sec:new_hists}
{\tt VBFNLO} offers two kinds of histograms, 1D and 2D.
Additional histograms can be created easily in {\tt VBFNLO} by modifying the file {\tt utilities/histograms.F} as follows.
\begin{enumerate}
\item In the subroutine {\tt InitHistograms} create a new histogram template with the following command, for the 1D case: 
\newline
\\
{\tt call CreateHist(HIST\_ID, "title", nobins, xmin, xmax)},\newline
\\
where {\tt HIST\_ID} is the integer identification of the 
histogram (make sure to use a different number for each histogram); 
{\tt title} the title of the histogram (will be used as plot title -- e.g. {\tt dS/dmjj: dijet mass}); 
{\tt nobins} the number of histogram bins; 
{\tt xmin} the lower bound of the histogram;
{\tt xmax} the upper bound of histogram.
Similarly, for the 2D case:
\newline
\\
{\tt call Create2dHist(HIST\_ID, "title", "x-axis label", "y-axis label",\newline 
nobins\_x, nobins\_y, xmin, xmax, ymin, ymax)},\newline
\\
where the number of bins and the bounds must be specified separately for the two axes. 
The axis labels {\tt "x-axis label"} and {\tt "y-axis label"} 
can read e.g. {\tt "eta\_jj"} and {\tt "m\_jj"}, respectively.
\item In the subroutine {\tt HistogramEvent}, for 1D case, 
calculate the desired distribution (or observable) value, named {\tt ObsValue} - e.g. dijet mass {\tt mjj}, then fill 
the new histogram using the command: 
\newline
\\
{\tt call FillHist(HIST\_ID, ObsValue, dw, NLO)},\newline
\\
where {\tt HIST\_ID} is the integer identification of the histogram, same number as in {\tt CreateHist};
{\tt dw} is the variable {\tt dw} (this is already calculated by the code, being the weight*cross section);
{\tt NLO} the variable {\tt NLO} (this is already filled by the code – flag determining the order of the calculation). 
Similarly, for a 2D histogram, the command reads
\newline
\\
{\tt call FillHist2d(HIST\_ID, ObsValue\_x, ObsValue\_y, dw, NLO)},\newline
\\
where {\tt ObsValue\_x} and {\tt ObsValue\_y} are the x-axis and y-axis observables, respectively. 
They can be, e.g. {\tt etajj} and {\tt mjj}, which must be calculated from the external-state momenta. 
\item The {\tt utilities} directory must be cleaned by using the {\tt make clean} command before {\tt VBFNLO} is 
remade and installed using the {\tt make} and {\tt make install} commands from the program directory.    
\end{enumerate}
Information about the final state particles is stored in the arrays {\tt jets}, {\tt leptons}, {\tt invisible} (neutrinos) and {\tt photons} as 
in the previous section.
\subsection[Implementation of new dynamical scales]
{Implementation of new dynamical scales}
\label{sec:new_scales}
A number of popular dynamical scale choices have been implemented and 
explained in \sect{sec:vbfnlodat}. They can be used as templates for implementing a new scale.
If the desired factorization or renormalization 
scale is not in this list, the user can easily add it as follows. 
\begin{enumerate}
\item One first looks into the interface subroutine {\tt Scales} in the file {\tt utilities/scales.F90} 
to find out where to implement the new scales of the process at hand. 
For example, for the class of VBF or VBS processes ($Hjj$, $Vjj$, $VVjj$), the scales are calculated in 
the subroutine {\tt calcVBFscales}, while it is {\tt calcVXjScales} for diboson and triboson processes.
\item One then implements the factorization and renormalization scales separately, e.g. in {\tt calcVBFscales}. 
The new identification numbers for {\tt ID\_MUF} and {\tt ID\_MUR} must be different from the ones already used. 
A good practice is to use the same value for both scales.
The factorization-scale implementation reads
\newline
\\
{\tt case(ID\_MUF\_NEW)} \newline
{\tt do L=1,Lmax}\newline
{\tt ...}\newline
{\tt mufsq(1,L) = ...}\newline
{\tt mufsq(2,L) = ...}\newline
{\tt enddo}\newline
\\
where {\tt Lmax} (already calculated by the program) 
is the number of different kinematic configurations, which is greater than unity when calculating the dipole-subtraction term at NLO; 
{\tt mufsq(1,L)} is the square of the factorization scale for the first hadron, 
{\tt mufsq(2,L)} is for the second hadron.
Similarly, for the renormalization scale, we have
\newline
\\
{\tt case(ID\_MUR\_NEW)} \newline
{\tt do L=1,Lmax}\newline
{\tt ...}\newline
{\tt mursq(1,L) = ...}\newline
{\tt als(1,L) = ...}\newline
{\tt als(2,L) = ...}\newline
{\tt ...}\newline
{\tt enddo}\newline
\\
where {\tt mursq(1,L)} is the square of the renormalization scale, 
the calculation of the strong coupling constants {\tt als(i,L)} 
is the same as for the already implemented cases. 
The values of {\tt mufsq(i,L)}, {\tt mursq(i,L)} and {\tt als(i,L)} 
are then transfered to other routines via the common block {\tt cscales} 
in {\tt scales.inc}. 
\item The {\tt utilities} directory must be cleaned by using the {\tt make clean} command before {\tt VBFNLO} is 
remade and installed using the {\tt make} and {\tt make install} commands from the program directory. 
The new scales are then ready to be used by specifying the values of the identification numbers 
in {\tt vbfnlo.dat}.  
\end{enumerate}
Dynamical scales are calculated from the momenta of the external-state particles. 
The following momenta are available and ready to be used
\begin{itemize}
\item {\tt p(i,j,L)}: 4-momenta of the partons.
\item {\tt v(i,j,L)}: 4-momenta of all other particles (charged leptons, neutrinos, photons, ...). 
The {\tt particle\_IDs(j)} can be used to select the desired momentum, as it 
returns the particle identification code according to the Monte Carlo particle numbering scheme. 
\item {\tt jets(i,j,L)}: array with the jet information (4-momenta, mass, $p_T$, rapidity $y$, azimuthal angle $\phi$).
\item {\tt leptons(i,j,L)}: array with the charged lepton information (4-momenta, mass, $p_T$, rapidity $y$, azimuthal angle $\phi$).
\item {\tt photons(i,j,L)}: array with the photon information (4-momenta, mass, $p_T$, rapidity $y$, azimuthal angle $\phi$).
\end{itemize}
Differently from the momenta in Section~\ref{sec:new_cuts}, the momenta here depend additionally on the index {\tt L} marking the 
kinematic configurations.
It is always a good practice to print out the value of an unknown variable to get more information about it.

\newpage

\section{\textsc{Changes}}

The release \textsc{Version 3.0} includes some changes that alter previous results:

\subsection{NLO calculation of VBF-$Hjjj$}

A bug has been found in the virtual and real-emission parts of VBF-$Hjjj$
production, which leads to a decrease of the NLO cross section of roughly 10\%.

\subsection{NLO calculation of VBF-$Z\gamma jj$ with anomalous couplings}

A bug has been found in the real-emission part of VBF-$Z\gamma jj$
production when anomalous couplings were switched on.

\subsection{K-matrix unitarisation}

Following the introduction of the dimension~8 operator $\mathcal{O}_{S,2}$, the
K-matrix unitarisation procedure for $\mathcal{O}_{S,0}$ has been reworked.
The relation to the parameters $\alpha_4$ and $\alpha_5$ of the electroweak
chiral Lagrangian is now process-universal. To account for the
isospin-conservation of these operators, which is used in the K-matrix
procedure, only the combination $\mathcal{O}_{S,0}\equiv\mathcal{O}_{S,2}$ can
be unitarised, as well as $\mathcal{O}_{S,1}$.

\subsection{Calculation of histogram-bin errors}

A bug has been found in the calculation of the statistical errors of histogram bins, 
which leads to an increase of the errors. The mean value of the cross section (bin-wise) is unchanged. 
The statistical error of the total cross section was correctly calculated, hence not affected by this change. 

\subsection{Previous changes -- version 2.7.1}

The release \textsc{Version 2.7.1} includes some changes that alter previous results:

\subsubsection{NLO calculations of $W^-jj$, $W^-Zjj$, $W^-\gamma jj$ production}

A bug in the PDF convolution of the QCD-induced production processes $W^-jj$,
$W^-Zjj$ and $W^-\gamma jj$ has been fixed. The size of the effect,
which can be observed both in the real-emission part of the NLO
calculation and the corresponding LO+jet processes, can reach several
percent, depending on the parameter settings.

\subsubsection{VBF cuts for $HHjj$}

VBF cuts now also apply to VBF production of a Higgs pair.

\subsection{Previous changes -- version 2.7.0}

The release \textsc{Version 2.7.0} includes some changes that alter previous results:

\subsubsection{Running scales in NLO calculations}

In release 2.7.0 a problem with certain dynamical renormalization and factorization
scales has been fixed, which lead to wrong results at NLO QCD in the subtraction part of the
real emission calculation. In particular, the scale choices ``$\min(p_T(j_i))$'' (ID=2)
and ``minimal transverse energy of the bosons'' (ID=7) did not give correct results
in previous releases.

\subsubsection{Jet cuts in VBF/gluon fusion Higgs boson production with $H\to b\bar{b}$}

Starting with {\sc Vbfnlo 2.7.0} the jet cuts will be applied to the
$b$-quarks from Higgs boson decay.

\subsubsection{NLO calculation of $W^+W^-Z$ production}

A bug in the calculation of the virtual contributions for $WWZ$
production has been fixed while comparing with the results of
Ref.~\cite{Nhung:2013jta}.  The new results are roughly one per cent
smaller, and agree between both codes for squared amplitudes at the
level of the machine precision and for integrated cross sections at the
per mill level.

\subsubsection{LO and NLO calculation of $W^+jj$ / $Zjj$ production in VBF}

The particle--anti-particle--assignment in $W^+jj$, $Zjj$ has been fixed.
In case of $W^+jj$ production this leads to an increase in cross section of roughly one per cent with basic cuts.

\subsubsection{Event output}

Several bugs have been fixed concerning the event output to Les Houches or HepMC files:
\begin{itemize}
 \item The color information in the event output for $W^-W^- jj$ in VBF has been fixed.
 \item The momenta assignment in the event output for $W^-W^+W^-$ has been corrected.
\end{itemize}

\subsection{Previous changes -- version 2.6.3}

The release \textsc{Version 2.6.3} includes some changes that alter previous results:

\subsubsection{Event output}

Several bugs have been fixed concerning the event output to Les Houches or HepMC files:
\begin{itemize}
 \item Fixed bugs in output of particle IDs for processes 191 and 43xx: In previous versions
       some particles had the particle ID 0.
 \item Parton and beam particle IDs have been fixed for process 260.
 \item The tau mass can now be included in the event output of all processes. Furthermore,
       several bugs have been fixed in the existing implementation of tau mass inclusion.
 \item Several bugs have been fixed in the helicity output.
\end{itemize}

\subsubsection{Calculation of the $H\to gg$ partial width}

Higher-order corrections to the $H\to gg$ partial width have been included which lead to slightly smaller
branching ratios for all other decay channels. Therefore cross sections of processes involving Higgs bosons
can be up to a few per cent smaller.

\subsubsection{Electroweak corrections in the VBF $Hjj$ processes}

Some bugs have been fixed in the calculation of electroweak corrections for the processes 10x.

\subsection{Previous changes -- version 2.6.2}

The release \textsc{Version 2.6.2} included some changes that altered previous results:

\subsubsection{Distributions and cross section for $H\to WW$ in VBF and gluon fusion}

Due to a bug which has been fixed in the lepton assignment for $H\to WW/ZZ \to 4\ell$ distributions
(and cross sections after the $m_{\ell\ell}$ cut) were off for the 1x5 and 4105 processes
in the previous versions. This bug has been fixed in v.2.6.2.

\subsubsection{Les Houches event output for processes with more than one phase space}

The fraction of events coming from the different phase spaces was not sampled correctly 
in previous versions. Processes affected are $W\gamma$, $Z\gamma$, $W\gamma j$,
$Z\gamma j$, $ZZZ$, $WW\gamma$, $WZ\gamma$, $W\gamma\gamma$, $Z\gamma\gamma$,
$\gamma\gamma\gamma$ and $W\gamma\gamma j$.

\subsubsection{$ZZZ$ production}

The QCD real emission part of the NLO computation gave no reliable result in previous versions
due to a bug in the dipole subtraction. This bug has been fixed in v.2.6.2.

\subsubsection{Form factor in $\gamma\,jj$ production with anomalous couplings}

 For the $\gamma\,jj$ production in VBF (procID 150) a different form factor is used.

\subsection{Previous changes -- version 2.6.1}

The release \textsc{Version 2.6.1} included some changes that altered previous results:

\subsubsection{Anomalous Higgs couplings}

A bug was found and fixed in the implementation of the {\tt TREEFACZ} and {\tt TREEFACW}, 
the factors which multiply the SM $HZZ$ and $HWW$ couplings.  Note that this bug was only 
present in \textsc{Version 2.6.0}, not in earlier versions.  Additionally, a small bug was 
found and fixed in the coefficient of the input {\tt FB\_ODD} in the $a_{3}^{HZZ}$ coupling.

\subsubsection{Symmetry factor $ZZ \rightarrow \ell^{+} \ell^{-} \ell^{+} \ell^{-}$ }

In the processes $pp \rightarrow H \gamma jj \rightarrow ZZ \gamma jj \rightarrow \ell^{+} 
\ell^{-} \ell^{+} \ell^{-} \gamma jj$ (ID 2106) and gluon fusion $pp \rightarrow H jj \rightarrow 
ZZ jj \rightarrow \ell^{+} \ell^{-} \ell^{+} \ell^{-} jj $ (ID 4106) a symmetry factor was missing 
when identical final-state leptons were chosen.

\subsection{Previous changes -- version 2.6.0}

The release \textsc{Version 2.6.0} included some changes 
that alter results (events, cross sections and distributions) from earlier versions.

\subsubsection{Allowed width of virtuality}
In the phase-space generators, the allowed range of the virtuality of a resonance of intermediate vector bosons has been increased.  This mainly affects processes where an intermediate $Z$ boson decays into a pair of neutrinos -- i.e.
\begin{itemize}
 \item \ $pp \rightarrow Hjj \rightarrow ZZjj \rightarrow \ell^{+} \ell^{-} \nu \overline{\nu} jj$ via vector boson fusion (process ID 107) and gluon fusion (process ID 4107)
 \item $pp \rightarrow Hjjj \rightarrow ZZjjj \rightarrow \ell^{+} \ell^{-} \nu \overline{\nu} jjj$ (process ID 117)
 \item $pp \rightarrow H\gamma jj \rightarrow ZZ\gamma jj \rightarrow \ell^{+} \ell^{-} \nu \overline{\nu} \gamma jj$ (process ID 2107)
 \item $pp \rightarrow ZZjj \rightarrow \ell^{+} \ell^{-} \nu \overline{\nu} jj$ (process ID 211)
\end{itemize}
This not only affects the cross sections for these processes, but also means that the events produced by \textsc{Vbfnlo-2.6.0} will differ from those produced by \textsc{Vbfnlo-2.5}, even if the same random numbers are used.

\subsubsection{Matrix element $H \rightarrow ZZ \rightarrow 4\ell$}
A bug was found and fixed in the implementation of the matrix element
calculating the decay $H \rightarrow ZZ \rightarrow 4\ell$.  

\subsubsection{Anomalous couplings}
Several changes have been made to the implementation of the anomalous couplings.  For Higgs production via vector boson fusion (process IDs 100-107) the variable {\tt TREEFAC}, which multiplies the Standard Model contribution to the tree-level $HVV$ couplings, has been corrected and altered -- now, separate factors for $HZZ$ and $HWW$ are input ({\tt TREEFACZ} and {\tt TREEFACW} respectively). 

When working with anomalous $HVV$ couplings two types of formfactor can be applied which model effective, momentum dependent $HVV$ vertices, motivated by new physics entering with a large scale $\Lambda$ at loop level.  Corrections to the $HVV$ formfactor $F_{2}$ (see Eq.~\ref{eq:ff2}) have been made.   The implementation of the parameterization described by {\tt PARAMETR3} -- where the input determining the anomalous couplings is in terms of the dimension-6 operators ($\mathcal{O}_{W}$, $\mathcal{O}_{B}$, $\mathcal{O}_{WW}$ and $\mathcal{O}_{BB}$) have also been altered. 

If anomalous triple (and quartic) gauge boson couplings are being studied, a formfactor given by
\begin{equation}
 F = \left(1 + \frac{s}{\Lambda^{2}} \right)^{-p},
\end{equation}
can be applied in order to preserve unitarity, where $\Lambda$ is the scale of new physics.  
The momentum dependence of the applied formfactor (i.e.\ $s$) is now universal for each phase-space point, with the invariant mass of the produced bosons as the scale.  This ensures the proper cancellations for anomalous contributions.  The values of the formfactor scales $\Lambda$ and $p$ can be set to different values for each input describing the triboson couplings.  In the parameterization {\tt TRIANOM = 2}, the formfactor scales for $\Delta \kappa_{\gamma}$ and $\Delta \kappa_{Z}$ are now separately set, and the consistency of related parameters  (i.e.\ $\Delta g_{1}^{Z}$, $\Delta \kappa_{\gamma}$ and $\Delta \kappa_{Z}$) is enforced when formfactors are applied.

When processes involving resonant Higgs diagrams (e.g.\ $WWW$ production) are studied with anomalous couplings, the Higgs width is now calculated with the appropriate anomalous $HVV$ couplings (the anomalous $HVV$ couplings in the production amplitudes were taken into account in previous versions of \textsc{Vbfnlo}).  Various corrections have also been made to the anomalous triboson couplings in diboson plus jet processes (these were incorporated into the intermediate release \textsc{Vbfnlo} 2.5.3).

\subsubsection{VBF Higgs boson production in association with three jets}
A small bug was found and fixed in the calculation of the processes $pp \rightarrow H jjj$, with process IDs 110-117.

\subsection{Previous changes -- version 2.5.0}

The previous version of \textsc{Vbfnlo -- 2.5.0} -- was altered in such a
way that some results differ from \textsc{Vbfnlo-2.0} and below.  These changes are
described briefly here, and in more detail on the \textsc{Vbfnlo} website, {\tt
\url{https://ific.uv.es/vbfnlo/}}.

\subsubsection{{\tt EWSCHEME}}

The implementation of options {\tt EWSCHEME = 1} and {\tt 4}, which are
described in Section~\ref{sec:physics}, was altered.  The new
implementation is hopefully more transparent, and is described in this manual. 

\subsubsection{Gluon fusion}

Since version 2.5.0, in gluon fusion processes the bottom-quark
mass $M_{b}(M_{H})$ is used throughout the calculation.  New, more stable,
tensor routines for the boxes and pentagons were implemented, reducing the
number of omitted points with bad numerical accuracy.


\newpage

\section{\textsc{Checks}}

Extensive checks for the LO and the real emission amplitudes, as well as for the
total LO cross sections, have been performed for all processes implemented in
\textsc{Vbfnlo}.  Born amplitudes and real emission diagrams have been compared
with the fully automatically generated results provided by
\textsc{MadGraph}\cite{Stelzer:1994ta}. Complete agreement has been found in
each case.  Moreover, total LO cross sections with a minimal set of cuts agree
with the respective results obtained by \textsc{MadEvent}\footnote{\tt
  \url{http://madgraph.physics.illinois.edu}}~\cite{Maltoni:2002qb,Alwall:2007st},
\textsc{Sherpa}\footnote{\tt \url{https://sherpa.hepforge.org/}}~\cite{Gleisberg:2008ta}
and/or \textsc{Helac-Phegas}\footnote{\tt
\url{https://helac-phegas.web.cern.ch/helac-phegas/}}
\cite{Kanaki:2000ey,Papadopoulos:2005ky,Cafarella:2007pc}. 

LHA event files for the LO processes have been tested with
\textsc{Herwig++}\footnote{\tt \url{https://herwig.hepforge.org/}}
\cite{Bahr:2008pv},  a general purpose Monte Carlo event generator for the
simulation of hard lepton-lepton and hadron-hadron collisions.

As a final and very important test, comparisons with already published results
have been made. The NLO results for Higgs boson production via VBF agree with those
produced by the code \textsc{Hawk}\footnote{
{\tt \href{http://omnibus.uni-freiburg.de/~sd565/programs/hawk/hawk.html}{http://omnibus.uni-freiburg.de/\~{}sd565/programs/hawk/hawk.html}}}.  In
Ref.~\cite{Adam:2008aa}, a tuned comparison of LO and NLO QCD results for Higgs 
boson production via vector boson fusion at the LHC has been performed. Three
different calculations have been cross checked: \textsc{Vbfnlo}, the results of
Refs.~\cite{Ciccolini:2007jr,Ciccolini:2007ec}, and the \textsc{VV2H}
program\footnote{\tt \url{http://tiger.web.psi.ch/vv2h/}}. For the dominant
$t$- and $u$-channel contributions which are implemented in \textsc{Vbfnlo},
good agreement has been found.  
Diboson processes, including $WH$, as well as $W$ and $Wj$ production have been
checked against \textsc{Mcfm}\footnote{{\tt
\url{https://mcfm.fnal.gov/}}}~\cite{Campbell:1999ah,Campbell:2011bn,Campbell:2011cu}.
The results for $WH$+jet production have been compared with Ref.~\cite{JiJuan:2010ga}.
For the triboson processes a comparison for the
production of on-shell gauge bosons without leptonic decays has been performed
with the results presented in Ref.~\cite{Binoth:2008kt}. Additionally,
the tree-level results for $W^\pm\gamma\gamma$ have been compared to the
results with an on-shell $W$ boson of Ref.~\cite{Baur:2010zf}. Again, in
all cases good agreement has been found.  Triple photon production has been tested against 
\textsc{FeynArts}, \textsc{FormCalc} and \textsc{HadCalc}~\cite{Hahn:1998yk,Hahn:2006qw,Rauch:2008fy}.
QCD-induced $\gamma jj$ production has been compared to \textsc{MadGraph5\_aMC@NLO}~\cite{Alwall:2014hca} 
using the model \texttt{sm-no\_b\_mass} and agreement at the $0.1\%$ level was found.
Processes with anomalous quartic gauge couplings have been compared with \textsc{MadGraph/MadEvent} and
\textsc{Whizard}\footnote{\tt
\url{https://whizard.hepforge.org/}}~\cite{Kilian:2007gr}
in Ref.~\cite{Degrande:2013rea}.
Results for the $\cal{CP}$-odd and $\cal{CP}$-even Higgs boson production via gluon
fusion have been tested against
\textsc{FeynArts} and \textsc{FormCalc}~\cite{Hahn:2000kx,Hahn:2001rv}.  

All fermion loops have been checked against results obtained with \textsc{FeynArts}, \textsc{FormCalc} and \textsc{LoopTools}. 
and the gluon-gluon loops have also been checked against \textsc{gg2VV}\footnote{\tt 
\url{https://gg2VV.hepforge.org/}}~\cite{Binoth:2008pr,Kauer:2012ma}.


\newpage

\section{\textsc{Outlook}}

Additional processes and features will become available in the near future and will be included in the code version on the \textsc{Vbfnlo} website.  


\section*{Acknowledgments}
We are very grateful to %
Ken Arnold, %
Manuel B\"ahr, %
Johannes Bellm, %
Giuseppe Bozzi, %
Martin Brieg, %
Christoph Englert, %
Bastian Feigl, %
Jessica Frank, %
Florian Geyer, %
Nicolas Greiner, %
Christoph Hackstein, %
Vera Hankele, %
Barbara J\"ager, %
Nicolas Kaiser, %
Gunnar Kl\"amke, %
Carlo Oleari, %
Sophy Palmer, %
Stefan Prestel, %
Heidi Rzehak, %
Franziska Schissler, %
Oliver Schlimpert, %
Michael Spannowsky and %
Malgorzata Worek %
who, as earlier members of the
\textsc{Vbfnlo} collaboration, made important contributions to the original
release.  We also gratefully acknowledge the collaboration of Stefan Kallweit
and Georg Weiglein in calculation of radiative corrections for specific processes.  
TF would like to thank the North American Foundation for The University of
Manchester and George Rigg for their financial support.


\newpage

\appendix

\section{\textsc{Operators for anomalous gauge boson couplings}}

\subsection{List of operators implemented in \textsc{Vbfnlo}}
\label{listofoperators}

This is a list of the full set of operators included the effective Lagrangian
which is used for calculations with anomalous gauge boson couplings.
The full set can be used in the diboson, triboson and $VVjj$ in VBF production
processes.

The dimension-6 operators are constructed according to Refs.~\cite{Hagiwara:1993qt,Hagiwara:1993ck} and
the dimension-8 operators are taken from~\cite{Eboli:2006wa,Eboli:2016kko}. The operators $\mathcal{O}_{T,3}$, $\mathcal{O}_{T,4}$ and $\mathcal{O}_{M,5hc}$ are constructed accordingly to complete the sets of $\mathcal{O}_{T}$ and $\mathcal{O}_{M}$ operators as presented in Refs.~\cite{Murphy:2020rsh,Li:2020gnx}. See also Refs.~\cite{Degrande:2013rea,Rauch:2016pai} for relations between different sets of definitions. More details on the implementation of these operators can be found in 
Refs.~\cite{anomWW, anomVVV, anomVBF}.

\vskip6pt
\noindent
With the building blocks (following the notation of~\cite{Hagiwara:1993qt,Hagiwara:1993ck})
\begin{eqnarray}
 \widehat{W}_{\mu\nu} &=& i g T^{a} W^{a}_{\mu\nu} \nonumber \\
 \widehat{B}_{\mu\nu} &=& i g' Y B_{\mu \nu} \nonumber \\
  D_\mu &=& \partial_\mu + i g T^{a} W^{a}_\mu  + i g' Y B_\mu,
\end{eqnarray}
we can construct the following operators, where $g$ and $g'$ are the SU(2) 
and U(1) gauge couplings, and $T^{a}$ are the SU(2) generators.
We obtain slightly different expressions for the dimension-8 operators compared to~\cite{Eboli:2006wa},
as they use different field strength tensor expressions ($\widehat{W}_{\mu\nu} = T^{a} W^{a}_{\mu\nu}$
and $\widehat{B}_{\mu\nu} = B_{\mu \nu}$). The conversion factors for the coupling
strengths $f_i$ between~\cite{Eboli:2006wa} and our implementation can be found in
Appendix~\ref{ancpl_conversion}.

\vskip6pt
\noindent
The explicit form of the $\cal{CP}$-even operators is
\begin{eqnarray}
\mathcal{O}_{W} &=& (D_{\mu} \Phi)^{\dagger} \widehat{W}^{\mu \nu} (D_{\nu} \Phi) \nonumber \\
\mathcal{O}_{B} &=& (D_{\mu} \Phi)^{\dagger} \widehat{B}^{\mu \nu} (D_{\nu} \Phi) \nonumber \\
\mathcal{O}_{WWW} &=& Tr \left[ \widehat{W}_{\mu\nu} \widehat{W}^{\nu\rho} \widehat{W}_\rho^{\mbox{ } \mu} \right] \nonumber \\
\mathcal{O}_{WW} &=& \Phi^{\dagger} \widehat{W}_{\mu\nu} \widehat{W}^{\mu\nu} \Phi \nonumber \\
\mathcal{O}_{BB} &=& \Phi^{\dagger} \widehat{B}_{\mu\nu} \widehat{B}^{\mu\nu} \Phi .
\end{eqnarray}
If we replace one field strength tensor with the corresponding dual field strength tensor,
we obtain the $\cal{CP}$-odd part of the Lagrangian
\begin{eqnarray}
\mathcal{O}_{\widetilde{W}} &=& (D_{\mu} \Phi)^{\dagger} \widehat{\widetilde{W}}^{\mu \nu} (D_{\nu} \Phi) \nonumber \\
\mathcal{O}_{\widetilde{B}} &=& (D_{\mu} \Phi)^{\dagger} \widehat{\widetilde{B}}^{\mu \nu} (D_{\nu} \Phi) \nonumber \\
\mathcal{O}_{\widetilde{W}WW} &=& Tr \left[ \widehat{\widetilde{W}}_{\mu\nu} \widehat{W}^{\nu\rho} \widehat{W}_\rho^{\mbox{ } \mu} \right] \nonumber \\
\mathcal{O}_{\widetilde{W}W} &=& \Phi^{\dagger} \widehat{\widetilde{W}}_{\mu\nu} \widehat{W}^{\mu\nu} \Phi \nonumber \\
\mathcal{O}_{\widetilde{B}B} &=& \Phi^{\dagger} \widehat{\widetilde{B}}_{\mu\nu} \widehat{B}^{\mu\nu} \Phi \nonumber \\
\mathcal{O}_{D\widetilde{W}} &=& Tr \left( [D_\mu, \widehat{\widetilde{W}}_{\nu\rho} ] [D^\mu, \widehat{W}^{\nu\rho}] \right) \nonumber \\
\mathcal{O}_{B\widetilde{W}} &=& \Phi^{\dagger} \widehat{B}_{\mu\nu} \widehat{\widetilde{W}}^{\mu\nu} \Phi \, .
\end{eqnarray}
The operators $\mathcal{O}_{B\widetilde{W}}$ and $\mathcal{O}_{D\widetilde{W}}$
are included only in their $\cal{CP}$-odd version, as their $\cal{CP}$-even counterparts affect the gauge boson two-point functions and
are already extremely constrained by electroweak precision tests.
Furthermore, only 4 of these 7 operators are linearly independent. For example $\mathcal{O}_{\widetilde{W}}$, $\mathcal{O}_{B\widetilde{W}}$ 
and $\mathcal{O}_{D\widetilde{W}}$ can be written as
\begin{eqnarray}
\mathcal{O}_{\widetilde{W}} &=& \mathcal{O}_{\widetilde{B}} - \frac{1}{2} \mathcal{O}_{\widetilde{W}W} + \frac{1}{2} \mathcal{O}_{\widetilde{B}B}  \nonumber \\
\mathcal{O}_{B\widetilde{W}} &=& -2 \, \mathcal{O}_{\widetilde{B}} - \mathcal{O}_{\widetilde{B}B} \nonumber \\
\mathcal{O}_{D\widetilde{W}} &=& -4 \, \mathcal{O}_{\widetilde{W}WW} \, .
\end{eqnarray}

\vskip6pt
\noindent
The dimension-8 operators can be split into three categories:
\begin{enumerate}
 \item Operators using $D_\mu \Phi$:
  \begin{eqnarray}
   \mathcal{O}_{S,0} &=& \left[ (D_\mu \Phi)^\dagger D_\nu \Phi \right] \times
      \left[ ( D^\mu \Phi)^\dagger D^\nu \Phi \right] \nonumber\\
   \mathcal{O}_{S,1} &=& \left[ ( D_\mu \Phi )^\dagger D^\mu \Phi  \right] \times
      \left[ ( D_\nu \Phi )^\dagger D^\nu \Phi \right] \nonumber\\
   \mathcal{O}_{S,2} &=& \left[ (D_\mu \Phi)^\dagger D_\nu \Phi \right] \times
      \left[ ( D^\nu \Phi)^\dagger D^\mu \Phi \right]
  \end{eqnarray}

 \item Operators using $D_\mu \Phi$, $\hat{W}_{\mu \nu}$ and $\hat{B}_{\mu \nu}$:
  \begin{eqnarray}
   \mathcal{O}_{M,0} &=& \hbox{Tr}\left[ \hat{W}_{\mu\nu} \hat{W}^{\mu\nu} \right] \times 
      \left[ ( D_\beta \Phi )^\dagger D^\beta \Phi \right] \nonumber\\
   \mathcal{O}_{M,1} &=& \hbox{Tr}\left[ \hat{W}_{\mu\nu} \hat{W}^{\nu\beta} \right] \times  
      \left[ ( D_\beta \Phi )^\dagger D^\mu \Phi \right] \nonumber\\
   \mathcal{O}_{M,2} &=& \left[ \hat{B}_{\mu\nu} \hat{B}^{\mu\nu} \right ] \times  
      \left [ ( D_\beta \Phi )^\dagger D^\beta \Phi \right ] \nonumber\\
   \mathcal{O}_{M,3} &=& \left[ \hat{B}_{\mu\nu} \hat{B}^{\nu\beta} \right ] \times  
      \left [ ( D_\beta \Phi )^\dagger D^\mu \Phi \right ] \nonumber\\
   \mathcal{O}_{M,4} &=& \left[ ( D_\mu \Phi )^\dagger \hat{W}_{\beta\nu} 
      D^\mu \Phi  \right] \times \hat{B}^{\beta\nu} \nonumber\\
   \mathcal{O}_{M,5} &=& \left[ ( D_\mu \Phi )^\dagger \hat{W}_{\beta\nu} 
     D^\nu \Phi  \right] \times \hat{B}^{\beta\mu} \nonumber\\
    \mathcal{O}_{M,5hc} &=& \left[ ( D_\mu \Phi )^\dagger \hat{W}_{\beta\mu}
      D^\nu \Phi  \right] \times \hat{B}^{\beta\nu} \nonumber\\
      \mathcal{O}_{M,6} &=& \left[ ( D_\mu \Phi )^\dagger \hat{W}_{\beta\nu} 
      \hat{W}^{\beta\nu} D^\mu \Phi \right] \nonumber\\
   \mathcal{O}_{M,7} &=& \left[ ( D_\mu \Phi )^\dagger \hat{W}_{\beta\nu} 
     \hat{W}^{\beta\mu} D^\nu \Phi  \right]
  \end{eqnarray}

 \item Operators using $\hat{W}_{\mu \nu}$ and $\hat{B}_{\mu \nu}$:
  \begin{eqnarray}
   \mathcal{O}_{T,0} &=& \hbox{Tr}\left[ \hat{W}_{\mu\nu} \hat{W}^{\mu\nu} \right] \times 
      \hbox{Tr}\left[ \hat{W}_{\alpha\beta} \hat{W}^{\alpha\beta} \right] \nonumber\\
   \mathcal{O}_{T,1} &=& \hbox{Tr}\left[ \hat{W}_{\alpha\nu} \hat{W}^{\mu\beta} \right] 
      \times \hbox{Tr}\left[ \hat{W}_{\mu\beta} \hat{W}^{\alpha\nu} \right] \nonumber\\
   \mathcal{O}_{T,2} &=& \hbox{Tr}\left[ \hat{W}_{\alpha\mu} \hat{W}^{\mu\beta} \right] 
   \times   \hbox{Tr}\left[ \hat{W}_{\beta\nu} \hat{W}^{\nu\alpha} \right] \nonumber\\
\mathcal{O}_{T,3} &=& \hbox{Tr}\left[ \hat{W}^{\mu\nu} \hat{W}^{\alpha\beta} \right]
      \times   \hbox{Tr}\left[ \hat{W}_{\nu\alpha} \hat{W}^{\beta\mu} \right] \nonumber\\
\mathcal{O}_{T,4} &=& \hbox{Tr}\left[ \hat{W}^{\mu\nu} \hat{W}^{\alpha\beta} \right]
      \times   \hbox{Tr}\left[ \hat{B}_{\nu\alpha} \hat{B}^{\beta\mu} \right] \nonumber\\   
   \mathcal{O}_{T,5} &=& \hbox{Tr}\left[ \hat{W}_{\mu\nu} \hat{W}^{\mu\nu} \right]\times 
      \hat{B}_{\alpha\beta} \hat{B}^{\alpha\beta} \nonumber\\
   \mathcal{O}_{T,6} &=& \hbox{Tr}\left[ \hat{W}_{\alpha\nu} \hat{W}^{\mu\beta} \right] 
      \times \hat{B}_{\mu\beta} \hat{B}^{\alpha\nu} \nonumber\\
   \mathcal{O}_{T,7} &=& \hbox{Tr}\left[ \hat{W}_{\alpha\mu} \hat{W}^{\mu\beta} \right] 
      \times \hat{B}_{\beta\nu} \hat{B}^{\nu\alpha} \nonumber\\
   \mathcal{O}_{T,8} &=& \hat{B}_{\mu\nu} \hat{B}^{\mu\nu}  \hat{B}_{\alpha\beta} \hat{B}^{\alpha\beta} \nonumber\\
   \mathcal{O}_{T,9} &=& \hat{B}_{\alpha\mu} \hat{B}^{\mu\beta} \hat{B}_{\beta\nu} \hat{B}^{\nu\alpha}
  \end{eqnarray}
\end{enumerate}

\newpage
\noindent
Each new operator introduces a new coupling strength $f_i$, so the complete effective Lagrangian
containing effects from dimension six and eight operators is

\vbox{
\begin{eqnarray}
\mathcal{L}_{eff} &=& \mathcal{L}_{SM} \nonumber\\
        &+& \frac{f_{W}}{\Lambda^{2}} \mathcal{O}_{W} + \frac{f_{B}}{\Lambda^{2}} \mathcal{O}_{B} 
    + \frac{f_{WWW}}{\Lambda^{2}} \mathcal{O}_{WWW} + \frac{f_{WW}}{\Lambda^{2}} \mathcal{O}_{WW}
        + \frac{f_{BB}}{\Lambda^{2}} \mathcal{O}_{BB} \nonumber\\
        &+& \frac{f_{Wt}}{\Lambda^{2}} \mathcal{O}_{\widetilde{W}} + \frac{f_{Bt}}{\Lambda^{2}} \mathcal{O}_{\widetilde{B}}
    + \frac{f_{WWWt}}{\Lambda^{2}} \mathcal{O}_{\widetilde{W}WW}
        + \frac{f_{WWt}}{\Lambda^{2}} \mathcal{O}_{\widetilde{W}W} + \frac{f_{BBt}}{\Lambda^{2}} \mathcal{O}_{\widetilde{B}B} \nonumber\\
        &+& \frac{f_{DWt}}{\Lambda^{2}} \mathcal{O}_{D\widetilde{W}} + \frac{f_{BWt}}{\Lambda^{2}} \mathcal{O}_{B\widetilde{W}} \nonumber\\
    &+& \frac{f_{S0}}{\Lambda^4} \mathcal{O}_{S,0} + \frac{f_{S1}}{\Lambda^4} \mathcal{O}_{S,1} + \frac{f_{S2}}{\Lambda^4} \mathcal{O}_{S,2} \nonumber\\
    &+& \frac{f_{M0}}{\Lambda^4} \mathcal{O}_{M,0} + \frac{f_{M1}}{\Lambda^4} \mathcal{O}_{M,1}
    + \frac{f_{M2}}{\Lambda^4} \mathcal{O}_{M,2} + \frac{f_{M3}}{\Lambda^4} \mathcal{O}_{M,3} \nonumber\\
    &+& \frac{f_{M4}}{\Lambda^4} \mathcal{O}_{M,4} + \frac{f_{M5}}{\Lambda^4} \mathcal{O}_{M,5}+ \frac{f_{M5hc}}{\Lambda^4} \mathcal{O}_{M,5hc}
    + \frac{f_{M6}}{\Lambda^4} \mathcal{O}_{M,6} + \frac{f_{M7}}{\Lambda^4} \mathcal{O}_{M,7} \nonumber\\
    &+& \frac{f_{T0}}{\Lambda^4} \mathcal{O}_{T,0} + \frac{f_{T1}}{\Lambda^4} \mathcal{O}_{T,1}
    + \frac{f_{T2}}{\Lambda^4} \mathcal{O}_{T,2} + \frac{f_{T3}}{\Lambda^4} \mathcal{O}_{T,3} \nonumber\\
    &+& \frac{f_{T4}}{\Lambda^4} \mathcal{O}_{T,4} + \frac{f_{T5}}{\Lambda^4} \mathcal{O}_{T,5} +\frac{f_{T6}}{\Lambda^4} \mathcal{O}_{T,6} + \frac{f_{T7}}{\Lambda^4} \mathcal{O}_{T,7} \,
 +\frac{f_{T8}}{\Lambda^4} \mathcal{O}_{T,8} + \frac{f_{T9}}{\Lambda^4} \mathcal{O}_{T,9} \,.
  \label{leff}
\end{eqnarray}
}

\subsection{Conventions for anomalous gauge boson coupling parameters in \textsc{Vbfnlo}}
\label{ancpl_conversion}

As mentioned in Appendix~\ref{listofoperators}, our definition of the field strengths is
slightly different than the one of \cite{Eboli:2006wa}. For comparisons with
programs that use the definition from \cite{Eboli:2006wa} (for example MadGraph5~\cite{Alwall:2011uj} with
the UFO file written by \'Eboli et 
al.\footnote{available from \tt \url{https://feynrules.irmp.ucl.ac.be/wiki/AnomalousGaugeCoupling}}), 
you should rescale the coupling strengths by
\begin{eqnarray}
 f_{S,0,1}   &=&                            f_{S,0,1}^\textrm{\'Eboli}  \\
 f_{M,0,1}   &=& - \frac{1}{g^2}      \cdot f_{M,0,1}^\textrm{\'Eboli}  \\
 f_{M,2,3}   &=& - \frac{4}{g'^2}     \cdot f_{M,2,3}^\textrm{\'Eboli}  \\
 f_{M,4,5}   &=& - \frac{2}{g g'}     \cdot f_{M,4,5}^\textrm{\'Eboli}  \\
 f_{M,6,7}   &=& - \frac{1}{g^2}      \cdot f_{M,6,7}^\textrm{\'Eboli}  \\
 f_{T,0,1,2} &=&   \frac{1}{g^4}      \cdot f_{T,0,1,2}^\textrm{\'Eboli}\\
 f_{T,5,6,7} &=&   \frac{4}{g^2 g'^2} \cdot f_{T,5,6,7}^\textrm{\'Eboli}\\
 f_{T,8,9}   &=&   \frac{16}{g'^4}    \cdot f_{T,8,9}^\textrm{\'Eboli}
\end{eqnarray}
where $g$ and $g'$ are the SU(2) and U(1) gauge couplings.
The numerical values of $g, g$ used in the calculation are printed out at
the beginning of the \textsc{Vbfnlo} run.


\newpage

\section{Process list}
\label{app:proc_list}
The following is a complete list of all processes available in \textsc{Vbfnlo},
including any Beyond the Standard Model (BSM) effects that are implemented. 
{
\footnotesize
\setlength\LTleft{0pt plus \textwidth minus \textwidth}
\setlength\LTright{0pt plus \textwidth minus \textwidth}
\begin{longtable}{clccccccccc}
\textsc{ProcId} & \textsc{Process} & \rot{\textsc{BLHA}} & \rot{semi-leptonic decay} & \rot{VBF process} & \rot{anom.\ gauge couplings} & \rot{anom.\ Higgs couplings} & \rot{Two-Higgs model} & \rot{Kaluza-Klein model} & \rot{Spin-2 model} & \rot{MSSM} \\
&\\
\hline
\endhead
&\\*
\bf 100 & $p \overset{\mbox{\tiny{(--)}}}{p} \to H \, jj$ &\bsmoptionsBLHA{BHMF}\\*
\bf 101 & $p \overset{\mbox{\tiny{(--)}}}{p} \to H \, jj\to \gamma\gamma \, jj$ &\bsmoptionsBLHA{HMF}\\*
\bf 102 & $p \overset{\mbox{\tiny{(--)}}}{p} \to H \, jj\to \mu^+\mu^- \, jj$ &\bsmoptionsBLHA{HMF}\\*
\bf 103 & $p \overset{\mbox{\tiny{(--)}}}{p} \to H \, jj\to \tau^+\tau^- \, jj$ &\bsmoptionsBLHA{HMF}\\*
\bf 104 & $p \overset{\mbox{\tiny{(--)}}}{p} \to H \, jj\to b\bar{b} \, jj$ &\bsmoptionsBLHA{HMF}\\*
\bf 105 & $p \overset{\mbox{\tiny{(--)}}}{p} \to H \, jj\to W^{+}W^{-} \, jj\to \ell_{1}^+\nu_{\ell_{1}} \ell_{2}^- \bar{\nu}_{\ell_{2}} \,jj$ &\bsmoptionsBLHA{HMF}\\*
\bf 106 & $p \overset{\mbox{\tiny{(--)}}}{p} \to H \, jj\to ZZ \, jj\to \ell_{1}^+ \ell_{1}^- \ell_{2}^+ \ell_{2}^- \,jj$ &\bsmoptionsBLHA{HMF}\\*
\bf 107 & $p \overset{\mbox{\tiny{(--)}}}{p} \to H \, jj\to ZZ \, jj\to \ell_{1}^+ \ell_{1}^- \nu_{\ell_{2}}  \bar{\nu}_{\ell_{2}} \,jj$ &\bsmoptionsBLHA{HMF}\\*
\bf 108 & $p \overset{\mbox{\tiny{(--)}}}{p} \to H \, jj\to W^{+}W^{-} \, jj\to q\bar{q} \, \ell^- \bar{\nu}_{\ell} \,jj$ &\bsmoptionsBLHA{LHMF}\\*
\bf 109 & $p \overset{\mbox{\tiny{(--)}}}{p} \to H \, jj\to W^{+}W^{-} \, jj\to \ell^+\nu_{\ell} \, q\bar{q}  \,jj$ &\bsmoptionsBLHA{LHMF}\\*
\bf 1010 & $p \overset{\mbox{\tiny{(--)}}}{p} \to H \, jj\to ZZ \, jj\to q\bar{q} \, \ell^+ \ell^- \,jj$ &\bsmoptionsBLHA{LHMF}\\*
&\\*
\hline
&\\*
\bf 110 & $p \overset{\mbox{\tiny{(--)}}}{p} \to H \, jjj$ &\bsmoptionsBLHA{F}\\*
\bf 111 & $p \overset{\mbox{\tiny{(--)}}}{p} \to H \, jjj\to \gamma\gamma \, jjj$ &\bsmoptionsBLHA{F}\\*
\bf 112 & $p \overset{\mbox{\tiny{(--)}}}{p} \to H \, jjj\to \mu^+\mu^- \, jjj$ &\bsmoptionsBLHA{F}\\*
\bf 113 & $p \overset{\mbox{\tiny{(--)}}}{p} \to H \, jjj\to \tau^+\tau^- \, jjj$ &\bsmoptionsBLHA{F}\\*
\bf 114 & $p \overset{\mbox{\tiny{(--)}}}{p} \to H \, jjj\to b\bar{b} \, jjj$ &\bsmoptionsBLHA{F}\\*
\bf 115 & $p \overset{\mbox{\tiny{(--)}}}{p} \to H \, jjj\to W^+W^- \, jjj\to \ell_{1}^{+}\nu_{\ell_{1}} \ell_{2}^- \bar{\nu}_{\ell_{2}} \,jjj$ &\bsmoptionsBLHA{F}\\*
\bf 116 & $p \overset{\mbox{\tiny{(--)}}}{p} \to H \, jjj\to ZZ \, jjj\to \ell_{1}^+ \ell_{1}^- \ell_{2}^+ \ell_{2}^- \,jjj$ &\bsmoptionsBLHA{F}\\*
\bf 117 & $p \overset{\mbox{\tiny{(--)}}}{p} \to H \, jjj\to ZZ \, jjj\to \ell_{1}^+ \ell_{1}^- \nu_{\ell_{2}} \bar{\nu}_{\ell_{2}} \,jjj$ &\bsmoptionsBLHA{F}\\*
&\\*
\hline
&\\*
\bf 120 & $p \overset{\mbox{\tiny{(--)}}}{p} \to Z \, jj \to \ell^{+} \ell^{-} \, jj$ &\bsmoptionsBLHA{BVF}\\*
\bf 121 & $p \overset{\mbox{\tiny{(--)}}}{p} \to Z  \, jj\to \nu_\ell \bar{\nu}_\ell \, jj$ &\bsmoptionsBLHA{BVF}\\*
\bf 130 & $p \overset{\mbox{\tiny{(--)}}}{p} \to W^{+} \,  jj\to \ell^{+} \nu_\ell \, jj$ &\bsmoptionsBLHA{BVF}\\*
\bf 140 & $p \overset{\mbox{\tiny{(--)}}}{p} \to W^{-} \, jj\to \ell^{-} \bar{\nu}_\ell  \, jj$ &\bsmoptionsBLHA{BVF}\\*
\bf 150 & $p \overset{\mbox{\tiny{(--)}}}{p} \to \gamma \, jj$ &\bsmoptionsBLHA{BVF}\\*
&\\*
\hline
&\\*
\bf 191 & $p \overset{\mbox{\tiny{(--)}}}{p} \to S_{2}  \, jj\to \gamma \gamma \, jj$ &\bsmoptionsBLHA{SF}\\*
\bf 195 & $p \overset{\mbox{\tiny{(--)}}}{p} \to S_{2}  \, jj\to W^{+}W^{-} \, jj\to \ell_{1}^+\nu_{\ell_{1}} \ell_{2}^- \bar{\nu}_{\ell_{2}} \,jj$ &\bsmoptionsBLHA{SF}\\*
\bf 196 & $p \overset{\mbox{\tiny{(--)}}}{p} \to S_{2}  \, jj\to ZZ \, jj\to \ell_{1}^+ \ell_{1}^- \ell_{2}^+ \ell_{2}^- \,jj$ &\bsmoptionsBLHA{SF}\\*
\bf 197 & $p \overset{\mbox{\tiny{(--)}}}{p} \to S_{2}  \, jj\to  ZZ \, jj\to \ell_{1}^+ \ell_{1}^- \nu_{\ell_{2}} \bar{\nu}_{\ell_{2}} \,jj$ &\bsmoptionsBLHA{SF}\\*
&\\*
\hline
&\\*
\bf 160 & $p \overset{\mbox{\tiny{(--)}}}{p} \to HH \,  jj$ &\bsmoptionsBLHA{F}\\*
\bf 161 & $p \overset{\mbox{\tiny{(--)}}}{p} \to HH \,  jj \to b\bar{b}\tau^+\tau^- \, jj$ &\bsmoptionsBLHA{F}\\*
\bf 162 & $p \overset{\mbox{\tiny{(--)}}}{p} \to HH \,  jj \to b\bar{b}\gamma\gamma \, jj$ &\bsmoptionsBLHA{F}\\*
&\\*
\hline
&\\*
\bf 2100 & $p \overset{\mbox{\tiny{(--)}}}{p} \to H \gamma \, jj$ &\bsmoptionsBLHA{F}\\*
\bf 2101 & $p \overset{\mbox{\tiny{(--)}}}{p} \to H \gamma \, jj\to \gamma\gamma \gamma \, jj$ &\bsmoptionsBLHA{F}\\*
\bf 2102 & $p \overset{\mbox{\tiny{(--)}}}{p} \to H \gamma \, jj\to \mu^+\mu^- \gamma \, jj$ &\bsmoptionsBLHA{F}\\*
\bf 2103 & $p \overset{\mbox{\tiny{(--)}}}{p} \to H \gamma \, jj\to \tau^+\tau^- \gamma \, jj$ &\bsmoptionsBLHA{F}\\*
\bf 2104 & $p \overset{\mbox{\tiny{(--)}}}{p} \to H \gamma \, jj\to b\bar{b} \gamma \, jj$ &\bsmoptionsBLHA{F}\\*
\bf 2105 & $p \overset{\mbox{\tiny{(--)}}}{p} \to H \gamma \, jj\to W^+W^- \gamma \, jj\to \ell_{1}^+\nu_{\ell_{1}} \ell_{2}^- \bar{\nu}_{\ell_{2}} \gamma \,jj$ &\bsmoptionsBLHA{F}\\*
\bf 2106 & $p \overset{\mbox{\tiny{(--)}}}{p} \to H \gamma \, jj\to ZZ \gamma \, jj\to \ell_{1}^+ \ell_{1}^- \ell_{2}^+ \ell_{2}^- \gamma \,jj$ &\bsmoptionsBLHA{F}\\*
\bf 2107 & $p \overset{\mbox{\tiny{(--)}}}{p} \to H \gamma \, jj\to ZZ \gamma \, jj\to \ell_{1}^+ \ell_{1}^- \nu_{\ell_{2}}  \bar{\nu}_{\ell_{2}} \gamma \,jj$ &\bsmoptionsBLHA{F}\\*
&\\*
\hline
&\\*
\bf 200 & $p \overset{\mbox{\tiny{(--)}}}{p} \to W^{+}W^{-} \, jj \to \ell_{1}^{+} \nu_{\ell_{1}} \ell_{2}^{-} \bar{\nu}_{\ell_{2}} \, jj$ &\bsmoptionsBLHA{BVTKSF}\\*
\bf 201 & $p \overset{\mbox{\tiny{(--)}}}{p} \to W^{+}W^{-} \, jj \to q \bar{q} \, \ell^{-} \bar{\nu}_{\ell} \, jj$ &\bsmoptionsBLHA{VTLF}\\*
\bf 202 & $p \overset{\mbox{\tiny{(--)}}}{p} \to W^{+}W^{-}  \, jj\to \ell^{+} \nu_\ell \, q \bar{q} \, jj$ &\bsmoptionsBLHA{VTLF}\\*
\bf 210 & $p \overset{\mbox{\tiny{(--)}}}{p} \to ZZ  \, jj\to \ell_{1}^{+} \ell_{1}^{-} \ell_{2}^{+} \ell_{2}^{-} \, jj$ &\bsmoptionsBLHA{BVTKSF}\\*
\bf 211 & $p \overset{\mbox{\tiny{(--)}}}{p} \to ZZ  \, jj\to \ell_{1}^{+} \ell_{1}^{-} \nu_{\ell_{2}} \bar{\nu}_{\ell_{2}} \, jj$ &\bsmoptionsBLHA{BVTKSF}\\*
\bf 212 & $p \overset{\mbox{\tiny{(--)}}}{p} \to ZZ  \, jj\to q \bar{q} \, \ell^{+} \ell^{-} \, jj$ &\bsmoptionsBLHA{VTLF}\\*
\bf 220 & $p \overset{\mbox{\tiny{(--)}}}{p} \to W^{+}Z \,  jj\to \ell_{1}^{+} \nu_{\ell_{1}} \ell_{2}^{+} \ell_{2}^{-} \, jj$ &\bsmoptionsBLHA{BVTKSF}\\*
\bf 221 & $p \overset{\mbox{\tiny{(--)}}}{p} \to W^{+}Z \,  jj\to q \bar{q} \, \ell^{+} \ell^{-} \, jj$ &\bsmoptionsBLHA{VTLF}\\*
\bf 222 & $p \overset{\mbox{\tiny{(--)}}}{p} \to W^{+}Z \,  jj\to \ell^{+} \nu_{\ell} \, q \bar{q} \, jj$ &\bsmoptionsBLHA{VTLF}\\*
\bf 230 & $p \overset{\mbox{\tiny{(--)}}}{p} \to W^{-}Z \, jj\to \ell_{1}^{-} \bar{\nu}_{\ell _{1}} \ell_{2}^{+} \ell_{2}^{-} \, jj$ &\bsmoptionsBLHA{BVTKSF}\\*
\bf 231 & $p \overset{\mbox{\tiny{(--)}}}{p} \to W^{-}Z \, jj\to q \bar{q} \, \ell^{+} \ell^{-} \, jj$ &\bsmoptionsBLHA{VTLF}\\*
\bf 232 & $p \overset{\mbox{\tiny{(--)}}}{p} \to W^{-}Z \, jj\to \ell^{-} \bar{\nu}_{\ell} \, q \bar{q} \, jj$ &\bsmoptionsBLHA{VTLF}\\*
\bf 240 & $p \overset{\mbox{\tiny{(--)}}}{p} \to \gamma\gamma \, jj$ &\bsmoptionsBLHA{BVF}\\*
\bf 250 & $p \overset{\mbox{\tiny{(--)}}}{p} \to W^{+}W^{+} \,  jj\to \ell_{1}^{+} \nu_{\ell_{1}} \ell_{2}^{+} \nu_{\ell_{2}} \, jj$ &\bsmoptionsBLHA{BVTF}\\*
\bf 251 & $p \overset{\mbox{\tiny{(--)}}}{p} \to W^{+}W^{+} \,  jj\to q \bar{q} \, \ell^{+} \nu_{\ell} \, jj$ &\bsmoptionsBLHA{VTLF}\\*
\bf 260 & $p \overset{\mbox{\tiny{(--)}}}{p} \to W^{-}W^{-} \,  jj\to \ell_{1}^{-} \bar{\nu}_{\ell_{1}} \ell_{2}^{-} \bar{\nu}_{\ell_{2}} \, jj$ &\bsmoptionsBLHA{BVTF}\\*
\bf 261 & $p \overset{\mbox{\tiny{(--)}}}{p} \to W^{-}W^{-} \,  jj\to q \bar{q} \, \ell^{-} \bar{\nu}_{\ell} \, jj$ &\bsmoptionsBLHA{VTLF}\\*
\bf 270 & $p \overset{\mbox{\tiny{(--)}}}{p} \to W^{+}\gamma \, jj\to \ell^{+} \nu_{\ell} \gamma \, jj$ &\bsmoptionsBLHA{BVF}\\*
\bf 280 & $p \overset{\mbox{\tiny{(--)}}}{p} \to W^{-}\gamma \, jj\to \ell^{-} \bar{\nu}_{\ell} \gamma \, jj$ &\bsmoptionsBLHA{BVF}\\*
\bf 290 & $p \overset{\mbox{\tiny{(--)}}}{p} \to Z\gamma \, jj\to \ell^{+} \ell^{-} \gamma \, jj$ &\bsmoptionsBLHA{BVF}\\*
\bf 291 & $p \overset{\mbox{\tiny{(--)}}}{p} \to Z\gamma \, jj\to \nu_{\ell}\bar{\nu}_{\ell} \gamma \, jj$ &\bsmoptionsBLHA{BVF}\\*
&\\*
\hline
&\\*
\bf 3120 & $p \overset{\mbox{\tiny{(--)}}}{p} \to Z \,  jj\to \ell^{+} \ell^{-} \, jj$  &\bsmoptionsBLHA{}\\*
\bf 3121 & $p \overset{\mbox{\tiny{(--)}}}{p} \to Z \,  jj\to \nu_{\ell} \bar{\nu}_{\ell} \, jj$    &\bsmoptionsBLHA{}\\*
\bf 3130 & $p \overset{\mbox{\tiny{(--)}}}{p} \to W^{+} \,  jj\to \ell^{+} \nu_\ell \, jj$ &\bsmoptionsBLHA{}\\*
\bf 3140 & $p \overset{\mbox{\tiny{(--)}}}{p} \to W^{-} \, jj\to \ell^{-} \bar{\nu}_\ell  \, jj$ &\bsmoptionsBLHA{}\\*
&\\*
\hline
&\\*
\bf 3210 & $p \overset{\mbox{\tiny{(--)}}}{p} \to ZZ \,  jj\to \ell_{1}^{+} \ell_{1}^{-} \ell_{2}^{+} \ell_{2}^{-} \, jj$   &\bsmoptionsBLHA{}\\*
\bf 3211 & $p \overset{\mbox{\tiny{(--)}}}{p} \to ZZ \,  jj\to \ell_{1}^{+} \ell_{1}^{-} \nu_{\ell_{2}} \bar{\nu}_{\ell_{2}} \, jj$  &\bsmoptionsBLHA{}\\*
\bf 3220 & $p \overset{\mbox{\tiny{(--)}}}{p} \to W^{+}Z \,  jj\to \ell_{1}^{+} \nu_{\ell_{1}} \ell_{2}^{+} \ell_{2}^{-} \, jj$ &\bsmoptionsBLHA{}\\*
\bf 3230 & $p \overset{\mbox{\tiny{(--)}}}{p} \to W^{-}Z \, jj\to \ell_{1}^{-} \bar{\nu}_{\ell _{1}} \ell_{2}^{+} \ell_{2}^{-} \, jj$ &\bsmoptionsBLHA{}\\*
\bf 3240 & $p \overset{\mbox{\tiny{(--)}}}{p} \to \gamma\gamma \,  jj$  &\bsmoptionsBLHA{}\\*
\bf 3250 & $p \overset{\mbox{\tiny{(--)}}}{p} \to W^{+}W^{+} \,  jj\to \ell_{1}^{+} \nu_{\ell_{1}} \ell_{2}^{+} \nu_{\ell_{2}} \, jj$ &\bsmoptionsBLHA{}\\*
\bf 3260 & $p \overset{\mbox{\tiny{(--)}}}{p} \to W^{-}W^{-} \,  jj\to \ell_{1}^{-} \bar{\nu}_{\ell_{1}} \ell_{2}^{-} \bar{\nu}_{\ell_{2}} \, jj$ &\bsmoptionsBLHA{}\\*
\bf 3270 & $p \overset{\mbox{\tiny{(--)}}}{p} \to W^{+}\gamma \,  jj\to \ell^{+} \nu_{\ell} \gamma \, jj$ &\bsmoptionsBLHA{}\\*
\bf 3280 & $p \overset{\mbox{\tiny{(--)}}}{p} \to W^{-}\gamma \, jj\to \ell^{-} \bar{\nu}_{\ell} \gamma \, jj$ &\bsmoptionsBLHA{}\\*
\bf 3290 & $p \overset{\mbox{\tiny{(--)}}}{p} \to Z\gamma \,  jj\to \ell^{+} \ell^{-} \gamma \, jj$  &\bsmoptionsBLHA{}\\*
\bf 3291 & $p \overset{\mbox{\tiny{(--)}}}{p} \to Z\gamma \,  jj\to \nu_{\ell} \bar{\nu}_{\ell} \gamma\, jj$    &\bsmoptionsBLHA{}\\*
&\\*
\hline
&\\*
\bf 1330 & $p \overset{\mbox{\tiny{(--)}}}{p} \to W^+ \to \ell^+\nu_{\ell} $ &\bsmoptionsBLHA{}\\*
\bf 1340 & $p \overset{\mbox{\tiny{(--)}}}{p} \to W^- \to \ell^- \bar{\nu}_{\ell} $ &\bsmoptionsBLHA{}\\*
\bf 1630 & $p \overset{\mbox{\tiny{(--)}}}{p} \to W^+ \, j \to \ell^+\nu_{\ell} \, j $ &\bsmoptionsBLHA{}\\*
\bf 1640 & $p \overset{\mbox{\tiny{(--)}}}{p} \to W^- \, j\to \ell^- \bar{\nu}_{\ell} \, j $ &\bsmoptionsBLHA{}\\*
&\\*
\hline
&\\*
\bf 300 & $p \overset{\mbox{\tiny{(--)}}}{p} \to W^{+}W^{-} \to \ell_{1}^{+} \nu_{\ell_{1}} \ell_{2}^{-}\bar{\nu}_{\ell_{2}} $ &\bsmoptionsBLHA{BHV}\\*
\bf 301 & $p \overset{\mbox{\tiny{(--)}}}{p} \to W^{+}W^{-} \to q \bar{q} \, \ell^{-}\bar{\nu}_{\ell} $ &\bsmoptionsBLHA{HVL}\\*
\bf 302 & $p \overset{\mbox{\tiny{(--)}}}{p} \to W^{+}W^{-} \to \ell^{+} \nu_{\ell} \, q \bar{q} $ &\bsmoptionsBLHA{HVL}\\*
\bf 310 & $p \overset{\mbox{\tiny{(--)}}}{p} \to W^{+}Z \to  \ell_{1}^{+} \nu_{\ell_1}  \ell_{2}^{+} \ell_{2}^{-} $ &\bsmoptionsBLHA{BV}\\*
\bf 312 & $p \overset{\mbox{\tiny{(--)}}}{p} \to W^{+}Z \to  q \bar{q} \, \ell^{+} \ell^{-} $ &\bsmoptionsBLHA{VL}\\*
\bf 313 & $p \overset{\mbox{\tiny{(--)}}}{p} \to W^{+}Z \to  \ell^{+} \nu_{\ell} \, q \bar{q} $ &\bsmoptionsBLHA{VL}\\*
\bf 320 & $p \overset{\mbox{\tiny{(--)}}}{p} \to W^{-}Z \to \ell_{1}^{-} \bar{\nu}_{\ell_{1}}  \ell_{2}^{+} \ell_{2}^{-} $ &\bsmoptionsBLHA{BV}\\*
\bf 322 & $p \overset{\mbox{\tiny{(--)}}}{p} \to W^{-}Z \to q \bar{q} \, \ell^{+} \ell^{-} $ &\bsmoptionsBLHA{VL}\\*
\bf 323 & $p \overset{\mbox{\tiny{(--)}}}{p} \to W^{-}Z \to \ell^{-} \bar{\nu}_{\ell} \, q \bar{q} $ &\bsmoptionsBLHA{VL}\\*
\bf 330 & $p \overset{\mbox{\tiny{(--)}}}{p} \to ZZ \to \ell_{1}^{-} \ell_{1}^{+}  \ell_{2}^{-} \ell_{2}^{+} $ &\bsmoptionsBLHA{BH}\\*
\bf 331 & $p \overset{\mbox{\tiny{(--)}}}{p} \to ZZ \to q \bar{q} \, \ell^{-} \ell^{+} $ &\bsmoptionsBLHA{HL}\\*
\bf 340 & $p \overset{\mbox{\tiny{(--)}}}{p} \to W^{+}\gamma \to \ell_{1}^{+} \nu_{\ell_1} \gamma $ &\bsmoptionsBLHA{BV}\\*
\bf 350 & $p \overset{\mbox{\tiny{(--)}}}{p} \to W^{-}\gamma \to \ell_{1}^{-} \bar{\nu}_{\ell_{1}} \gamma $ &\bsmoptionsBLHA{BV}\\*
\bf 360 & $p \overset{\mbox{\tiny{(--)}}}{p} \to Z\gamma \to \ell_{1}^{-} \ell_{1}^{+}  \gamma $ &\bsmoptionsBLHA{BH}\\*
\bf 370 & $p \overset{\mbox{\tiny{(--)}}}{p} \to \gamma\gamma $ &\bsmoptionsBLHA{BH}\\*
&\\*
\hline
&\\*
\bf 1300 & $p \overset{\mbox{\tiny{(--)}}}{p} \to W^+ H \to \ell^+\nu_{\ell} H $ &\bsmoptionsBLHA{V}\\*
\bf 1301 & $p \overset{\mbox{\tiny{(--)}}}{p} \to W^+ H \to \ell^+\nu_{\ell} \gamma\gamma $ &\bsmoptionsBLHA{V}\\*
\bf 1302 & $p \overset{\mbox{\tiny{(--)}}}{p} \to W^+ H \to \ell^+\nu_{\ell} \mu^+\mu^- $ &\bsmoptionsBLHA{V}\\*
\bf 1303 & $p \overset{\mbox{\tiny{(--)}}}{p} \to W^+ H \to \ell^+\nu_{\ell} \tau^+\tau^- $ &\bsmoptionsBLHA{V}\\*
\bf 1304 & $p \overset{\mbox{\tiny{(--)}}}{p} \to W^+ H \to \ell^+\nu_{\ell} b\bar{b} $ &\bsmoptionsBLHA{V}\\*
\bf 1305 & $p \overset{\mbox{\tiny{(--)}}}{p} \to W^+ H \to W^+ W^{+}W^{-} \to \ell_{1}^+\nu_{\ell_{1}} \ell_{2}^+\nu_{\ell_{2}} \ell_{3}^- \bar{\nu}_{\ell_{3}}$ &\bsmoptionsBLHA{V}\\*
\bf 1306 & $p \overset{\mbox{\tiny{(--)}}}{p} \to W^+ H \to W^+ ZZ \to \ell_{1}^+\nu_{\ell_{1}} \ell_{2}^+ \ell_{2}^- \ell_{3}^+ \ell_{3}^-$ &\bsmoptionsBLHA{V}\\*
\bf 1307 & $p \overset{\mbox{\tiny{(--)}}}{p} \to W^+ H \to W^+ ZZ \to \ell_{1}^+\nu_{\ell_{1}} \ell_{2}^+ \ell_{2}^- \nu_{\ell_{3}}  \bar{\nu}_{\ell_{3}}$ &\bsmoptionsBLHA{V}\\*
&\\*
\hline
&\\*
\bf 1310 & $p \overset{\mbox{\tiny{(--)}}}{p} \to W^- H \to \ell^- \bar{\nu}_{\ell} H $ &\bsmoptionsBLHA{V}\\*
\bf 1311 & $p \overset{\mbox{\tiny{(--)}}}{p} \to W^- H \to \ell^- \bar{\nu}_{\ell} \gamma\gamma $ &\bsmoptionsBLHA{V}\\*
\bf 1312 & $p \overset{\mbox{\tiny{(--)}}}{p} \to W^- H \to \ell^- \bar{\nu}_{\ell} \mu^+\mu^- $ &\bsmoptionsBLHA{V}\\*
\bf 1313 & $p \overset{\mbox{\tiny{(--)}}}{p} \to W^- H \to \ell^- \bar{\nu}_{\ell} \tau^+\tau^- $ &\bsmoptionsBLHA{V}\\*
\bf 1314 & $p \overset{\mbox{\tiny{(--)}}}{p} \to W^- H \to \ell^- \bar{\nu}_{\ell} b\bar{b} $ &\bsmoptionsBLHA{V}\\*
\bf 1315 & $p \overset{\mbox{\tiny{(--)}}}{p} \to W^- H \to W^- W^{+}W^{-} \to \ell_{1}^-\bar{\nu}_{\ell_{1}} \ell_{2}^+\nu_{\ell_{2}} \ell_{3}^- \bar{\nu}_{\ell_{3}}$ &\bsmoptionsBLHA{V}\\*
\bf 1316 & $p \overset{\mbox{\tiny{(--)}}}{p} \to W^- H \to W^- ZZ \to \ell_{1}^-\bar{\nu}_{\ell_{1}} \ell_{2}^+ \ell_{2}^- \ell_{3}^+ \ell_{3}^-$ &\bsmoptionsBLHA{V}\\*
\bf 1317 & $p \overset{\mbox{\tiny{(--)}}}{p} \to W^- H \to W^- ZZ \to \ell_{1}^-\bar{\nu}_{\ell_{1}} \ell_{2}^+ \ell_{2}^- \nu_{\ell_{3}}  \bar{\nu}_{\ell_{3}}$ &\bsmoptionsBLHA{V}\\*
&\\*
\hline
&\\*
\bf 600 & $p \overset{\mbox{\tiny{(--)}}}{p}  \to W^{+} W^{-} j \to \ell_{1}^{+}\nu_{\ell_1} \ell_{2}^{-} \bar{\nu}_{\ell_2} j $ & \bsmoptionsBLHA{H}\\*
\bf 601 & $p \overset{\mbox{\tiny{(--)}}}{p}  \to W^{+} W^{-} j \to q\bar{q} \ell^{-} \bar{\nu}_{\ell} j $ & \bsmoptionsBLHA{L}\\*
\bf 602 & $p \overset{\mbox{\tiny{(--)}}}{p}  \to W^{+} W^{-} j \to \ell^{+}\nu_{\ell} q\bar{q} j $ & \bsmoptionsBLHA{L}\\*
\bf 610 & $p \overset{\mbox{\tiny{(--)}}}{p}  \to W^{-} \gamma j \to \ell^{-} \bar \nu_{\ell} \gamma j $ &\bsmoptionsBLHA{V}\\*
\bf 620 & $p \overset{\mbox{\tiny{(--)}}}{p}  \to W^{+} \gamma j  \to \ell^{+} \nu_{\ell} \gamma j $ &\bsmoptionsBLHA{V}\\*
\bf 630 & $p \overset{\mbox{\tiny{(--)}}}{p}  \to W^{-} Z j \to \ell_{1}^{-} \bar \nu_{\ell_1} \ell_{2}^{-} \ell_{2}^{+} j$ &\bsmoptionsBLHA{V}\\*
\bf 640 & $p \overset{\mbox{\tiny{(--)}}}{p}  \to W^{+} Z j \to  \ell_{1}^{+}\nu_{\ell_1} \ell_{2}^{-} \ell_{2}^{+}j $ &\bsmoptionsBLHA{V}\\*
\bf 650 & $p \overset{\mbox{\tiny{(--)}}}{p}  \to Z Z j \to \ell_{1}^{+} \ell_{1}^{-} \ell_{2}^{+} \ell_{2}^{-} j $ & \bsmoptionsBLHA{H}\\*
&\\*
\hline
&\\*
\bf 1600 & $p \overset{\mbox{\tiny{(--)}}}{p} \to W^+ H \, j \to \ell^+\nu_{\ell} H \, j $ &\bsmoptionsBLHA{V}\\*
\bf 1601 & $p \overset{\mbox{\tiny{(--)}}}{p} \to W^+ H \, j \to \ell^+\nu_{\ell} \gamma\gamma \, j $ &\bsmoptionsBLHA{V}\\*
\bf 1602 & $p \overset{\mbox{\tiny{(--)}}}{p} \to W^+ H \, j \to \ell^+\nu_{\ell} \mu^+\mu^- \, j $ &\bsmoptionsBLHA{V}\\*
\bf 1603 & $p \overset{\mbox{\tiny{(--)}}}{p} \to W^+ H \, j \to \ell^+\nu_{\ell} \tau^+\tau^- \, j $ &\bsmoptionsBLHA{V}\\*
\bf 1604 & $p \overset{\mbox{\tiny{(--)}}}{p} \to W^+ H \, j \to \ell^+\nu_{\ell} b\bar{b} \, j $ &\bsmoptionsBLHA{V}\\*
\bf 1605 & $p \overset{\mbox{\tiny{(--)}}}{p} \to W^+ H \, j \to W^+ W^{+}W^{-} \, j \to \ell_{1}^+\nu_{\ell_{1}} \ell_{2}^+\nu_{\ell_{2}} \ell_{3}^- \bar{\nu}_{\ell_{3}}\, j $ &\bsmoptionsBLHA{V}\\*
\bf 1606 & $p \overset{\mbox{\tiny{(--)}}}{p} \to W^+ H \, j \to W^+ ZZ \, j \to \ell_{1}^+\nu_{\ell_{1}} \ell_{2}^+ \ell_{2}^- \ell_{3}^+ \ell_{3}^-\, j $ &\bsmoptionsBLHA{V}\\*
\bf 1607 & $p \overset{\mbox{\tiny{(--)}}}{p} \to W^+ H \, j \to W^+ ZZ \, j \to \ell_{1}^+\nu_{\ell_{1}} \ell_{2}^+ \ell_{2}^- \nu_{\ell_{3}}  \bar{\nu}_{\ell_{3}}\, j $ &\bsmoptionsBLHA{V}\\*
&\\*
\hline
&\\*
\bf 1610 & $p \overset{\mbox{\tiny{(--)}}}{p} \to W^- H \, j \to \ell^- \bar{\nu}_{\ell} H \, j $ &\bsmoptionsBLHA{V}\\*
\bf 1611 & $p \overset{\mbox{\tiny{(--)}}}{p} \to W^- H \, j \to \ell^- \bar{\nu}_{\ell} \gamma\gamma \, j $ &\bsmoptionsBLHA{V}\\*
\bf 1612 & $p \overset{\mbox{\tiny{(--)}}}{p} \to W^- H \, j \to \ell^- \bar{\nu}_{\ell} \mu^+\mu^- \, j $ &\bsmoptionsBLHA{V}\\*
\bf 1613 & $p \overset{\mbox{\tiny{(--)}}}{p} \to W^- H \, j \to \ell^- \bar{\nu}_{\ell} \tau^+\tau^- \, j $ &\bsmoptionsBLHA{V}\\*
\bf 1614 & $p \overset{\mbox{\tiny{(--)}}}{p} \to W^- H \, j \to \ell^- \bar{\nu}_{\ell} b\bar{b} \, j $ &\bsmoptionsBLHA{V}\\*
\bf 1615 & $p \overset{\mbox{\tiny{(--)}}}{p} \to W^- H \, j \to W^- W^{+}W^{-} \, j \to \ell_{1}^-\bar{\nu}_{\ell_{1}} \ell_{2}^+\nu_{\ell_{2}} \ell_{3}^- \bar{\nu}_{\ell_{3}}\, j $ &\bsmoptionsBLHA{V}\\*
\bf 1616 & $p \overset{\mbox{\tiny{(--)}}}{p} \to W^- H \, j \to W^- ZZ \, j \to \ell_{1}^-\bar{\nu}_{\ell_{1}} \ell_{2}^+ \ell_{2}^- \ell_{3}^+ \ell_{3}^-\, j $ &\bsmoptionsBLHA{V}\\*
\bf 1617 & $p \overset{\mbox{\tiny{(--)}}}{p} \to W^- H \, j \to W^- ZZ \, j \to \ell_{1}^-\bar{\nu}_{\ell_{1}} \ell_{2}^+ \ell_{2}^- \nu_{\ell_{3}}  \bar{\nu}_{\ell_{3}}\, j $ &\bsmoptionsBLHA{V}\\*
&\\*
\hline
&\\*
\bf 400 & $p \overset{\mbox{\tiny{(--)}}}{p} \to W^{+}W^{-}Z \to \ell_{1}^{+}\nu_{\ell_{1}} \ell_{2}^{-} \bar{\nu}_{\ell_{2}} \ell_{3}^{+} \ell_{3}^{-} $ &\bsmoptionsBLHA{BVK}\\*
\bf 401 & $p \overset{\mbox{\tiny{(--)}}}{p} \to W^{+}W^{-}Z \to q \bar{q} \, \ell_{1}^{-} \bar{\nu}_{\ell_{1}} \ell_{2}^{+} \ell_{2}^{-} $ &\bsmoptionsBLHA{LV}\\*
\bf 402 & $p \overset{\mbox{\tiny{(--)}}}{p} \to W^{+}W^{-}Z \to \ell_{1}^{+}\nu_{\ell_{1}} \, q \bar{q} \, \ell_{2}^{+} \ell_{2}^{-} $ &\bsmoptionsBLHA{LV}\\*
\bf 403 & $p \overset{\mbox{\tiny{(--)}}}{p} \to W^{+}W^{-}Z \to \ell_{1}^{+}\nu_{\ell_{1}} \ell_{2}^{-} \bar{\nu}_{\ell_{2}} \, q \bar{q} $ &\bsmoptionsBLHA{LV}\\*
\bf 410 & $p \overset{\mbox{\tiny{(--)}}}{p} \to ZZW^{+} \to  \ell_{1}^{+} \ell_{1}^{-}  \ell_{2}^{+} \ell_{2}^{-} \ell_{3}^{+} \nu_{\ell_{3}} $ &\bsmoptionsBLHA{BVK}\\*
\bf 411 & $p \overset{\mbox{\tiny{(--)}}}{p} \to ZZW^{+} \to  \ell_{1}^{+} \ell_{1}^{-}  \ell_{2}^{+} \ell_{2}^{-} \, q \bar{q} $ &\bsmoptionsBLHA{LV}\\*
\bf 412 & $p \overset{\mbox{\tiny{(--)}}}{p} \to ZZW^{+} \to  q \bar{q} \, \ell_{1}^{+} \ell_{1}^{-} \ell_{2}^{+} \nu_{\ell_{2}} $ &\bsmoptionsBLHA{LV}\\*
\bf 420 & $p \overset{\mbox{\tiny{(--)}}}{p} \to ZZW^{-} \to \ell_{1}^{+} \ell_{1}^{-}  \ell_{2}^{+} \ell_{2}^{-} \ell_{3}^{-}  \bar{\nu}_{\ell_{3}}$ &\bsmoptionsBLHA{BVK}\\*
\bf 421 & $p \overset{\mbox{\tiny{(--)}}}{p} \to ZZW^{-} \to \ell_{1}^{+} \ell_{1}^{-}  \ell_{2}^{+} \ell_{2}^{-} \, q \bar{q} $ &\bsmoptionsBLHA{LV}\\*
\bf 422 & $p \overset{\mbox{\tiny{(--)}}}{p} \to ZZW^{-} \to q \bar{q} \, \ell_{1}^{+} \ell_{1}^{-} \ell_{2}^{-}  \bar{\nu}_{\ell_{2}}$ &\bsmoptionsBLHA{LV}\\*
\bf 430 & $p \overset{\mbox{\tiny{(--)}}}{p} \to W^{+}W^{-}W^{+} \to \ell_{1}^{+}\nu_{\ell_1} \ell_{2}^{-} \bar{\nu}_{\ell_2} \ell_{3}^{+}\nu_{\ell_{3}}$ &\bsmoptionsBLHA{BVK}\\*
\bf 431 & $p \overset{\mbox{\tiny{(--)}}}{p} \to W^{+}W^{-}W^{+} \to  q \bar{q} \, \ell_{1}^{-} \bar{\nu}_{\ell_1} \ell_{2}^{+}\nu_{\ell_{2}}$ &\bsmoptionsBLHA{LV}\\*
\bf 432 & $p \overset{\mbox{\tiny{(--)}}}{p} \to W^{+}W^{-}W^{+} \to \ell_{1}^{+}\nu_{\ell_1} \, q \bar{q} \, \ell_{2}^{+}\nu_{\ell_{2}}$ &\bsmoptionsBLHA{LV}\\*
\bf 440 & $p \overset{\mbox{\tiny{(--)}}}{p} \to W^{-}W^{+}W^{-} \to \ell_{1}^{-} \bar{\nu}_{\ell_1}\ell_{2}^{+}\nu_{\ell_2} \ell_{3}^{-} \bar{\nu}_{\ell_{3}} $ &\bsmoptionsBLHA{BVK}\\*
\bf 441 & $p \overset{\mbox{\tiny{(--)}}}{p} \to W^{-}W^{+}W^{-} \to \ell_{1}^{-} \bar{\nu}_{\ell_1} \, q \bar{q} \, \ell_{2}^{-} \bar{\nu}_{\ell_{2}} $ &\bsmoptionsBLHA{LV}\\*
\bf 442 & $p \overset{\mbox{\tiny{(--)}}}{p} \to W^{-}W^{+}W^{-} \to  q \bar{q} \, \ell_{1}^{+}\nu_{\ell_1} \ell_{2}^{-} \bar{\nu}_{\ell_{2}} $ &\bsmoptionsBLHA{LV}\\*
\bf 450 & $p \overset{\mbox{\tiny{(--)}}}{p} \to ZZZ \to \ell_{1}^{-} \ell_{1}^{+} \ell_{2}^{-} \ell_{2}^{+} \ell_{3}^{-} \ell_{3}^{+} $ &\bsmoptionsBLHA{BV}\\*
\bf 451 & $p \overset{\mbox{\tiny{(--)}}}{p} \to ZZZ \to  q \bar{q} \, \ell_{1}^{-} \ell_{1}^{+} \ell_{2}^{-} \ell_{2}^{+}  $ &\bsmoptionsBLHA{LV}\\*
&\\*
\hline
&\\*
\bf 460 & $p \overset{\mbox{\tiny{(--)}}}{p} \to W^{-}W^{+} \gamma \to \ell_{1}^{-} \bar{\nu}_{\ell_1} \ell_{2}^{+}\nu_{\ell_2} \gamma$ &\bsmoptionsBLHA{BV}\\*
\bf 461 & $p \overset{\mbox{\tiny{(--)}}}{p} \to W^{+}W^{-} \gamma \to  q \bar{q} \, \ell^{-}\bar{\nu}_{\ell} \gamma$ &\bsmoptionsBLHA{LV}\\*
\bf 462 & $p \overset{\mbox{\tiny{(--)}}}{p} \to W^{+}W^{-} \gamma \to \ell^{+} \nu_{\ell} \, q \bar{q} \, \gamma$ &\bsmoptionsBLHA{LV}\\*
\bf 470 & $p \overset{\mbox{\tiny{(--)}}}{p} \to Z Z \gamma \to \ell_{1}^{-} \ell_{1}^{+} \ell_{2}^{-} \ell_{2}^{+} \gamma$ &\bsmoptionsBLHA{BV}\\*
\bf 471 & $p \overset{\mbox{\tiny{(--)}}}{p} \to Z Z \gamma \to \ell^{-} \ell^{+} \, q \bar{q} \, \gamma$ &\bsmoptionsBLHA{LV}\\*
\bf 472 & $p \overset{\mbox{\tiny{(--)}}}{p} \to Z Z \gamma \to \ell_{1}^{-} \ell_{1}^{+} \nu_{\ell_2} \bar{\nu}_{\ell_2} \gamma$ &\bsmoptionsBLHA{B}\\*
\bf 480 & $p \overset{\mbox{\tiny{(--)}}}{p} \to W^{+} Z \gamma \to \ell_{1}^{+}\nu_{\ell_1} \ell_{2}^{-} \ell_{2}^{+} \gamma$ &\bsmoptionsBLHA{BV}\\*
\bf 481 & $p \overset{\mbox{\tiny{(--)}}}{p} \to W^{+} Z \gamma \to  q \bar{q} \, \ell^{-} \ell^{+} \gamma$ &\bsmoptionsBLHA{LV}\\*
\bf 482 & $p \overset{\mbox{\tiny{(--)}}}{p} \to W^{+} Z \gamma \to \ell^{+}\nu_{\ell} \, q \bar{q} \, \gamma$ &\bsmoptionsBLHA{LV}\\*
\bf 490 & $p \overset{\mbox{\tiny{(--)}}}{p} \to W^{-} Z \gamma \to \ell_{1}^{-} \bar{\nu}_{\ell_1} \ell_{2}^{-} \ell_{2}^{+} \gamma$ &\bsmoptionsBLHA{BV}\\*
\bf 491 & $p \overset{\mbox{\tiny{(--)}}}{p} \to W^{-} Z \gamma \to  q \bar{q} \, \ell^{-} \ell^{+} \gamma$ &\bsmoptionsBLHA{LV}\\*
\bf 492 & $p \overset{\mbox{\tiny{(--)}}}{p} \to W^{-} Z \gamma \to \ell^{-} \bar{\nu}_{\ell} \, q \bar{q} \, \gamma$ &\bsmoptionsBLHA{LV}\\*
\bf 500 & $p \overset{\mbox{\tiny{(--)}}}{p} \to W^{+} \gamma \gamma \to {\ell}^{+}\nu_{\ell} \gamma \gamma$ &\bsmoptionsBLHA{BV}\\*
\bf 510 & $p \overset{\mbox{\tiny{(--)}}}{p} \to W^{-} \gamma \gamma \to {\ell}^{-} \bar{\nu}_{\ell} \gamma \gamma$ &\bsmoptionsBLHA{BV}\\*
\bf 520 & $p \overset{\mbox{\tiny{(--)}}}{p} \to Z \gamma \gamma \to {\ell}^{-} {\ell}^{+} \gamma \gamma$ &\bsmoptionsBLHA{BV}\\*
\bf 521 & $p \overset{\mbox{\tiny{(--)}}}{p} \to Z \gamma \gamma \to \nu_{\ell} \bar{\nu}_{\ell} \gamma \gamma$ &\bsmoptionsBLHA{BV}\\*
\bf 530 & $p \overset{\mbox{\tiny{(--)}}}{p} \to \gamma \gamma \gamma $ &\bsmoptionsBLHA{B}\\*
&\\*
\hline
&\\*
\bf 800 & $p \overset{\mbox{\tiny{(--)}}}{p}  \to W^{+} \gamma \gamma j  \to \ell^{+} \nu_{\ell} \gamma \gamma j $ &\bsmoptionsBLHA{V}\\*
\bf 810 & $p \overset{\mbox{\tiny{(--)}}}{p}  \to W^{-} \gamma \gamma j \to \ell^{-} \bar \nu_{\ell} \gamma \gamma j $ &\bsmoptionsBLHA{V}\\*
&\\*
\hline
\end{longtable}
}

\clearpage
The gluon-fusion processes are given below.

{
\footnotesize
\setlength\LTleft{0pt plus \textwidth minus \textwidth}
\setlength\LTright{0pt plus \textwidth minus \textwidth}
\begin{longtable}{clccccccccc}

\textsc{ProcId} & \textsc{Process} &\rot{BLHA} &  \rot{gluon-fusion process} & \rot{semi-leptonic decay} & \rot{anom.\ Higgs couplings} & \rot{general 2HDM} & \rot{MSSM}\\
&\\
\hline
\endhead
&\\*
\bf 4100 & $p \overset{\mbox{\tiny{(--)}}}{p} \to H \, jj $ &\bsmgfoptionsBLHA{GTM}\\*
\bf 4101 & $p \overset{\mbox{\tiny{(--)}}}{p} \to H \, jj\to \gamma\gamma \, jj$ &\bsmgfoptionsBLHA{GM}\\*
\bf 4102 & $p \overset{\mbox{\tiny{(--)}}}{p} \to H \, jj\to \mu^+\mu^- \, jj$ &\bsmgfoptionsBLHA{GM}\\*
\bf 4103 & $p \overset{\mbox{\tiny{(--)}}}{p} \to H \, jj\to \tau^+\tau^- \, jj$ &\bsmgfoptionsBLHA{GM}\\*
\bf 4104 & $p \overset{\mbox{\tiny{(--)}}}{p} \to H \, jj\to b\bar{b} \, jj$ &\bsmgfoptionsBLHA{GM}\\*
\bf 4105 & $p \overset{\mbox{\tiny{(--)}}}{p} \to H \, jj\to W^{+}W^{-} \, jj\to \ell_{1}^+\nu_{\ell_{1}} \ell_{2}^- \bar{\nu}_{\ell_{2}} \,jj$ &\bsmgfoptionsBLHA{GHTM}\\*
\bf 4106 & $p \overset{\mbox{\tiny{(--)}}}{p} \to H \, jj\to ZZ \, jj\to \ell_{1}^+ \ell_{1}^- \ell_{2}^+ \ell_{2}^- \,jj$ &\bsmgfoptionsBLHA{GHTM}\\*
\bf 4107 & $p \overset{\mbox{\tiny{(--)}}}{p} \to H \, jj\to ZZ \, jj\to \ell_{1}^+ \ell_{1}^- \nu_{\ell_{2}}  \bar{\nu}_{\ell_{2}} \,jj$ &\bsmgfoptionsBLHA{GHTM}\\*
&\\*
\hline
&\\
\bf 4200 & $p \overset{\mbox{\tiny{(--)}}}{p} \to H \, jjj $ &\bsmgfoptionsBLHA{GTM}\\*
&\\
\hline
&\\*
\bf 4300 & $p \overset{\mbox{\tiny{(--)}}}{p} \to W^{+}W^{-} \to \ell_{1}^{+} \nu_{\ell_{1}} \ell_{2}^{-}\bar{\nu}_{\ell_{2}} $ &\bsmgfoptionsBLHA{GH}\\*
\bf 4301 & $p \overset{\mbox{\tiny{(--)}}}{p} \to W^{+}W^{-} \to q \bar{q} \, \ell^{-}\bar{\nu}_{\ell} $ &\bsmgfoptionsBLHA{GHL}\\*
\bf 4302 & $p \overset{\mbox{\tiny{(--)}}}{p} \to W^{+}W^{-} \to \ell^{+} \nu_{\ell} \, q \bar{q} $ &\bsmgfoptionsBLHA{GHL}\\*
\bf 4330 & $p \overset{\mbox{\tiny{(--)}}}{p} \to ZZ \to \ell_{1}^{-} \ell_{1}^{+}  \ell_{2}^{-} \ell_{2}^{+} $ &\bsmgfoptionsBLHA{GH}\\*
\bf 4331 & $p \overset{\mbox{\tiny{(--)}}}{p} \to ZZ \to q \bar{q} \, \ell^{-} \ell^{+} $ &\bsmgfoptionsBLHA{GHL}\\*
\bf 4360 & $p \overset{\mbox{\tiny{(--)}}}{p} \to Z\gamma \to \ell_{1}^{-} \ell_{1}^{+}  \gamma $ &\bsmgfoptionsBLHA{GH}\\*
\bf 4370 & $p \overset{\mbox{\tiny{(--)}}}{p} \to \gamma\gamma $ &\bsmgfoptionsBLHA{GH}\\*
&\\*
\hline
&\\*
\bf 4600 & $p \overset{\mbox{\tiny{(--)}}}{p} \to W^{+}W^{-} j \to \ell_{1}^{+} \nu_{\ell_{1}} \ell_{2}^{-}\bar{\nu}_{\ell_{2}} j $ & \bsmgfoptionsBLHA{GH}\\*
\bf 4601 & $p \overset{\mbox{\tiny{(--)}}}{p} \to W^{+}W^{-} j \to q\bar{q} \ell^{-}\bar{\nu}_{\ell} j $ & \bsmgfoptionsBLHA{GL}\\*
\bf 4602 & $p \overset{\mbox{\tiny{(--)}}}{p} \to W^{+}W^{-} j \to \ell^{+} \nu_{\ell} q\bar{q} j $ & \bsmgfoptionsBLHA{GL}\\*
\bf 4650 & $p \overset{\mbox{\tiny{(--)}}}{p} \to ZZ j \to \ell_{1}^{-} \ell_{1}^{+}  \ell_{2}^{-} \ell_{2}^{+} j $ & \bsmgfoptionsBLHA{GH}\\*
&\\* 
\hline
\end{longtable}
}


\newpage


\providecommand{\href}[2]{#2}

\end{document}
